\documentclass[12pt]{article}
\pdfoutput=1
\usepackage{myjheppub}
\usepackage{graphicx,booktabs,latexsym,amssymb,amsmath,bm,epsfig,psfrag,MnSymbol,float}
\hypersetup{pdfnewwindow=true}

\preprint{
\begin{flushright}	
	PSI-PR-15-09,
	TUM-HEP-1019/15,\\
	ZU-TH 33/15,
	DCPT/15/122,\\
	IPPP/15/61
\end{flushright}}

\title{Associated production of a top pair and a Higgs boson beyond NLO}

\author[a,b]{Alessandro Broggio,}
\author[c,d]{Andrea Ferroglia,}
\author[e]{Ben D. Pecjak,}
\author[a,f]{Adrian Signer,}
\author[g,h,i]{and Li Lin Yang}

\affiliation[a]{Paul Scherrer Institut,
CH-5232 Villigen PSI, Switzerland}
\affiliation[b]{Physik Department T31, Technische Universit\"at M\"unchen,
James Franck-Stra{\ss}e 1, D-85748 Garching, Germany}
\emailAdd{alessandro.broggio@tum.de}

\affiliation[c]{Physics Department, New York City College of Technology,
Brooklyn, NY 11201 USA}

\affiliation[d]{The Graduate School and University Center,
The City University of New York,
New York, NY 10016  USA}
\emailAdd{aferroglia@citytech.cuny.edu}

\affiliation[e]{Institute for Particle Physics Phenomenology, University of Durham,
DH1 3LE Durham, United Kingdom}
\emailAdd{ben.pecjak@durham.ac.uk}

\affiliation[f]{Physik-Institut, Universit\"at Z\"urich, 
Winterthurerstrasse 190, CH-8057 Z\"urich, Switzerland}
\emailAdd{adrian.signer@psi.ch}

\affiliation[g]{School of Physics and State Key Laboratory of Nuclear Physics and Technology,
Peking University, Beijing 100871, China}
\affiliation[h]{Collaborative Innovation Center of Quantum Matter, Beijing, China}
\affiliation[i]{Center for High Energy Physics, Peking University, Beijing 100871, China}
\emailAdd{yanglilin@pku.edu.cn}

\abstract{We consider soft gluon emission corrections to the
  production of a top-antitop pair in association with a Higgs boson
  at hadron colliders. In particular, we present a soft-gluon
  resummation formula for this production process and gather all
  elements needed to evaluate it at next-to-next-to-leading
  logarithmic order. We employ these results to obtain approximate
  next-to-next-to-leading order (NNLO) formulas, and implement them in
  a bespoke parton-level Monte Carlo program which can be used to
  calculate the total cross section along with arbitrary differential
  distributions.  We use this tool to study the phenomenological
  impact of the approximate NNLO corrections, finding that they
  increase the total cross section and the differential distributions
  which we evaluated in this work.}

\begin{document}
\newcommand{\Red}[1]{\textcolor{red}{#1}}
\newcommand{\Green}[1]{\textcolor{green}{#1}}
\newcommand{\Blue}[1]{\textcolor{blue}{#1}}
\newcommand{\Cyan}[1]{\textcolor{cyan}{#1}}
\newcommand{\Magenta}[1]{\textcolor{magenta}{#1}}
\newcommand{\alert}[1]{{\bf \Red{\boldmath[ #1 ]\unboldmath}}}

\newcommand{\be}{\begin{equation}}
\newcommand{\ee}{\end{equation}}

\newcommand{\nn}{\nonumber}
\def\ff{f\hspace{-0.3cm}f}

\maketitle


\section{Introduction}

The search for events in which a Higgs boson is produced in
association with a top-antitop quark pair ($t\bar{t} H$ production)
will be one of the experimental goals of Run 2 of the Large Hadron
Collider (LHC). While the Standard Model cross section for this
process is quite small ($\sim 0.6$~pb at a center of mass energy of
$14$~TeV), it provides important information about the coupling
between top quarks and Higgs bosons and, consequently, its measurement
can place severe constraints on Beyond the Standard Model
scenarios. It is therefore important to have precise theoretical
predictions for this process within the Standard Model.

The leading-order (LO) cross section for $t\bar{t}H$ production scales
as ${\cal O}(\alpha_s^2 \, \alpha)$.  The status of higher-order
perturbative calculations is as follows.  Next-to-leading-order (NLO)
QCD corrections to this process were evaluated by two different groups
in 2001-2003 \cite{Beenakker:2001rj, Beenakker:2002nc, Reina:2001sf,
  Reina:2001bc, Dawson:2002tg, Dawson:2003zu}. The calculation of
these corrections was repeated a few years ago with tools for the
automated calculation of NLO corrections \cite{Frederix:2011zi,
  Garzelli:2011vp}; in both papers the NLO corrections were interfaced
with parton showers and hadronization effects were studied. The weak
corrections of $\mathcal{O}(\alpha_s^2\, \alpha^2)$ were evaluated in
\cite{Yu:2014cka} (where also the QED corrections were considered) and
\cite{Frixione:2014qaa}. An additional study of the strong and
electroweak corrections to the associated production of a top-quark
pair and a massive boson ($Z$, $W$ or Higgs) was recently carried out
in \cite{Frixione:2015zaa}.   The NLO QCD and electroweak
corrections have been included in the {\tt POWHEG BOX} framework
\cite{Hartanto:2015uka}. Recently, a study of $t\bar{t}H$ production going
beyond stable top quarks was presented in \cite{Denner:2015yca}, where
differential cross sections have been computed, including the decay
of the top quarks as well as off-shell effects. Finally, the
soft gluon emission corrections to the total $t\bar{t} H$ cross
section in the production threshold limit, where the partonic
center-of-mass energy approaches $2m_t+m_H$, were evaluated up
next-to-leading logarithmic (NLL) accuracy, and they were matched to
NLO calculations \cite{Kulesza:2015vda}.

In this paper we add to the above literature by approximating the NNLO
QCD corrections to the total and differential $t\bar{t}H$ cross
sections using soft-gluon resummation. In contrast to
\cite{Kulesza:2015vda}, we study soft-gluon corrections in the limit
where the partonic center-of-mass energy approaches the invariant mass
of the $t\bar{t}H$ final state, which in turn can be arbitrarily
large.  This limit is well-suited for the calculation of differential
cross sections in addition to the total one.  It is the exact analogue
of the so-called pair-invariant mass (PIM) threshold limit used to
study top-quark pair production at NNLL and approximate NNLO in
\cite{Ahrens:2010zv}, and we will use this nomenclature throughout the
paper.  We obtain the approximate NNLO corrections from the
perturbative information contained in a soft-gluon resummation formula
valid to next-to-next-to-leading logarithmic (NNLL) accuracy.  The
derivation of this formula is based on the soft-collinear effective
theory (SCET) methods\footnote{See \cite{Becher:2014oda} for a first
  introduction to SCET.} used to study differential top-quark pair
production ($t\bar{t}$) cross sections at NNLL in \cite{Ahrens:2010zv,
  Ahrens:2011mw, Ahrens:2011uf} (see \cite{Beneke:2010da,
  Beneke:2011mq, Beneke:2012wb, Ahrens:2011px} for SCET-based studies
of the total $t\bar{t}$ cross section).  In fact, since both
$t\bar{t}$ and $t\bar{t}H$ production contain four colored partons,
the study of soft-gluon corrections to these processes is conceptually
identical and differs only because of the underlying kinematics.  In
particular, the soft-gluon resummation formula for both processes
contains three essential ingredients, all of which are matrices in the
color space needed to describe four-parton scattering: 1) a hard
function, related to virtual corrections; 2) a soft function, related
to real emission corrections in the soft limit; and 3) a soft
anomalous dimension, which governs the structure of certain all-order
soft-gluon corrections through the renormalization group (RG).

Of these three ingredients, both the NLO soft function
\cite{Ahrens:2010zv, Li:2014ula} and NLO soft anomalous dimension
\cite{Ferroglia:2009ep, Ferroglia:2009ii} needed for NNLL resummation
in processes involving two massless and two massive partons were
calculated in such a way that they are valid for arbitrary kinematics
(i.e. they do not use momentum conservation particular to two-to-two
kinematics) and can thus be adapted directly to $t\bar{t}H$ production or
indeed any $t\bar{t}$ production process in association with an
additional uncolored particle.  We perform such an adaptation here,
and find agreement with results obtained previously for $t\bar{t} W$
production in \cite{Li:2014ula}. The main technical challenge to
obtaining results at NNLL accuracy is thus the calculation of the hard
function to NLO, which unlike the soft function and soft anomalous
dimension is process dependent. We carry out this calculation for the
first time here, using a modified version of the one-loop providers
{\tt GoSam} \cite{Cullen:2011ac,Cullen:2014yla}, {\tt Openloops}
\cite{Cascioli:2011va} and {\tt MadLoop} \cite{Hirschi:2011pa}.  Our
result thus adds to the growing literature on hard functions for $2\to
3$ processes, i.e.  those obtained for $t\bar{t}W$ \cite{Li:2014ula}
production using {\tt MadLoop} and related calculations for massless
$2\to 3$ scattering presented in \cite{Farhi:2015jca}.  Our procedure
can be easily modified to include other $2\to 3$ processes.

Our results are formally valid at NNLL for differential distributions
in regions of phase space where the PIM soft limit is respected, which
is guaranteed to be the case only when the partonic center-of-mass
energy approaches the collider threshold energy.  However, due to the
mechanism of dynamical threshold enhancement \cite{Becher:2007ty,
  Ahrens:2010zv}, it is often the case that also observables sensitive
to other regions of phase space receive their dominant contributions
from soft-gluon corrections derived in the PIM threshold limit.
Obvious examples would be the cross section differential in the
invariant mass of the $t\bar{t}H$ final state at values far away from
the machine threshold, or the total cross section obtained by
integrating this distribution.  Moreover, given that results in the
PIM threshold limit are fully differential in the Mandelstam variables
characterizing the Born process, we can equally well use them to
estimate the NNLO corrections to any differential distribution which
is non-vanishing at Born level.

We take advantage of this fact in the present work by implementing our
results in an in-house parton level Monte Carlo, which can be used to
calculate arbitrary $t\bar{t}H$ differential distributions along with
the total cross section. To illustrate its use, we study approximate
NNLO corrections to the $p_T$ of the Higgs, the $p_T$ of the top
quark, the invariant mass of the $t\bar{t}$ pair, and the rapidities
of the top quark or Higgs boson, in addition to the total cross
section and differential cross section with respect to the $t\bar{t}H$
final state.  By matching our NNLO approximation in the PIM threshold
limit with the complete NLO calculation from
{\verb|MadGraph5_aMC@NLO|} \cite{Alwall:2014hca}, we obtain the
currently most complete result for QCD corrections to differential
$t\bar{t}H$ cross sections.  Such a procedure is very much in the
spirit of \cite{Broggio:2014yca}, and as in that work could be
extended to include the effects of top-quark decays by retaining
information on the spins of the final state particles.

The paper is organized as follows: in Section~\ref{sec:soft-limit} we
review the factorization properties of the partonic cross section in
the soft emission limit.  Furthermore, we discuss the evaluation of
the various components which contribute to the approximate NNLO
formulas derived in this work.  In Section~\ref{sec:approxNNLO} we
illustrate the structure of the approximate NNLO formulas obtained by
considering the soft limit of the partonic cross section.
Section~\ref{sec:pheno} contains numerical calculations of the total
$t \bar{t} H$ production cross section and of some differential
distributions for the LHC operating at center of mass energy of
13~TeV. The calculations include the approximate NNLO formulas
discussed in this work as well as the full set of NLO QCD
corrections. The residual perturbative uncertainty affecting these
results is discussed. Finally, we present our conclusions in
Section~\ref{sec:conclusions}.

\section{Soft-gluon resummation for $t\bar{t}H$ hadroproduction}
\label{sec:soft-limit}

We consider the partonic processes
\begin{align}
i(p_1) + j(p_2) &\longrightarrow
t(p_3) + \bar{t}(p_4) + H(p_5) + X \, ,
\end{align}
where the incoming partons $i,j\in \{q,\bar{q},g\}$ and $X$ is a
partonic final state. Furthermore, we define the Mandelstam invariants 
\begin{gather}
\hat{s} = (p_1 + p_2)^2 = 2 p_1 \cdot p_2 \, , 
\qquad \tilde{s}_{ij} = 2 p_i \cdot p_j
\, , (i=1,2\,; j=3,4) \, , 
\nn \\
s_{34} = (p_3 + p_4)^2 = \tilde{s}_{34} + 2 m_t^2 \,.
\end{gather}
The invariant mass of the $t\bar{t}H$ final state, 
\begin{align}
M^2 = (p_3 + p_4 + p_ 5)^2 \, ,
\end{align}
is of particular relevance to our work, since it enters in the
definition of the soft parameter $z$
\begin{align}
z = \frac{M^2}{\hat{s}}\, .
\end{align}
The PIM threshold limit (or, more simply, the soft limit) mentioned in the
introduction is defined as the limit where $z \to 1$, such that the
unobserved final state $X$ consists of soft partons only.  Note that,
in contrast to the production threshold limit, where the partonic
center-of-mass energy approaches $2m_t+m_H$, the PIM threshold limit
does not impose constraints on the velocity of massive particles in
the final state.  It is therefore well suited for the study of
differential cross sections.

The starting point for soft-gluon resummation is the factorization of
the partonic cross section in the soft limit.  One then obtains the
hadronic cross section for the collision process involving nucleons
$N_1$ and $N_2$ at center-of-mass energy $\sqrt{s}$ by the usual convolution
integral with parton distribution functions (PDFs).  The form of the factorization
of QCD corrections in the soft limit in the $t\bar{t}H$ case is identical
to the $t\bar{t}$ one, so we can simply quote the result for the cross
section in the soft limit by adapting that obtained for $t\bar{t}$
production using SCET methods in \cite{Ahrens:2010zv}. We write the
result for the total cross section as 
\begin{align}
\label{eq:soft-fact}
\sigma(s,m_t,m_H) &=\frac{1}{2 s} \int_{\tau_{\text{min}}}^1 \frac{d \tau}{\tau} \int_{\tau}^1 \frac{dz}{\sqrt{z}} \,
\sum_{ij} \, \ff_{ij}\left(\frac{\tau}{z},\mu\right)  \nonumber \\
& \quad \int  d \text{PS}_{t \bar{t} H} \text{Tr} \left[\mathbf{H}_{ij}(\{p_i\},\mu)  \, \mathbf{S}_{ij}\left( \frac{M(1-z)}{\sqrt{z}}, \{p_i\},\mu\right)\right] + {\cal O}(1-z) \, ,
\end{align}
where 
\begin{align}
\tau = \frac{M^2}{s} \, , \qquad \tau_{\rm min} = \frac{(2m_t+m_H)^2}{s} \, .
\end{align}
The content and notation of (\ref{eq:soft-fact}) is as follows.
First, the object ${\rm Tr}[\mathbf{H}_{ij} \mathbf{S}_{ij}]$ is proportional
to the spin and color averaged squared matrix element for
$t\bar{t}H+X_s$ production through two initial-state partons with
flavors $i$ and $j$, where $X_s$ is an unobserved final state
consisting of any number of soft gluons. The (matrix valued) hard
functions $\mathbf{H}_{ij}$ are related to color decomposed virtual
corrections to the underlying $2\to 3$ scattering process, and the
(matrix valued) soft functions $\mathbf{S}_{ij}$ are related to
color-decomposed real emission corrections in the soft limit. To
leading order in the soft limit, these soft real emission corrections
receive contributions from initial-state partons with flavor indices
$ij\in\{q\bar{q},\bar{q}q,gg\}$; throughout this work we will refer to the 
channels involving quarks with the generic term ``quark annihilation" channel, 
and the one involving gluons as the ``gluon fusion" channel.  Channels involving 
initial-state partons such as $qg$ and $\bar{q}g$ are subleading in the soft limit,
and shall be referred to generically as the ``$qg$" channel.  While the hard functions are simple
functions of their arguments, the soft functions depend on singular
(logarithmic) plus distributions of the form
\begin{equation}
P'_n(z) \equiv \left[\frac{1}{1-z}\ln^n \left(\frac{M^2(1-z)^2}{\mu^2 z}\right) \right]_+ \, ,
\end{equation}
as well as the Dirac delta function $\delta(1-z)$.

Second, the parton luminosity function is given by 
\begin{align}
\label{eq:ff}
\ff_{ij}\left(y, \mu\right) \equiv \int_y^1 \frac{dx_1}{x_1}   f_{i/N_1} (x_1,\mu) f_{j/N_2}\left(\frac{y}{x_1}, \mu \right) \, ,
\end{align}
where $f_{i/N}$ is the parton distribution function for parton with
flavor $i$ in nucleon $N$.

Finally, we write the phase-space integral for the $t\bar{t}H$ final
state in the soft limit (which is identical to the Born-level phase
space except that the total energy available is reduced from
$\sqrt{\hat{s}}$ to $M$ due to soft gluon emissions) as
\begin{align}
\label{eq:phase-space}
d \text{PS}_{t \bar{t} H} = & \frac{d^3 \vec{p}_3}{(2 \pi)^3 2 E_3} \frac{d^3 \vec{p}_4}{(2 \pi)^3 2 E_4}  \frac{d^3 \vec{p}_5}{(2 \pi)^3 2 E_5}
(2 \pi)^4 \delta^{(4)} (\bar{p}_1 +\bar{p}_2 -p_3 -p_4 -p_5) \,  \nn \\
= &
\frac{1}{(2\pi)^5}\frac{\kappa(s_{34},m_t^2,m_t^2)}{8 s_{34}}
\frac{\kappa(M^2, s_{34},m_h^2)}{8 M^2}
\Theta(s_{34} - 4m_t^2)
\nonumber \\ 
&\times \Theta\left((M-m_h)^2-s_{34}\right)
ds_{34}d\Omega_t^* d\Omega_H \, ,
\end{align}
where $\kappa$ is the K\'allen function
\begin{align}
\kappa(x,y,z)= \sqrt{x^2+y^2+z^2- 2 x y - 2 x z - 2 y z} \,.
\end{align}
The differential of the solid angle of the Higgs boson direction in
the laboratory frame is indicated by $d \Omega_H = d\cos\theta_H d
\phi_H$, while $\Omega_t^*$ is the solid angle of the top quark in the
$t \bar{t}$ rest frame.  The vectors $\bar{p}_1$ and $\bar{p}_2$ are
reduced momenta defined in such a way that
$(\bar{p}_1+\bar{p}_2)^2=M^2$.

 In order to calculate binned
differential cross sections using Monte-Carlo techniques we need
explicit parameterizations of the four-momenta $\bar{p}_1\dots p_5$ in
terms of the integration variables in (\ref{eq:phase-space}).  The
vectors $\bar{p}_1$ and $\bar{p}_2$ in the partonic center-of-mass
frame are written as
\begin{align}
  \bar{p}_1 = \frac{M}{2}\left(1, 0, 0, 1 \right) \, , \qquad
  \bar{p}_2 = \frac{M}{2}\left(1, 0, 0, - 1\right) \,,
\end{align}
where we took the $z$ axis to be in the direction of the incoming proton
$N_1$.  The top and antitop vectors in the $t\bar{t}$ rest frame can be
written as
\begin{align}
  p_3^* = & \left(E_3^*, k_3^* \sin\theta_t^* \cos\phi_t^*, k_3^*
    \sin\theta_t^* \sin\phi_t^*,
    k_3^* \cos\theta_t^* \right) \, ,\nonumber \\
  p_4^* = & \left(E_3^*,- k_3^* \sin\theta_t^* \cos\phi_t^*, -k_3^*
    \sin\theta_t^* \sin\phi_t^*, - k_3^* \cos\theta_t^* \right) \,
  , \label{eq:tbartmom}
\end{align}
where
\begin{align}
  E_3^* = \frac{\sqrt{s_{34}}}{2} \, , \quad
  k_3^* = \frac{\kappa(s_{34},m_t^2,m_t^2)}{2 \sqrt{s_{34}}} \, .
\end{align}
One can boost the top and antitop momenta in
(\ref{eq:tbartmom}) to the partonic center-of-mass frame by
using that the relative velocity between the two frames points
along the direction of flight of the $t\bar{t}$-pair in the
partonic rest frame and has magnitude $k^*_{12}/E_{12}^*$,
where $k_{12}^*$ and $E_{12}^*$ are the magnitudes of the
three-momentum and energy of the incoming parton pair in the
the $t\bar{t}$ rest frame, respectively:
\begin{align}
  E_{12}^*=\frac{M^2+s_{34}-m_H^2}{2 \sqrt{s_{34}}} \, , \quad
  k_{12}^*= \frac{\kappa(M^2,s_{34},m_H^2)}{2 \sqrt{s_{34}}} \,.
\end{align}
The four momentum of the Higgs boson $p_5$ can be easily written in
the partonic center-of-mass frame, in which the Higgs boson recoils against the
$t \bar{t}$ pair:
\begin{align}
p_5 =  & \left(E_{5} ,k_5 \sin \theta_H \cos \phi_H, k_5 \sin \theta_H \sin \phi_H, k_5 \cos \theta_H \right) \, ,
\end{align}	
with
\begin{align}
E_5 = \frac{M^2-s_{34} +m_H^2}{2 M} \, , \quad
k_5 = \frac{\kappa\left(M^2,s_{34},m_H^2\right)}{2 M} \, .
\end{align}
For the rapidity distributions we also need the momenta of the top quarks and Higgs boson in
the laboratory frame.  In order to implement the required boost we use
that the relative velocity between the partonic center-of-mass frame
and the laboratory frame is parallel to $p_1$ and has magnitude
$(x_1-x_2)/(x_1+x_2)$, where $x_1$ is the integration variable in the
definition of the luminosity (\ref{eq:ff}), and $x_2 = \tau/(z x_1)$,
with $\tau$ and $z$ the integration variables in (\ref{eq:soft-fact}).

We should emphasize that the factors of $\sqrt{z}$ in
(\ref{eq:soft-fact}) arise by isolating and keeping exact dependence
on the parton center-of-mass energy $\sqrt{\hat{s}} = M/\sqrt{z}$
during two steps in the derivation of the factorized differential
cross section. Both are related to the identification of the Fourier
transform of the position-space soft function defined in terms of
Wilson loops with the momentum-space object quoted in
(\ref{eq:soft-fact}), and can be understood by examining Eq.~(55) of
\cite{Ahrens:2010zv}.  The first of these is based on the observation
that the Fourier transform of the position-space soft function depends
on the total energy of the radiated soft partons in the center-of-mass
frame, which is equal to $\sqrt{\hat{s}}(1-z)$.  This explains the
form of the first argument of the soft functions $\mathbf{S}_{ij}$, and
keeping this $\sqrt{z}$ dependence is the $t\bar{t}H$ production
equivalent of the PIM$_{\rm SCET}$ scheme defined in
\cite{Ahrens:2010zv,Ahrens:2011mw} for $t\bar{t}$ production.  A
second factor of $\sqrt{\hat{s}}$ appears as an overall prefactor
between the position-space and Fourier-transformed functions and
explains the factor of $1/\sqrt{z}$ in the first line of
(\ref{eq:soft-fact}).\footnote{In \cite{Ahrens:2010zv} this particular
  prefactor of $\sqrt{\hat{s}}$ was instead set to $M$, so the
  prefactor in that work is $1/z$ instead of $1/\sqrt{z}$ as in
  (\ref{eq:soft-fact}).}  Since factorization in the soft limit is
valid as $z\to 1$, we could equally well set both of these factors of
$\sqrt{z}$ to unity, but we prefer to keep them as written since they
appear ``naturally" during the derivation and potentially account for
numerically important power corrections away from the soft limit.  We
study numerical ambiguities due to the prescription used for these
terms  in Section~\ref{sec:pheno}.

The final perturbative ingredient needed for soft-gluon resummation is
the soft anomalous dimension $\mathbf{\Gamma}_H$.  We define it through
the RG equation for the hard function, which reads (suppressing for
the moment the dependence on the channel $ij$)
\begin{align}
\frac{d}{d \ln \mu} \mathbf{H}\left(\{p_i\},\mu \right) = 
\mathbf{\Gamma}_H  \mathbf{H}\left(\{p_i\},\mu \right) + 
\mathbf{H}\left(\{p_i\},\mu \right)  
\mathbf{\Gamma}_H^\dagger \, ,
\label{eq:HRGE}
\end{align}
where $\mathbf{\Gamma}_H \equiv \mathbf{\Gamma}_H(\{p_i\},\mu)$.  The hard
function, soft function, and soft anomalous dimension in a given
production channel all have perturbative expansions in $\alpha_s$.  In
order to perform soft-gluon resummation at NNLL, one needs their
perturbative expansions to NLO.  We end this section by explaining how
we have extracted or calculated each of these NLO functions.

The results for the soft function and soft anomalous dimension to this
order can be read off from results in the literature.  The main step
in the calculation of the NLO soft function is obtaining the
phase-space type integrals ${\mathcal I}_{ij}$, defined as
\begin{equation}
{\mathcal I}_{ij}(\epsilon, x^0, \mu) = -\frac{\left(4 \pi \mu^2 \right)^\epsilon}{\pi^{(2-\epsilon)}} v_i \cdot v_j \int d^d k \frac{e^{- i k^0 x^0}}{v_i \cdot k \, v_j \cdot k}\,  (2 \pi) \theta\left(k^0 \right) \delta \left(k^2\right) \, ,
\end{equation}
where $v_i$ is the velocity vector of the parton carrying momentum
$p_i$.  These have been calculated in \cite{Li:2014ula}, and we have
performed an independent calculation and found complete agreement with
those results.  Rather than collecting the explicit results here, we
refer the reader to the list in Eq.~(33) of \cite{Li:2014ula}. Most of
the notation from that equation matches ours directly, and we
furthermore identify $\theta_3$ and $\beta_3$ with
\begin{align}
\cos \theta_3 &= \frac{\tilde{s}_{23}-\tilde{s}_{13}}{\sqrt{(\tilde{s}_{23} + \tilde{s}_{13})^2 - 4 m_t^2 \hat{s}}} \, , \qquad
\beta_3 = \sqrt{1 - \frac{4 m_t^2 \hat{s}}{(\tilde{s}_{23} + \tilde{s}_{13})^2}} \, ,
\end{align}
and similarly for $\beta_4$ and $\theta_4$ after obvious replacements.
The position space (or, after trivial substitutions, Laplace space)
soft function itself is then formed by calculating the ${\mathcal
  I}_{ij}$ for all possible attachments of gluons to partons $ij$ and
associating to each attachment a color matrix particular to the
partonic production channel. The momentum space function in
(\ref{eq:soft-fact}) is obtained through an integral transform; all
details related to the color matrices and integral transforms can be
found in \cite{Ahrens:2010zv}, and we shall not reproduce them here.

The soft anomalous dimensions in the $q\bar{q}$ channel and $gg$
channel were calculated to NLO in \cite{Ferroglia:2009ep,
  Ferroglia:2009ii}.  In our notation the results are
\begin{align}
\mathbf{\Gamma}_H^{q \bar{q}} &= \Bigl[C_F 
\gamma_{\text{cusp}} (\alpha_s) \left( \ln{\frac{\hat{s}}{\mu^2}} - i \pi\right) + C_F \gamma_{\text{cusp}}\left(\beta_{34}, \alpha_s \right) + 2 \gamma^q \left( \alpha_s\right)+ 2 \gamma^Q \left( \alpha_s\right)\Bigr] \mathbf{1} \nn \\
& +\frac{N_C}{2} \left[
\gamma_{\text{cusp}}(\alpha_s)
\left(\frac{1}{2}
\ln{\frac{\tilde{s}_{13}^2}{\hat{s} m_t^2}}
+
\frac{1}{2}
\ln{\frac{\tilde{s}_{24}^2}{\hat{s} m_t^2}}
+i \pi \right)
-\gamma_{\text{cusp}}(\beta_{34},\alpha_s)
 \right] \left(\begin{array}{cc}
 0 & 0  \\ 
 0 & 1
 \end{array}  \right) 
 \nn \\
&+ \gamma_{\text{cusp}}(\alpha_s)
 \left(\ln\frac{\tilde{s}_{13}}{\tilde{s}_{23}}
 -\ln\frac{\tilde{s}_{14}}{\tilde{s}_{24}}
  \right) \left[\left(\begin{array}{cc}
  0 & \frac{C_F}{2 N_C} \\ 
  1 & -\frac{1}{N_C} 
  \end{array}  \right)+
  \frac{\alpha_s}{4 \pi} g \left(\beta_{34}
  \right) 
\left(\begin{array}{cc}
0 & \frac{C_F}{2}  \\ 
- N_C & 0
\end{array}  \right)  
  \right] \,.
\end{align}
and
\begin{align}
\mathbf{\Gamma}_H^{gg} &= \Bigl[N_C 
\gamma_{\text{cusp}} (\alpha_s) \left( \ln{\frac{\hat{s}}{\mu^2}} - i \pi\right) + C_F \gamma_{\text{cusp}}\left(\beta_{34}, \alpha_s \right) + 2 \gamma^g \left( \alpha_s\right) +2 \gamma^Q \left( \alpha_s\right)\Bigr] \mathbf{1} \nn \\
& +\frac{N_c}{2} \left[
\gamma_{\text{cusp}}(\alpha_s)
\left(\frac{1}{2}
\ln{\frac{\tilde{s}_{13}^2}{\hat{s} m_t^2}}
+
\frac{1}{2}
\ln{\frac{\tilde{s}_{24}^2}{\hat{s} m_t^2}}
+i \pi \right)
-\gamma_{\text{cusp}}(\beta_{34},\alpha_s)
 \right] \left(\begin{array}{ccc}
 0 & 0 & 0 \\ 
 0 & 1 & 0 \\
 0 & 0 & 1
 \end{array}  \right) 
 \nn \\
&+ \gamma_{\text{cusp}}(\alpha_s)
 \left(\ln\frac{\tilde{s}_{13}}{\tilde{s}_{23}}
 -\ln\frac{\tilde{s}_{14}}{\tilde{s}_{24}}
  \right) \left[\left(\begin{array}{ccc}
  0 & \frac{1}{2} & 0 \\ 
  1 & -\frac{N_C}{4} & \frac{N_C^2-4}{4 N_C} \\
  0  & \frac{N_C}{4} & -\frac{N_C}{4}
  \end{array}  \right)+
  \frac{\alpha_s}{4 \pi} g \left(\beta_{34}
  \right) 
\left(\begin{array}{ccc}
0 & \frac{N_C}{2} & 0  \\ 
- N_C & 0 & 0 \\
 0 & 0 & 0
\end{array}  \right)  
  \right] \,.
\end{align} 
The perturbative expansions of all objects appearing in the soft
anomalous dimensions above can be found, for instance, in the Appendix
of \cite{Ahrens:2010zv}.

Finally, we must determine the hard function at NLO.  The definition
of and procedure for calculating the hard functions in the quark
annihilation and gluon fusion channels is in exact analogy with
\cite{Ahrens:2009uz,Ahrens:2010zv}.  In a nutshell, the hard function
is obtained by projecting out QCD amplitudes onto a particular color
basis.  The Higgs boson does not carry color charge, therefore the
color bases employed for the quark annihilation and gluon fusion
channels are chosen to be exactly the same as in \cite{Ahrens:2010zv}.
Since the calculation described here requires the hard function up to
NLO, we need to evaluate one-loop QCD amplitudes for both the quark
annihilation and gluon fusion partonic processes. After UV
renormalization, the one-loop amplitudes are still affected by IR
divergences, which appear as poles in the limit in which the
dimensional regulator $\varepsilon$ vanishes. In order to obtain the
finite amplitudes needed to build the hard functions, which are
finite, one needs to subtract these residual poles. This is done by
means of appropriate IR subtraction counterterms
\cite{Ferroglia:2009ep,Ferroglia:2009ii}, again following the same
procedure employed in \cite{Ahrens:2010zv}.

For most $2 \to 2$ processes with a limited number of mass scales
one-loop corrections can be easily evaluated analytically; this fact
allowed some of us to evaluate the NLO hard functions for top-quark
pair production analytically in \cite{Ahrens:2009uz}.  The evaluation
of the $2 \to 3$ amplitudes needed here is considerably more
involved. However, in the last decade a number of tools for the
automated numerical evaluation of multi-leg one-loop amplitudes became
available. Most of these tools are publicly available and many rely on
reduction techniques operating at the integrand level, such as
the Ossola-Papadopoulos-Pittau method \cite{Ossola:2006us}. For this
reason we decided to carry out the calculation of the NLO hard
function with three of these tools: {\tt GoSam}
\cite{Cullen:2011ac,Cullen:2014yla}, {\tt MadLoop}
\cite{Hirschi:2011pa} and {\tt Openloops} \cite{Cascioli:2011va}. All
of these tools required a certain level of customization in order to
make the calculation of the hard function possible.\footnote{For this
  reason we were in contact with several of the authors of these
  tools. In particular, we would like to acknowledge the very useful
  exchanges we had with Nicholas Greiner, Giovanni Ossola, Valentin
  Hirschi, and Philipp Maierhofer.} This approach can be easily
adapted to the calculation of NLO hard functions for other processes
of interest at the LHC.
The calculation was tested by checking that the coefficients of the
residual IR poles, evaluated numerically in several points of the
phase space by means of the automated codes listed above, were
correctly canceled by the appropriate IR subtraction counterterm. The
finite hard functions obtained with each one of the three automated
codes were then compared in a number of phase space points. Numerical
agreement to more than eight digits was found in all cases.

The {\tt GoSam} and {\tt Openloops} codes were interfaced with an
in-house Monte Carlo program which was written in order to evaluate
the total cross section and the differential distributions presented
in Section~\ref{sec:NumAnalyis}. (The program can also be easily
interfaced with {\tt MadLoop} if one prefers to use this particular
one-loop provider.) Running time can become an issue in a Monte Carlo
code, where one needs to evaluate the hard function at millions of
phase-space points.  The computer time needed to evaluate the hard
functions in the two partonic channels is similar if one uses either
{\tt MadLoop}, {\tt GoSam} or {\tt Openloops}, provided that the reduction is
carried out with {\tt CutTools} \cite{Ossola:2007ax} in {\tt
  Openloops}. {\tt GoSam} employs {\tt Ninja}
\cite{vanDeurzen:2013saa, Peraro:2014cba} as the default reduction
tool, although {\tt GoSam} can also be configured in such a way that
this particular step of the calculation of the one loop amplitudes is
done by means of {\tt Golem95} \cite{Binoth:2008uq} or {\tt Samurai}
\cite{Mastrolia:2010nb}. The calculation of the hard function is
considerably faster if {\tt Openloops} is run in combination with the
(private) {\tt Fortran} library {\tt Collier}
\cite{Denner:2002ii, Denner:2005nn, Denner:2010tr, Denner:2014gla}.\footnote{We are grateful to the authors of {\tt
    Collier} for allowing us to use a binary version of their code.}
The approximate NNLO predictions for the differential distributions
which can be found in Section~\ref{sec:NumAnalyis} were obtained by
employing {\tt Openloops} (in combination with {\tt Collier}) and/or
{\tt GoSam} as providers for the hard functions.

\section{Approximate NNLO formulas}
\label{sec:approxNNLO}

By combining the information encoded in the NLO hard function and soft
function with the solution of the RG equations that they satisfy, it
is possible to resum logarithms of the ratio between the hard scale
$\mu_h$ (which characterizes the hard function) and the soft scale
$\mu_s$ (which is characteristic of the soft emission) up to NNLL
accuracy. In particular, the (differential) hard-scattering kernel
\begin{equation}
C(z,\mu) \equiv \mbox{Tr} \left[ \mathbf{H} \left(\{p_i\}, \mu \right)  \mathbf{S}\left(\sqrt{\hat{s}}(1-z),\{p_i\}, \mu \right) \right] \, ,
\end{equation}
can be rewritten in resummed form
\begin{align}
C \left(z,\mu_f \right) =& \exp{4 a_{\gamma_\phi} (\mu_s,\mu_f)} 
\mbox{Tr} \biggl[ \mathbf{U}\left(\mu_f, \mu_h,\mu_s \right) \mathbf{H}(\{p_i\},\mu_h)\mathbf{U}^\dagger \left(\mu_f,\mu_h, \mu_s \right) \nonumber  \\
&\times 
 \tilde{\mathbf{s}}\left(\ln{\frac{M^2}{\mu_s}} +\partial_\eta ,\{p_i\},\mu_s\right) \biggr] 
\frac{e^{-2 \gamma_E \eta}}{\Gamma \left(2 \eta \right)} \frac{z ^{-\eta}}{(1-z)^{1-2 \eta}} \, . \label{eq:resummedC}
\end{align}
The definitions of the anomalous dimensions and evolution matrices, as
well as the Laplace transformed soft function $\tilde{\mathbf{s}}$
found in (\ref{eq:resummedC}) are the same as in
\cite{Ahrens:2010zv, Li:2014ula}. Here we simply stress that the hard
functions and soft functions are evaluated at their characteristic
scale, ($\mu_h$ and $\mu_s$, respectively) and are therefore free from
large logarithmic corrections. As such, they can be safely evaluated
up to a given order in perturbation theory. Large corrections
depending on the ratio of the hard and soft scale are resummed in the
evolution factors $\mathbf{U}$. While the all-order hard scattering
kernels do not depend on the hard and soft scales but only on the
factorization scale at which the PDFs are evaluated, all practical
implementations of (\ref{eq:resummedC}) show a residual dependence on
the hard and soft scales due to the truncation of the perturbative
expansions of the hard functions, soft functions and anomalous
dimensions. For this reason, when implementing resummed formulas, one
must choose the hard and soft scales judiciously and carefully
estimate the related theoretical uncertainty.

The fixed-order expansion of the cross section and the resummation of
soft emission effects are two complementary approaches to the precise
determination of physical observables. For this reason, one typically
wants to match resummed and fixed-order calculations in order to
account for all of the known effects when obtaining phenomenological
predictions. However, there are situations in which the perturbative
expansion in $\alpha_s$ is still justified, but soft gluon emission
effects provide the bulk of the corrections at a given perturbative
order. In those cases, one can use the resummed hard scattering
kernels in order to obtain approximate formulas which include all of
the terms proportional to plus distributions up to a given power of
$\alpha_s$ in fixed-order perturbation theory. To be specific, one can
write
\begin{equation}
C(z, \mu) = \alpha_s^2 \left[C^{(0)}(\mu) + \frac{\alpha_s}{4 \pi} C^{(1)}(z, \mu) + \left(\frac{\alpha_s}{4 \pi} \right)^2 C^{(2)}(z, \mu) + {\mathcal O}\left(\alpha_s^3 \right) \right] \, , \label{eq:Cexpansion}
\end{equation} 
where we have set $\mu_f=\mu_r=\mu$, with $\mu_r$ the renormalization scale.\footnote{Note that it is possible
to keep these two scales separate using the RG equations for $\alpha_s$.}
The NNLO term in (\ref{eq:Cexpansion}) has the following structure
\begin{equation}
C^{(2)} \left(z,\mu \right) = \sum_{i=0}^3 D_i(\mu) P_i(z)  + C_0(\mu) \delta(1-z) + R(z,\mu)\,, \label{eq:C2struct}
\end{equation}
where the $P_n$ distributions are defined as 
\begin{equation}
P_n(z) \equiv \left[\frac{\ln^n(1-z)}{1-z} \right]_+ \, .
\end{equation}
In (\ref{eq:Cexpansion},\ref{eq:C2struct}) we dropped all arguments
with the exception of $\mu$ and $z$. The approximate NNLO formulas for
the partonic cross sections which we obtain in this work include the
complete set of functions $D_i$, some of the scale dependent terms in the
function $C_0$ as well as partial information on the function $R(z)$
which is non singular in the $z \to 1$ limit.
In particular, here we follow exactly the same procedure employed
  in \cite{Ahrens:2011mw, Broggio:2013uba}. That is, the terms
  included in $R(z)$ arise from the transformation of logarithms in
  Laplace space back to momentum space. A complete list of those
  transformations for PIM kinematics can be found for example in
  Eq.~(33) of \cite{Broggio:2013uba}. As pointed out in
  \cite{Ahrens:2011mw}, the $C_0$ term is ambiguous; in fact, in order
  to completely determine the coefficients multiplying the delta
  functions in the NNLO hard-scattering kernels, one would need to
  know the complete NNLO hard and soft matrices. Only the
  scale-dependent part of $C_0$ can be exactly determined, and one
  needs to specify which contributions are included there. One
  contribution to $C_0$ comes from the conversion of powers of Laplace-space
  logarithms according to Eq.~(33) of \cite{Broggio:2013uba}. Since
  these formula are exact, they are not a source of ambiguity for
  $C_0$ and those terms are included. Further contributions to $C_0$
  arise from \emph{i)} the product of the one-loop hard function with
  the one-loop soft function in Laplace space, \emph{ii)} the product
  of the tree-level hard function with the two-loop soft function in
  Laplace space, and \emph{iii)} the product of the two-loop hard
  function with the tree-level soft function in Laplace space. The
  contribution in \emph{i)} is known exactly and therefore included
  while the term in \emph{ii)} is unknown and dropped. One can
  reconstruct the scale dependent part of the contribution
  \emph{iii)}. However, it was observed in \cite{Ahrens:2010zv,
    Ahrens:2011mw, Broggio:2013uba} that by including these extra
  $\mu$-dependent terms one runs the risk of artificially reducing the
  scale dependence, rendering it an ineffective means of estimating
  theoretical uncertainties. Therefore, here again we follow
  \cite{Ahrens:2011mw, Broggio:2013uba} and drop completely the
  contributions of the two-loop hard function.

The information obtained from approximate NNLO formulas can be added
to the complete NLO calculation of a given observable in order to
obtain what we refer to as approximate NNLO predictions for a physical
quantity.  The matching of the approximate NNLO calculation to
complete NLO calculations is straightforward; for example, for the
total cross section one finds
\begin{align}
\sigma^{\text{NLO+ approx NNLO}} = \sigma^{\text{NLO}}+ \sigma^{\text{approx. NNLO}} - \sigma^{\text{approx. NLO}} \, , \label{eq:matching}
\end{align}
where the subtraction of the last term avoids double counting of NLO
terms proportional to plus distributions and delta functions. It must
be observed that all of the terms on the r.h.s. of (\ref{eq:matching})
must be evaluated with NNLO PDFs. To avoid lengthy superscripts, in
the following we indicate matched NLO + approx. NNLO calculations with
the symbol ``nNLO''. In contrast to resummed calculations, nNLO
calculations show a residual dependence on the factorization scale
only. As usual, the residual dependence of the observable on the
factorization scale can be exploited in order to study and estimate
the theoretical uncertainty affecting physical predictions.

The use of approximate formulas offers an additional advantage: the
numerical evaluations of the total cross section and distributions to
approximate NNLO accuracy require shorter running times than the
evaluation of the corresponding resummed formulas. For this reason, in
this work we present predictions based upon approximate NNLO formulas.

\section{Numerical analysis \label{sec:NumAnalyis}}
\label{sec:pheno}

In this section we present results obtained from the numerical
evaluation of the nNLO formulas and discuss their implications.  We
cover the total cross section in Section~\ref{sec:xs} and differential
distributions in Section~\ref{sec:dxs}.

A central issue is that the soft limit $z\to 1$ is only guaranteed
to provide accurate predictions for observables where
$\hat{s}\to s$, with $\sqrt{s}$ the collider energy; an example would
be the case where $M\to \sqrt{s}$.  More realistic observables such as
the total cross section or differential distributions at their peaks
are also sensitive to regions of phase space far away from $z\to
1$. Thus, in order for corrections in the soft limit to be dominant
also in those cases, the mechanism of dynamical threshold
enhancement \cite{Becher:2007ty, Ahrens:2010zv} must occur. This
simply means that the parton luminosities appearing in
(\ref{eq:soft-fact}) should drop off quickly enough away from the
integration region where $z\to 1$, that an expansion under the
integrand of the partonic cross section in the soft limit is
justified.

In order to address this issue we begin both of the following
subsections with a comparison of approximate NLO calculations, valid
in the soft limit, with the full NLO calculation.  Approximate NLO
calculations are obtained by re-expanding the NNLL resummed partonic
cross section to NLO; consequently they reproduce completely all of
the terms singular in the $z\to 1$ limit in the NLO partonic cross
section, but they miss terms which are subleading in the soft limit.
We verify in all cases that the soft approximation works quite well at
NLO.  This obviously does not immediately imply that the same holds at
higher orders, but is an important sanity check nonetheless.  After
these initial studies at NLO we then present the main results of this
paper, namely numerical results from the NNLO approximations.  We will
see that these NNLO corrections tend to enhance both the total cross
section and differential distributions to the top of the NLO
uncertainty band, and also greatly decrease the uncertainties
associated with scale variations.  In fact, the residual uncertainties
due to scale variation alone at approximate NNLO are so small that it
is rather doubtful that they reflect the true theoretical uncertainty
associated both with even higher-order soft gluon corrections and with
terms subleading in the soft limit.  In this section we address this
issue and discuss a way to obtain a more conservative estimate of the
theoretical uncertainty affecting approximate NLO and nNLO
calculations. 

The study that follows is meant to be illustrative rather than 
exhaustive.  Therefore, we consider only one LHC energy,
namely $\sqrt{s} = 13$~TeV, and do not apply any cuts on the
momenta of the  final state particles. We carry out all of our
calculations with MSTW 2008 PDFs \cite{Martin:2009iq}, along with the additional input
parameters shown in Table~\ref{tab:tabGmu}.  Throughout the analysis
we need exact NLO results for the total cross section and differential
distributions.  All of the numbers at NLO accuracy reported
below are obtained from the code \verb|MadGraph5_aMC@NLO|
\cite{Alwall:2014hca}, which for convenience we indicate with
\verb|MG5| in the rest of this work.

  Finally, we conclude the introduction to this section by pointing
  out that in our analysis we keep the renormalization and
  factorization scales equal. However, we explicitly calculated the
  NLO cross section varying independently the renormalization and
  factorization scales in the range $[\mu_0/2, 2\mu_0]$, where $\mu_0$
  indicates one of the central values of the
  renormalization/factorization scale employed in this work. (The
  values chosen for $\mu_0$ are explicitly indicated in each
  calculation discussed below.) For all of the choices of $\mu_0$
  which we consider in this work, the separate variation of the
  factorization and renormalization scales does not lead to larger
  uncertainty bands with respect to the ones obtained by setting the
  factorization and renormalization scales equal and by varying this
  single scale in the range $[\mu_0/2, 2\mu_0]$. In particular, we
  always find that the smallest NLO cross section is obtained by
  setting the renormalization and factorization scales equal to $2
  \mu_0$, while the largest cross section is obtained by setting the
  two scales equal to $\mu_0/2$. Therefore, setting the two scales
  equal to each other does not underestimate the theoretical
  uncertainty at NLO, compared to individual variations.  For this
  reason, we feel justified in setting the renormalization and
  factorization scales equal also in the nNLO analysis, which greatly
  reduces the amount of running time required to obtain nNLO
  predictions. Separate variations, but unlikely to increase the final error bands we advocate in Section 4.

\subsection{Total cross section}
\label{sec:xs}

\begin{table}
	\begin{center}
		\def\arraystretch{1.3}
	\begin{tabular}{|c|c||c|c|}
		\hline $G_\mu$ & $1.16639 \times 10^{-5}\,\, \text{GeV}^{-2}$ & $m_t$ & $172.5$~GeV\\ 
		\hline $M_Z$ &  $91.188$~GeV & $m_H$ & $125$~GeV \\ 
		\hline $1/\alpha$ & $132.507$ & $\alpha_s \left(M_Z\right)$ & from MSTW2008 PDFs \\ 
		\hline 
	\end{tabular} 
	\caption{Input parameters employed throughout the calculation. \label{tab:tabGmu}}
	\end{center}
\end{table}
%

\begin{table}
	\begin{center}
		\def\arraystretch{1.3}
		\begin{tabular}{|c|c|c|c|}
			\hline $\mu_0$~[GeV] & NLO  {\tt MG5} [fb] & NLO no $qg$ channel {\tt MG5} [fb] &NLO approx. [fb]\\ 
			\hline $235$ & $515.5^{+30.6}_{-49.4}$ & $499.5^{+0.0}_{-30.1}$ &$486.3^{+0.0}_{-47.4}$\\ 
			\hline 
		\end{tabular} 
		\caption{Total cross section at the LHC with $\sqrt{s} = 13$~TeV and MSTW2008 NLO PDFs. The uncertainties reflect scale variation only. \label{tab:CS13}}	
	\end{center}
\end{table}

Table~\ref{tab:CS13} shows the numerical values of the Standard
Model total cross section for the production of a top-antitop pair in
association with a Higgs boson,
using the central value ($\mu_0$) for the factorization scale ($\mu$) employed in 
\cite{Beenakker:2001rj, Beenakker:2002nc}, namely
\begin{align}
\mu = \mu_0 = \frac{2 m_t + m_H}{2} = 235~\text{GeV}\, .
\end{align}
In addition to the complete NLO calculation, we show the NLO cross
section without the contribution of the quark-gluon channel, which
opens up at NLO. This channel is formally subleading in the soft
emission limit and is therefore absent in approximate NLO
calculations. However, it is important to keep in mind that the
quark-gluon channel is accounted for by the matching procedure in nNLO
calculations consider later on. Consequently, physical quantities evaluated to nNLO
include the same quark-gluon channel contributions included in NLO
calculations.  By looking at Table~\ref{tab:CS13}, we observe that the
approximate NLO calculation, based exclusively on the soft emission
limit, captures 97.4 \% of the full NLO calculation without the
contribution of the quark-gluon channel. Very similar results are
found for $\sqrt{s} = 7$~TeV and $\sqrt{s} = 14$~TeV.

\begin{figure}
\begin{center}	
	\includegraphics[width=13cm]{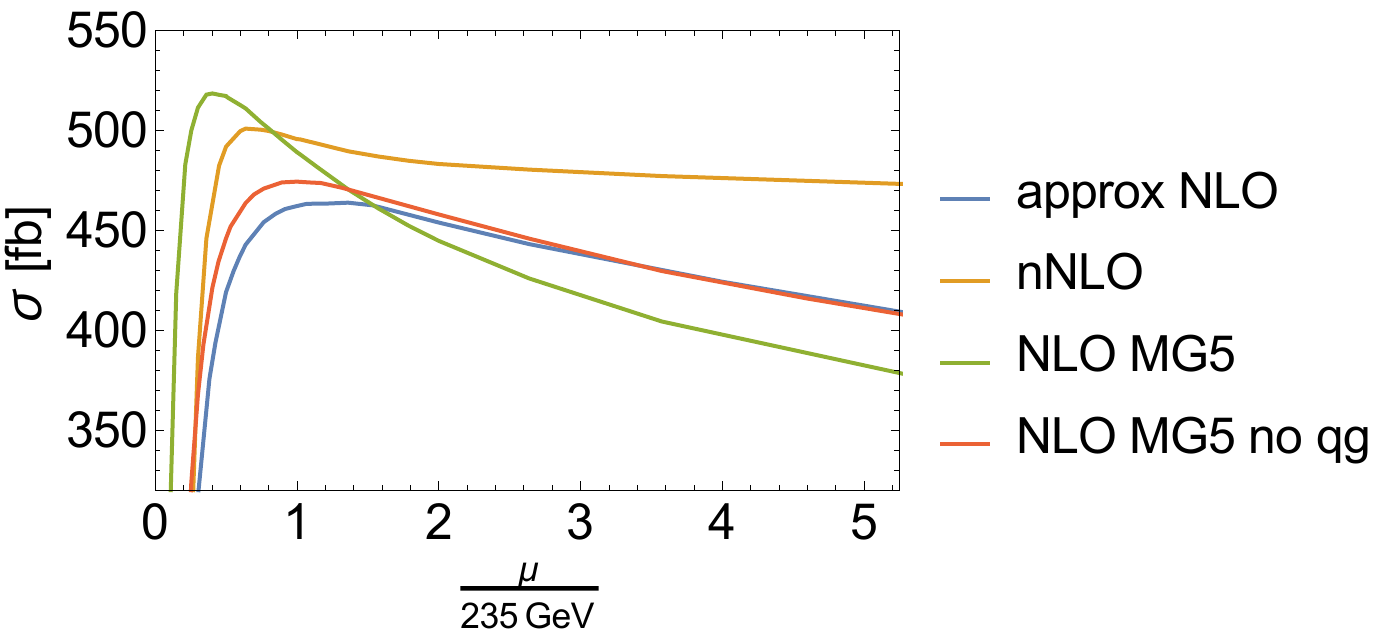}
\end{center}	
\caption{Scale dependence of the total cross section. The curves represent the 
	NLO cross section evaluated with {\tt MG5} by excluding the quark-gluon channel contribution (red line), the complete NLO cross section evaluated with {\tt MG5} (green line), the nNLO cross section (orange line), and the approximate NLO 
	cross section (light blue line). 
 In this figure, all perturbative orders are evaluated with NNLO MSTW2008 PDFs.	 \label{fig:scaledep}}
\end{figure}

The residual theoretical uncertainty is estimated by evaluating the
cross section also at $\mu = 235/2 =117.5$~GeV and at $\mu =
2\times235 = 470$~GeV. 
By looking at Table~\ref{tab:CS13} one can see
that the effect of the quark-gluon channel, which is not
  included in the last two columns on the right, is quite large on
the scale variation. Furthermore, while the complete NLO cross section
is a monotonically decreasing function of $\mu$ in the range
$[117-470]$~GeV, the NLO cross section has a maximum close to $\mu =
235$~GeV if the quark-gluon channel is excluded. This fact explains
the $+0.0$ in the scale variation in the third column of
Table~\ref{tab:CS13}. This behavior is reproduced by the approximate
NLO calculation (rightmost column of Table~\ref{tab:CS13}). A similar
situation was found in the study of top-quark pair production
\cite{Ahrens:2011mw}. This kind of behavior is even more pronounced if
NNLO PDFs are employed, as can be seen from Figure~\ref{fig:scaledep}.
%
%
In view of the steep decrease of all the curves in
Figure~\ref{fig:scaledep} for values of the ratio $\mu/
(235~\text{GeV})$ smaller than one, it is reasonable to choose a value
for the central scale $\mu_0$ larger than $235$~GeV.\footnote{The
  steep decrease of the cross section for small values of the
  factorization scale is an unphysical effect, in fact by choosing a
  factorization scale of the order $10-20$ GeV one can even obtain
  negative values for the NLO total cross section. This effect can be
  cured either by incorporating resummation effects or by choosing a
  dynamic scale. Our goal in this work is to validate a method for the
  calculation of the approximate NNLO cross section, therefore we do
  not analyze this aspect further.}  As an example, we choose $\mu_0 = 620$~GeV
($\mu_0/ (235~\text{GeV}) \sim 2.64$), which is a value close to the
maximum of the distribution differential with respect to the total
final state invariant mass $M$. We have checked that the location of this
maximum is not very sensitive to the LHC
energy. Table~\ref{tab:CS13nlo620} shows that also when one chooses
$\mu_0 = 620$~GeV the approximate NLO calculation reproduces to a very
good extent the NLO corrections if one excludes the contribution of
the quark-gluon channel from the latter.
\begin{table}
	\begin{center}
		\def\arraystretch{1.3}
		\begin{tabular}{|c|c|c|c|}
			\hline $\mu_0$~[GeV] & NLO  {\tt MG5} [fb] & NLO no $qg$ channel {\tt MG5} [fb] &NLO approx. [fb]\\ 
			\hline $620$ & $445.7^{+51.4}_{-51.4}$ & $467.1^{+28.1}_{-41.0}$ &$464.5^{+22.2}_{-38.1}$\\ 
			\hline 
		\end{tabular} 
		\caption{Total cross section at the LHC with $\sqrt{s} =
                  13$~TeV and MSTW2008 NLO PDFs. The uncertainties
                  reflect scale variation
                  only. \label{tab:CS13nlo620}}
	\end{center}
\end{table}

\begin{table}
	\begin{center}
		\def\arraystretch{1.3}
		\begin{tabular}{|c|c|c|c|}
			\hline $\mu_0$~[GeV] & LO   [fb] & NLO  {\tt MG5} [fb] &nNLO [fb] \\ 
			\hline $620$ & $317.2^{+97.4}_{-69.2}$ & $445.7^{+51.4}_{-51.4}$ &$479.8^{+10.3}_{-7.1}$\\ 
			\hline $235$ & $464.2^{+164.4}_{-112.1}$ & $515.2^{+30.6}_{-49.4}$ &$495.6^{+0.0}_{-12.5}$\\ 
			\hline 
		\end{tabular} 
		\caption{Total cross section at the LHC with $\sqrt{s} =
                  13$~TeV. Each order is evaluated with the MSTW2008
                  PDFs at the corresponding perturbative order
                  (meaning, e.g. NNLO PDFs for the nNLO
                  calculation). The uncertainties reflect scale
                  variation only.  \label{tab:CSat620}}
	\end{center}
\end{table}
\begin{table}
	\begin{center}
		\def\arraystretch{1.3}
		\begin{tabular}{|c|c|c|c|}
			\hline $\mu_0$~[GeV] &  NLO  {\tt MG5} [fb] & approx. NLO [fb] & nNLO [fb]  \\ 
			\hline $620$ & $445.7^{+51.4}_{-51.4}$ & $442.4^{+44.3}_{-44.3}$ &$467.2^{+22.9}_{-22.9}$  \\ 
			\hline $235$ & $515.2^{+30.6}_{-49.4}$ & $462.6^{+23.7}_{-23.7}$ &$481.6^{+14.0}_{-14.0}$  \\ 
			\hline 
		\end{tabular} 
		\caption{Total cross section at the LHC with $\sqrt{s} = 13$~TeV with an estimate of the error associated to the scale variation and to the formally subleading terms, as explained in the text.  Each order is evaluated with the MSTW2008 PDFs at the corresponding perturbative order.
                  \label{tab:LBCS}}
	\end{center}
\end{table}

\begin{figure}
	\begin{center}
		\begin{tabular}{cc}
			\includegraphics[width=7cm]{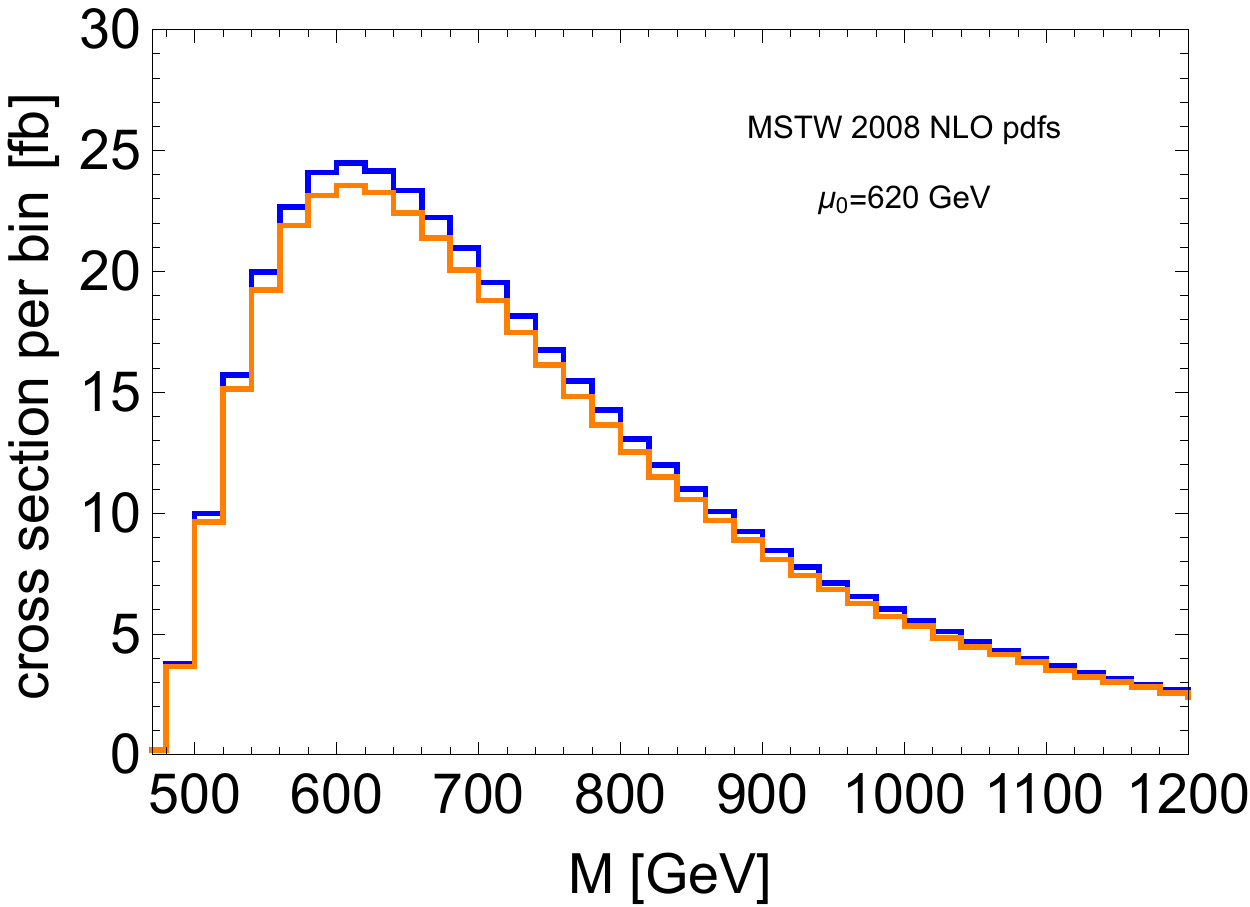} & \includegraphics[width=7cm]{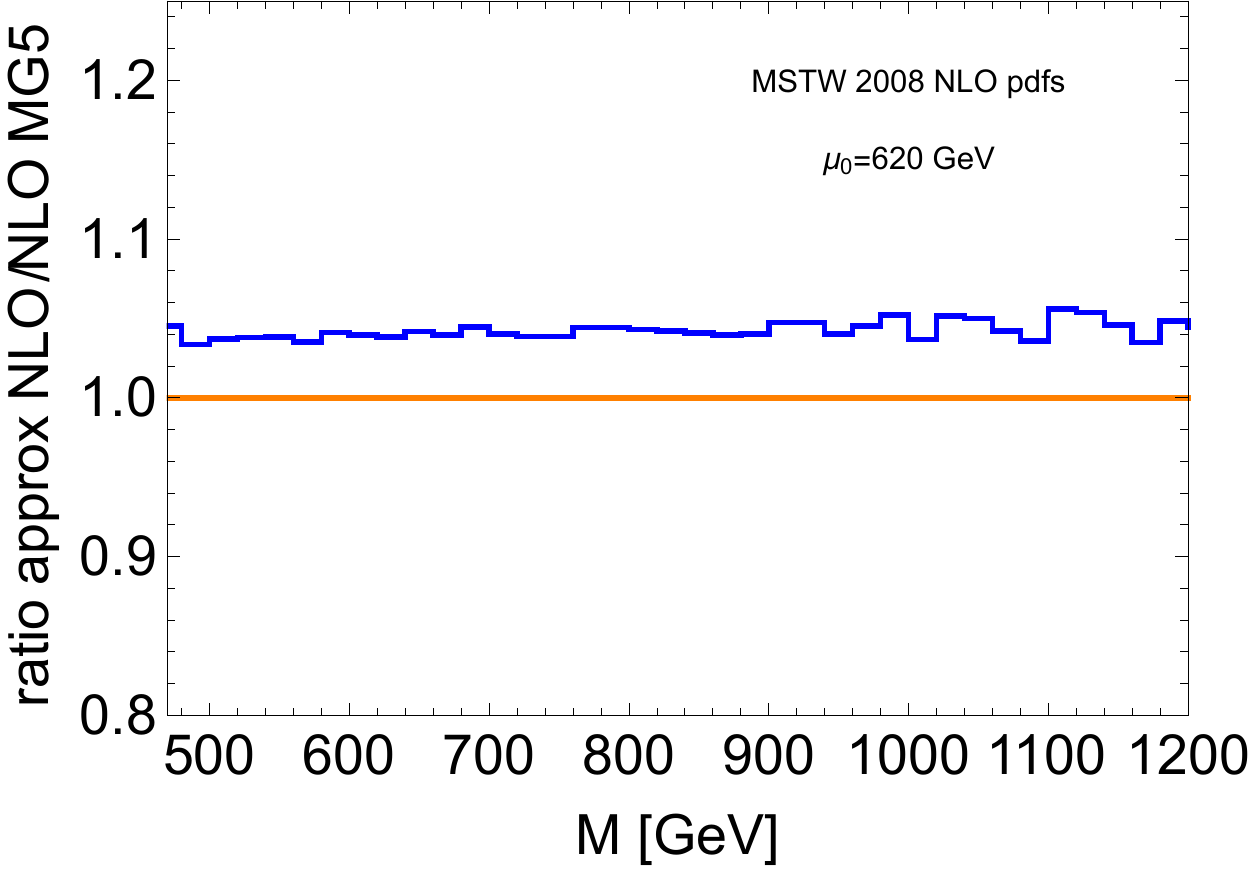} \\
			\includegraphics[width=7cm]{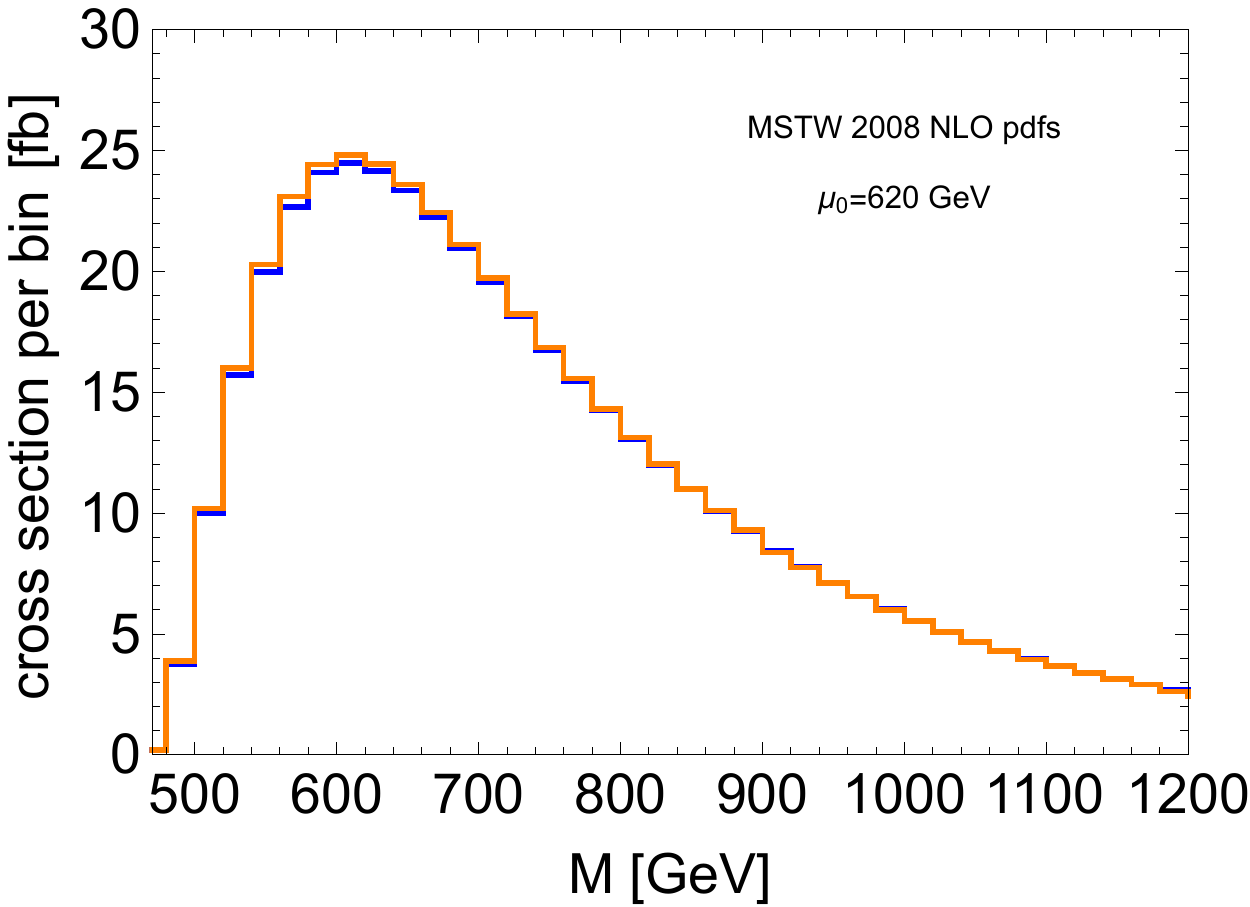} & \includegraphics[width=7cm]{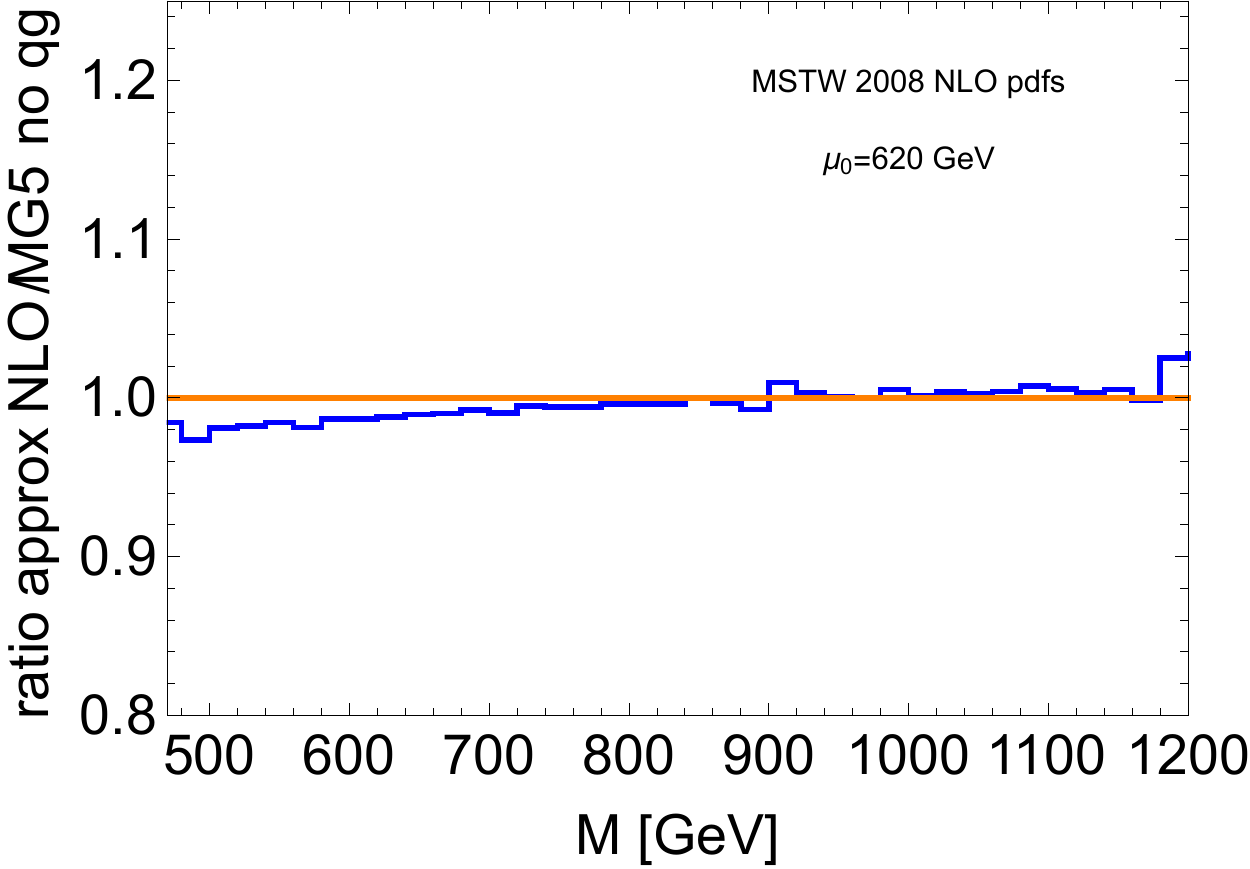} \\
		\end{tabular}
	\end{center}
	\caption{Distribution differential in the invariant mass $M$
          of the $t\bar{t}H$ system. In the panels on the l.h.s. the
          light (orange) line represents the \texttt{MG5} results
          while the darker (blue) line corresponds to the
          approximate NLO results. The panels on the r.h.s. show the
          ratio between the approximate NLO result and the
          \texttt{MG5} result bin by bin. In the upper panels,
          \texttt{MG5} was used to calculate the full NLO corrections,
          while in the lower panels the \texttt{MG5} result does not
          include the contribution of the quark-gluon channel. The
          factorization/renormalization scale is fixed to $\mu_0 =
          620$~GeV. \label{fig:MdistNLO}}
\end{figure}

\begin{figure}
	\begin{center}
		\begin{tabular}{cc}
			\includegraphics[width=7cm]{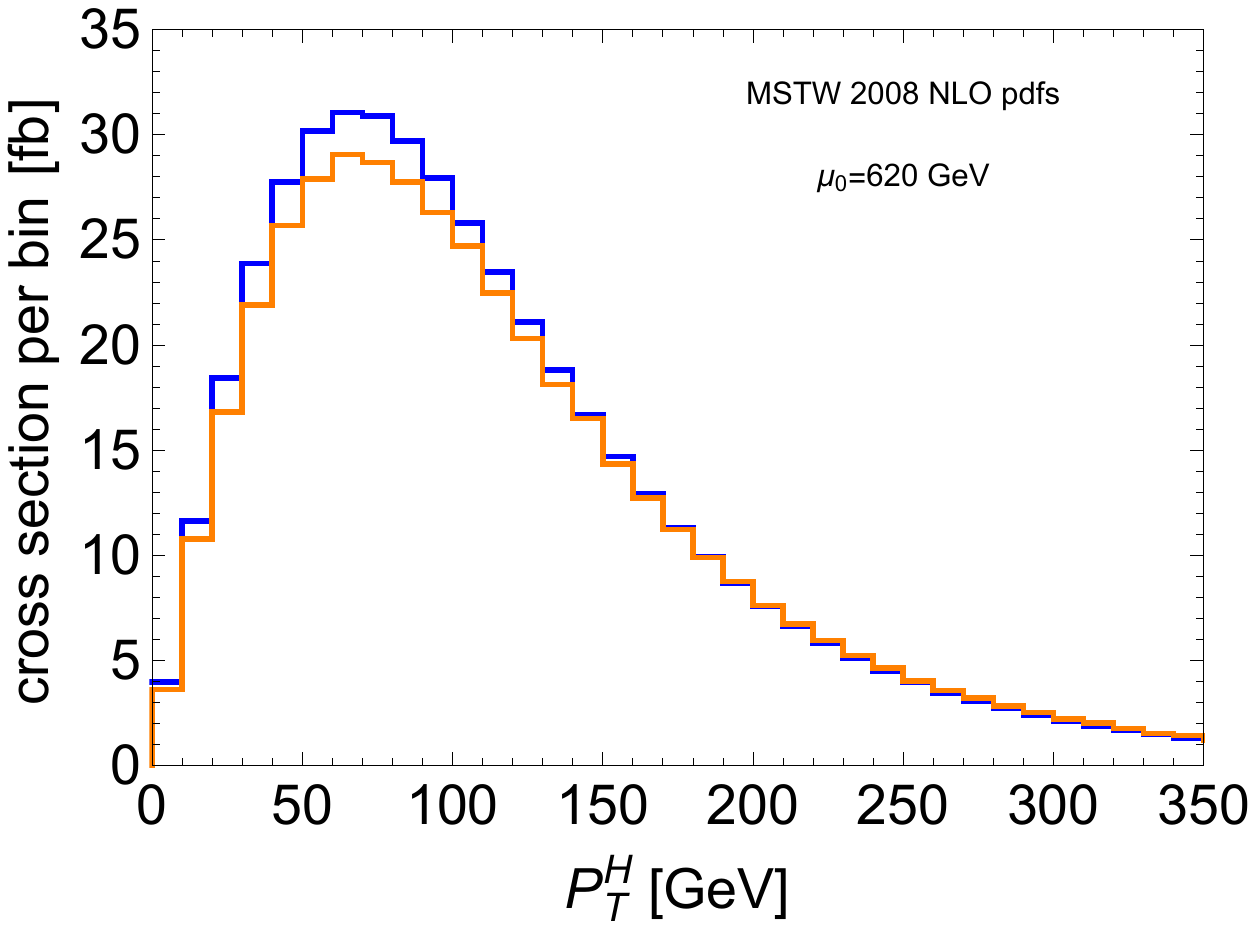} & \includegraphics[width=7cm]{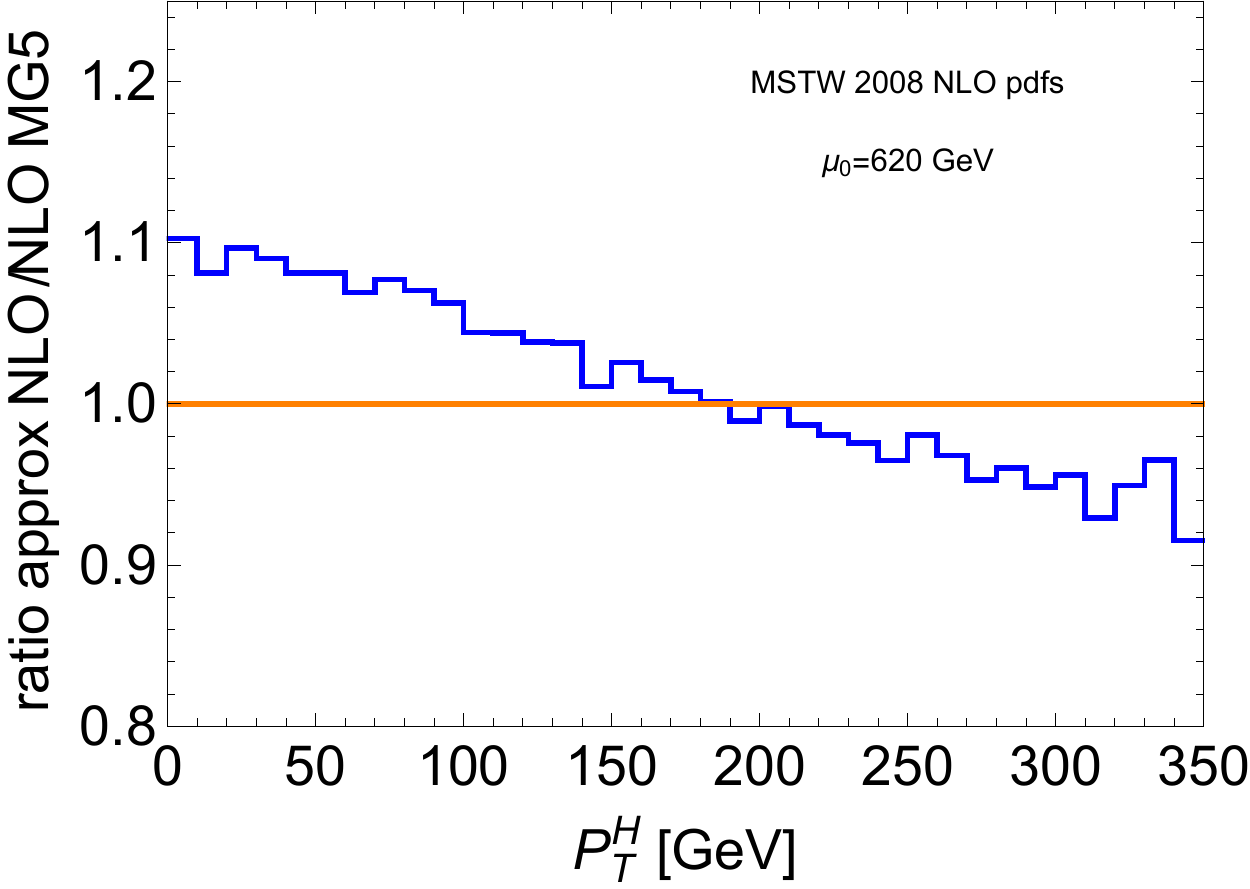} \\
			\includegraphics[width=7cm]{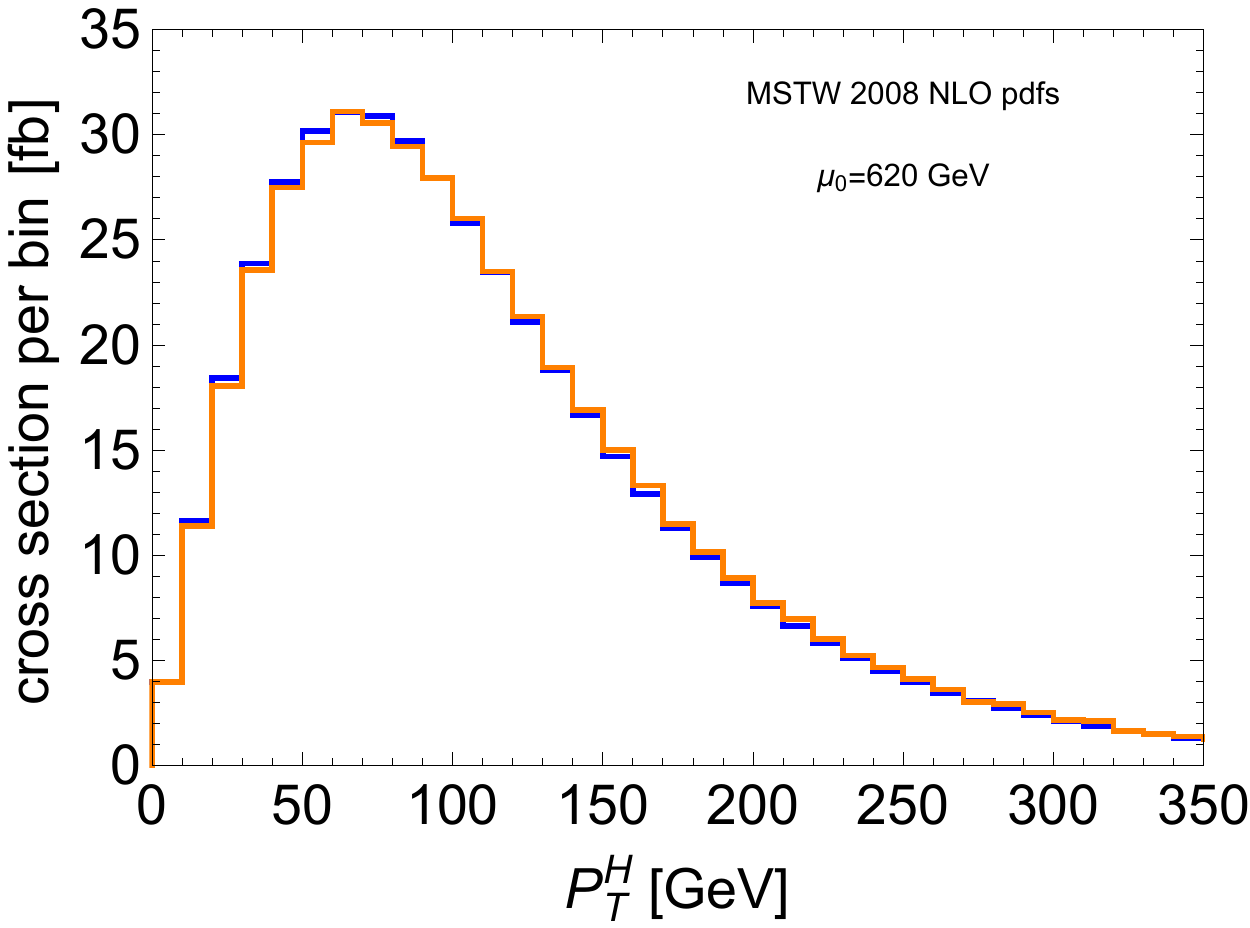} & \includegraphics[width=7cm]{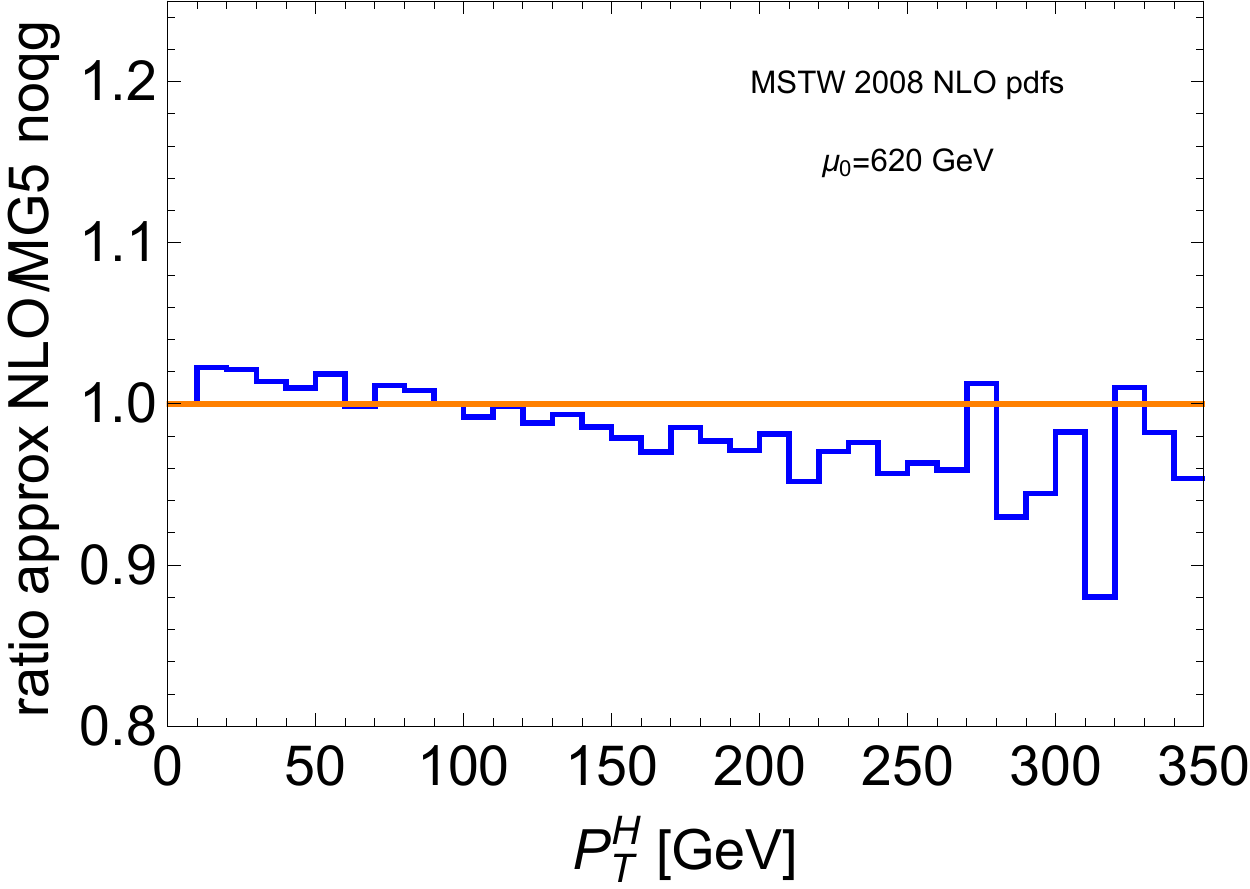} \\
		\end{tabular}
	\end{center}
	\caption{Distribution differential in $p_T^H$. In the panels
          on the l.h.s. the light (orange) line represents the
          \texttt{MG5} results while the darker (blue) line
          corresponds to the approximate NLO results. The panels on
          the r.h.s. show the ratio between the approximate NLO result
          and the \texttt{MG5} result bin by bin. In the upper panels,
          \texttt{MG5} was used to calculate the full NLO corrections,
          while in the lower panels the \texttt{MG5} result does not
          include the contribution of the quark-gluon channel. The
          factorization/renormalization scale is fixed to $\mu_0 =
          620$~GeV.  \label{fig:pHfig}}
\end{figure}

\begin{figure}[ht]
	\begin{center}
		\begin{tabular}{cc}
			\includegraphics[width=7cm]{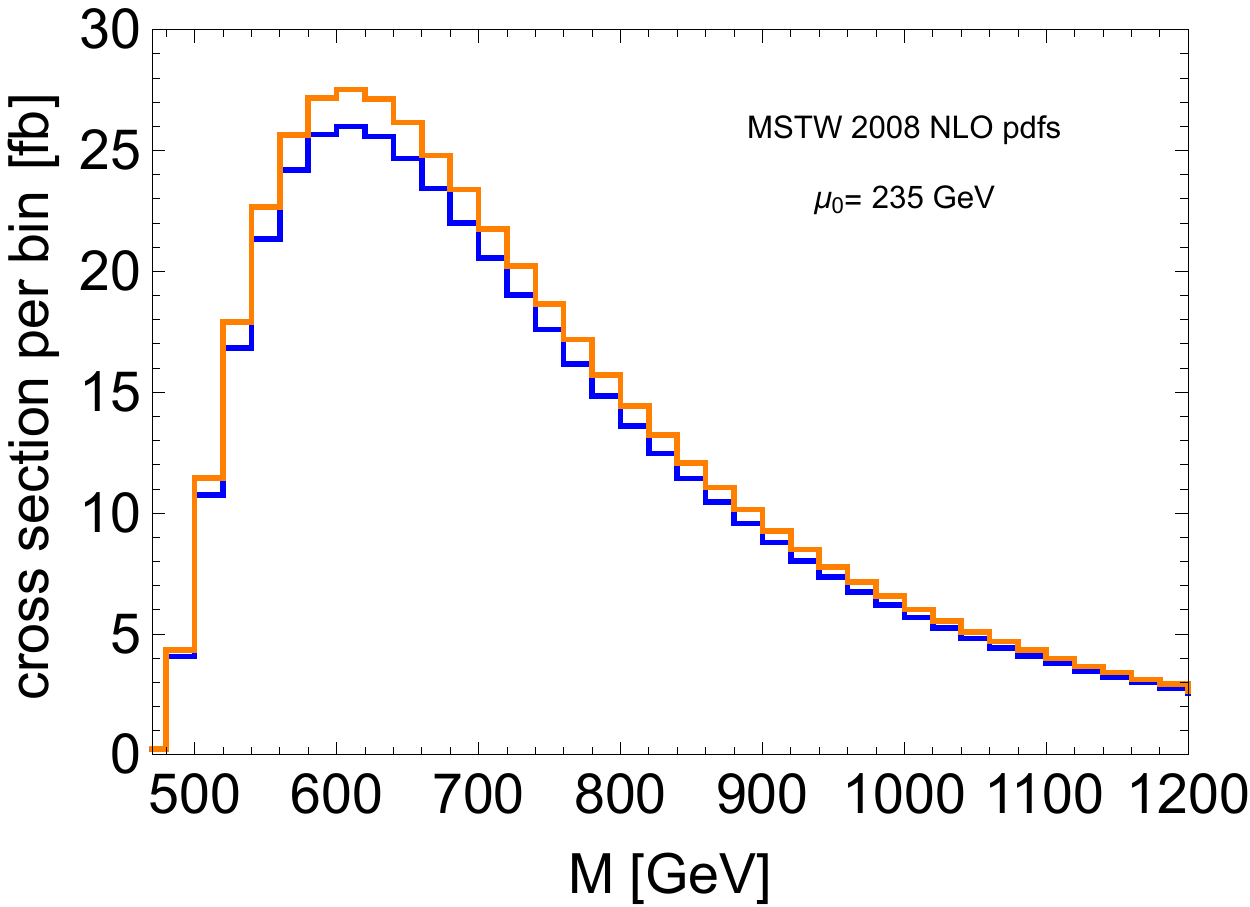} & \includegraphics[width=7cm]{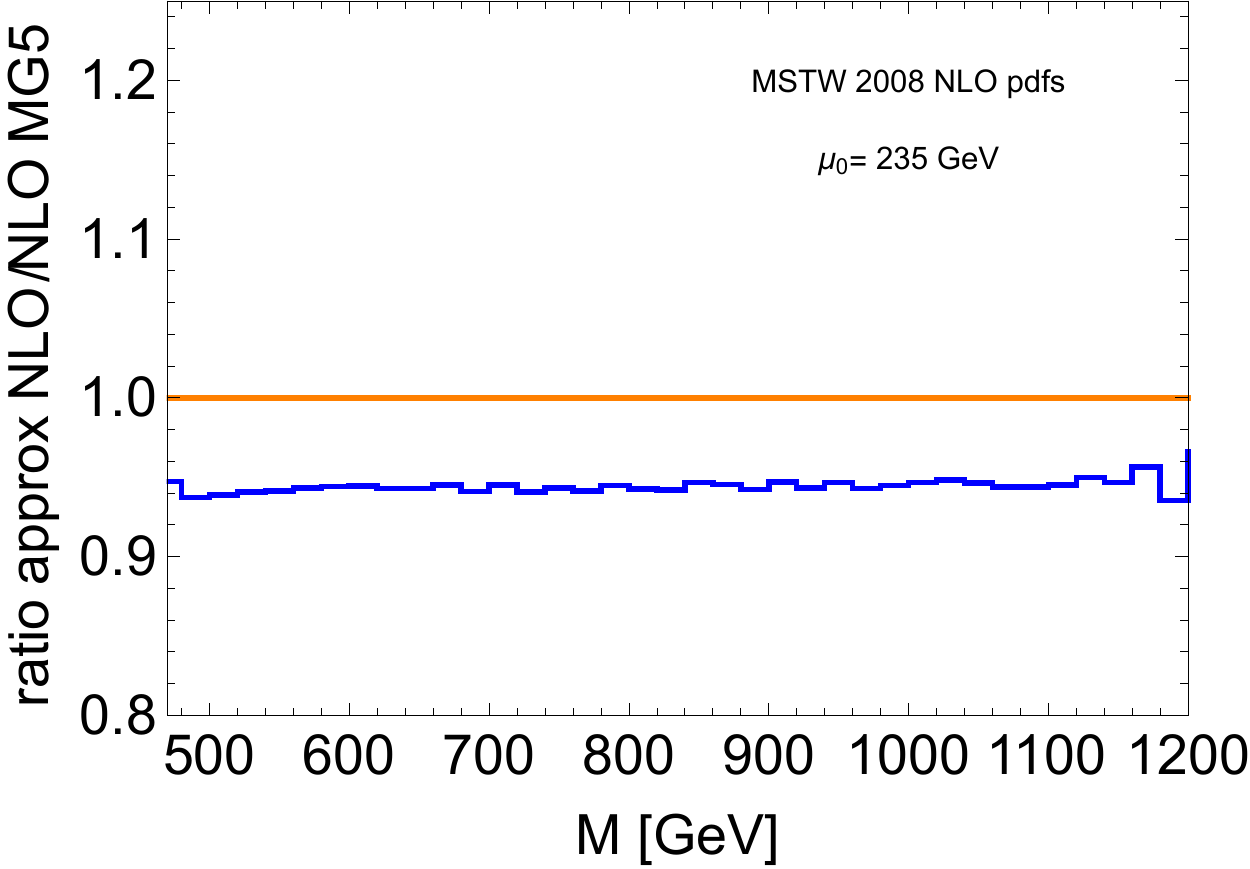} \\
			\includegraphics[width=7cm]{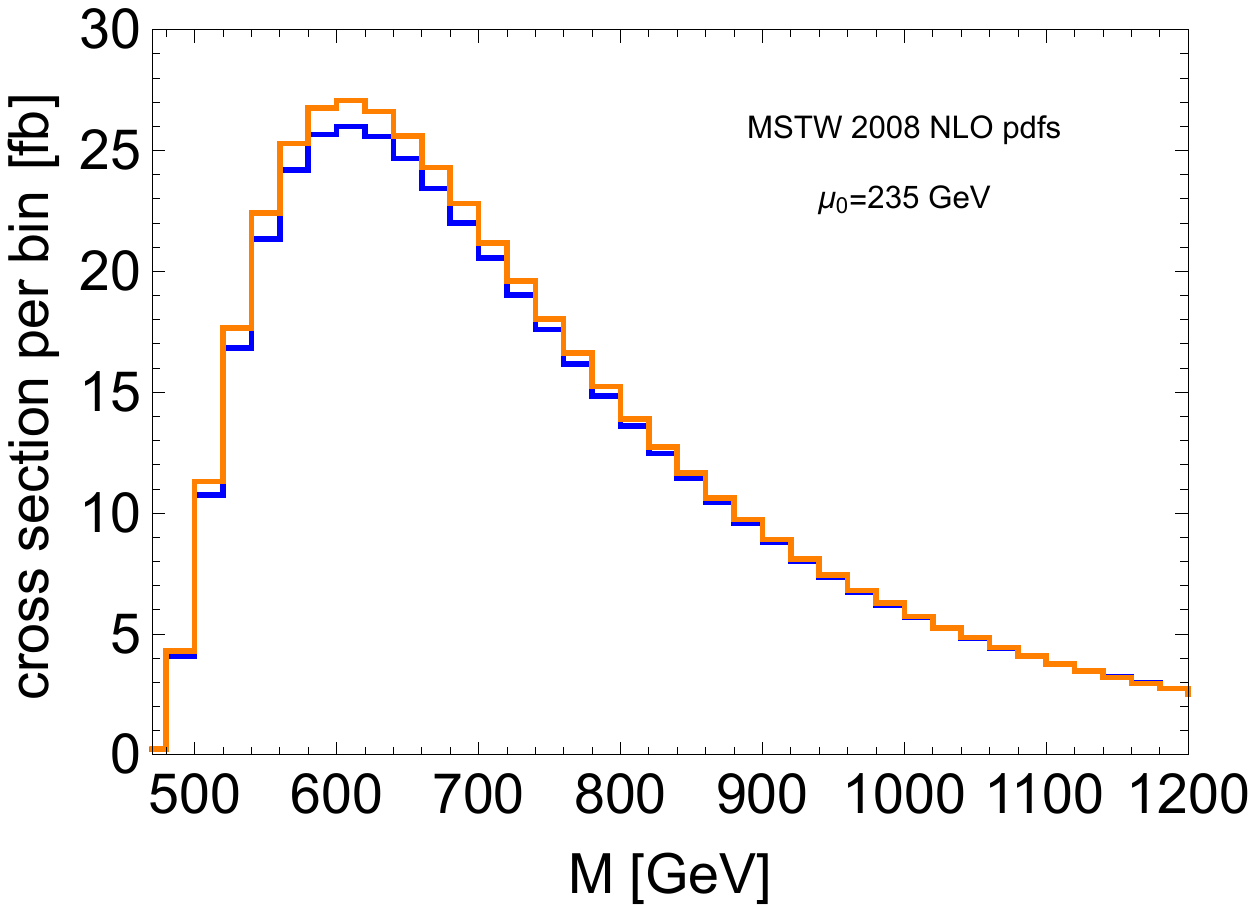} & \includegraphics[width=7cm]{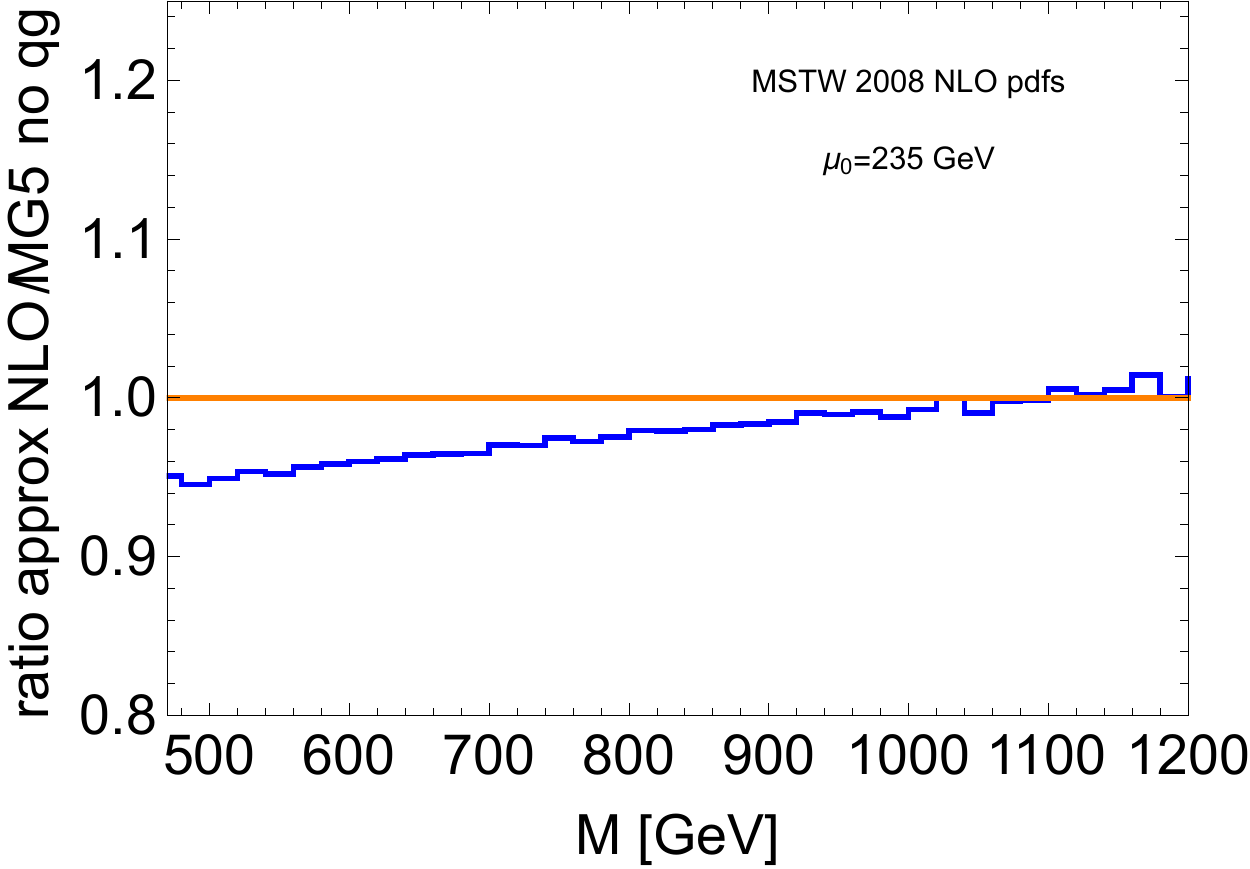} \\
		\end{tabular}
	\end{center}
	\caption{Same as in Fig.~\ref{fig:MdistNLO}, but with the renormalization and factorization scales fixed at $\mu_0 = 235$~GeV.  \label{fig:MdistNLO235}}
\end{figure}

\begin{figure}[ht]
	\begin{center}
		\begin{tabular}{cc}
			\includegraphics[width=7cm]{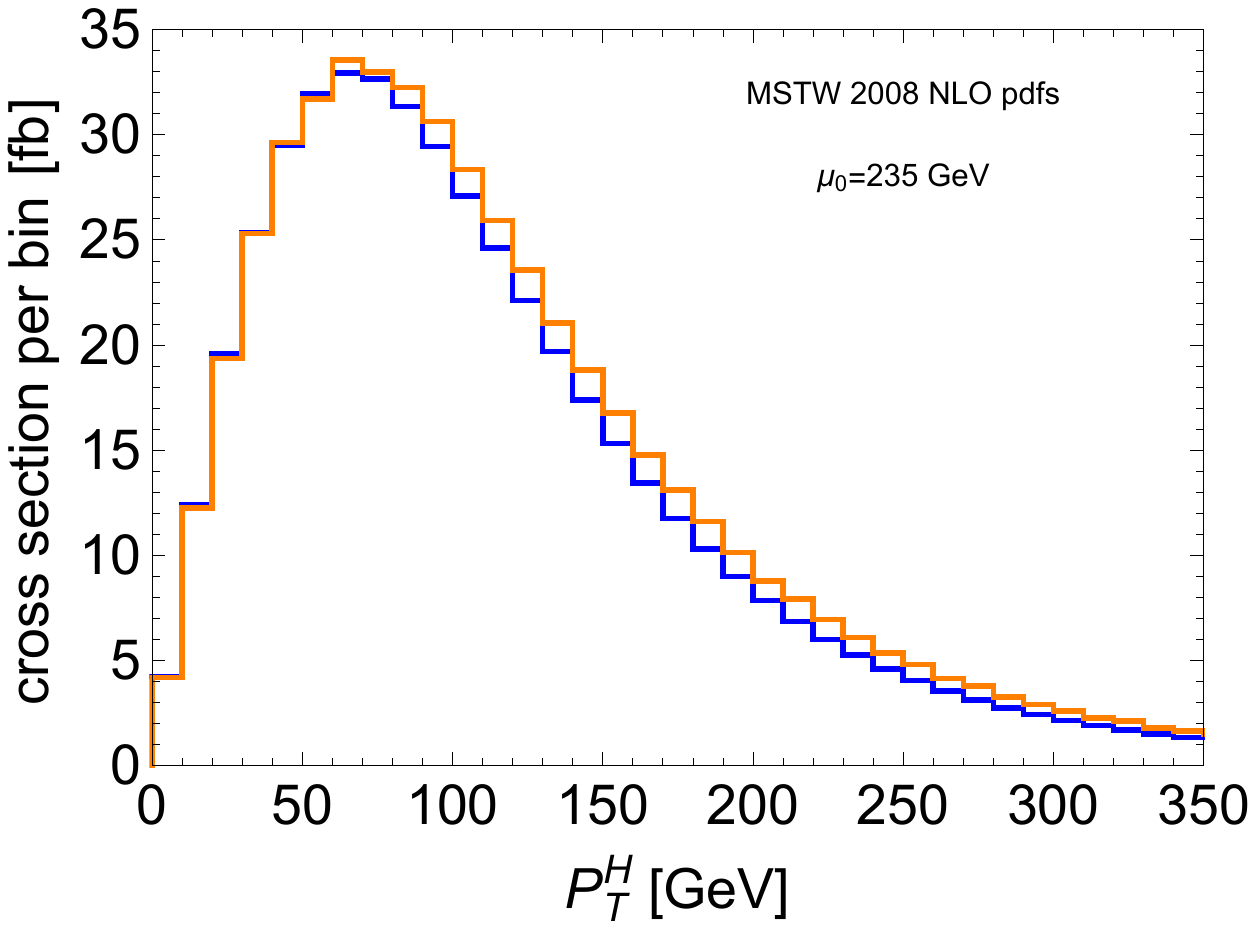} & \includegraphics[width=7cm]{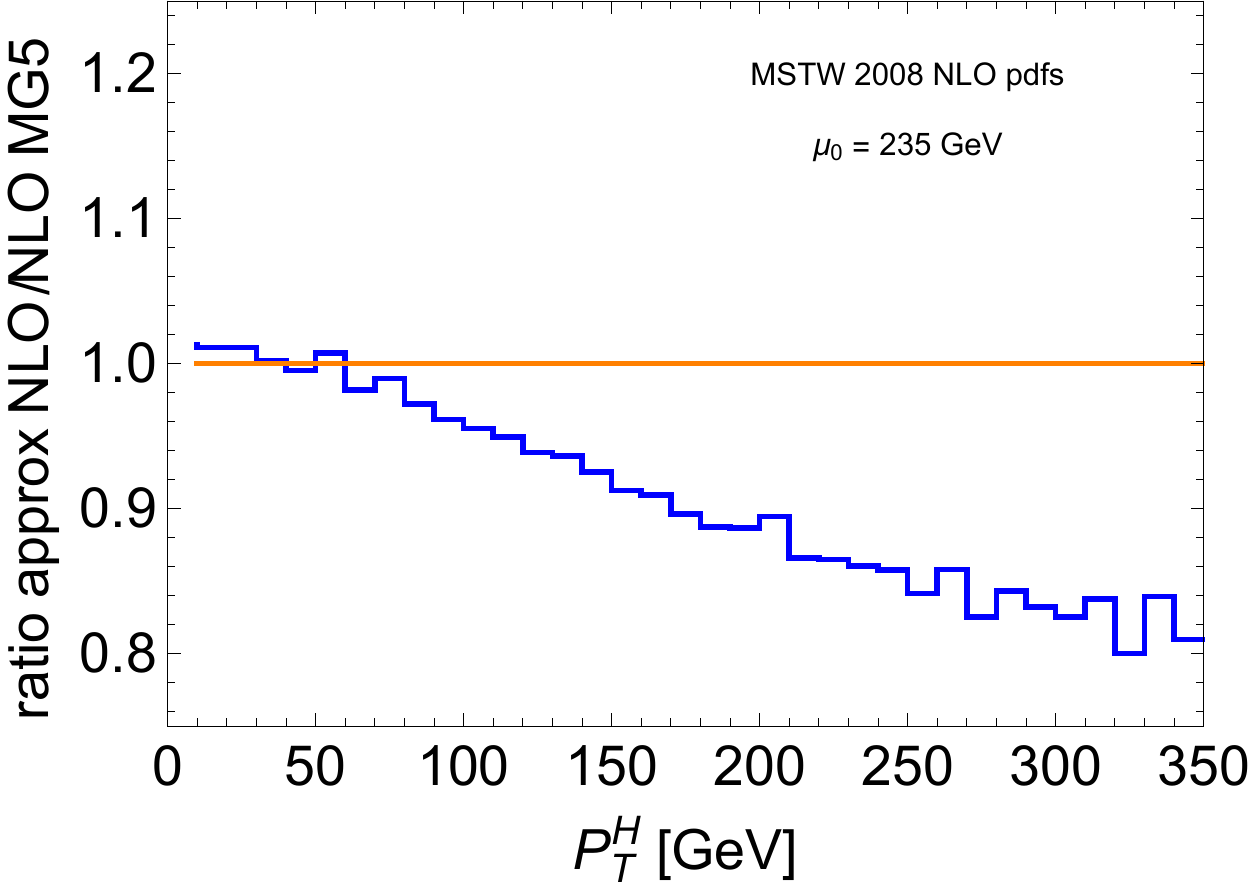} \\
			\includegraphics[width=7cm]{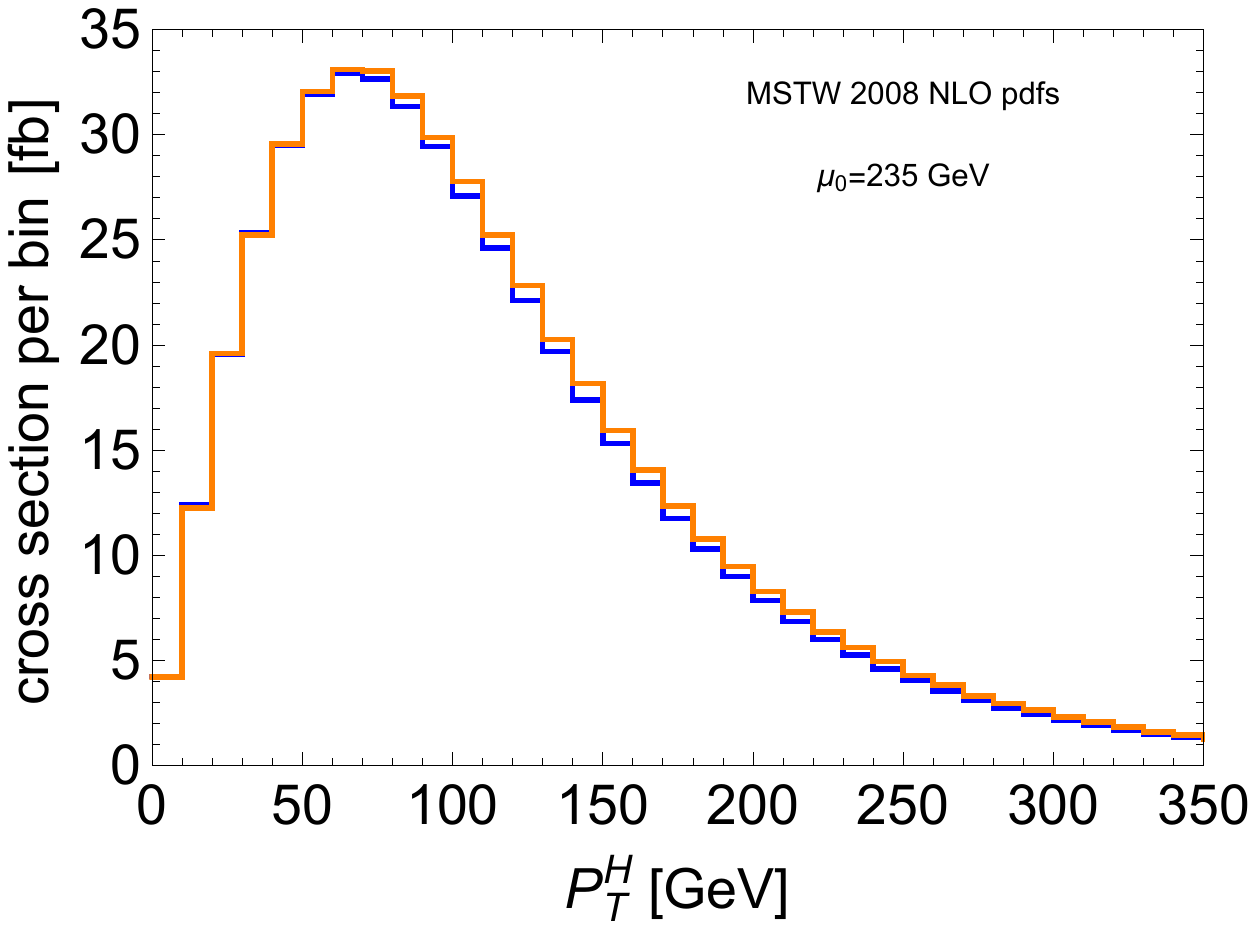} & \includegraphics[width=7cm]{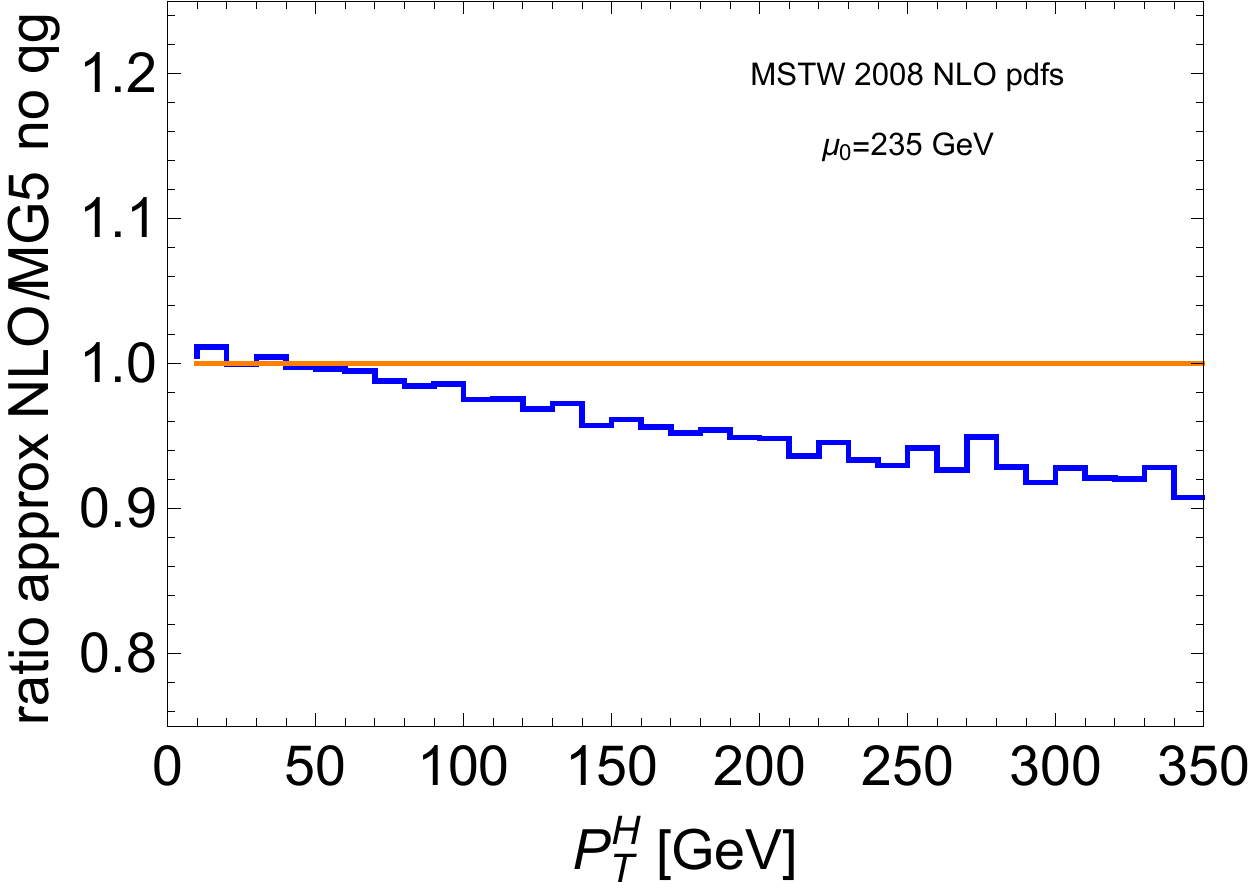} \\
		\end{tabular}
	\end{center}
	\caption{Same as in Fig.~\ref{fig:pHfig}, but with the renormalization and factorization scales fixed at $\mu_0 = 235$~GeV.  \label{fig:pHfig235}}
\end{figure}
%


The total cross section at LO, NLO and nNLO calculated at $\mu_0 = 620$~
GeV can be found in Table~\ref{tab:CSat620}. If one accounts for the
approximate NNLO corrections, the central value of the cross section
increases by about  8~\% with respect to the NLO calculation, while the
scale uncertainty is reduced by a factor of $5$.
For completeness, in Table~\ref{tab:CSat620} we report the LO, NLO and nNLO total cross section obtained by setting $\mu_0 = 235$~ GeV. While we do not advocate the use of $\mu_0 = 235$~ GeV for the reasons discussed above, we observe that the range of values for the nNLO cross section obtained with $\mu_0 = 235$~ GeV, namely $[483.1-495.6]$~fb, is reasonably close to the one obtained by setting $\mu_0 = 620$~GeV, which is $[472.7-490.1]$~fb. As already pointed out above for the NLO cross section, the nNLO cross section has a maximum close to $\mu =
	235$~GeV, and this explains the $+0.0$ in the scale variation in the third row of Table~\ref{tab:CSat620}.

So far, we have estimated uncertainties associated with unknown 
corrections beyond nNLO through scale variations.  The motivation 
for this is that such scale variations induce changes in the result 
which are beyond the accuracy of the nNLO calculation, that is, 
both beyond NNLO and also subleading soft terms even at NNLO (since the scale dependence of
the approximate NNLO corrections is not exact). Indeed, this method is commonly accepted for 
standard fixed-order calculations.  However, one might question if that 
method is sufficient here, given that it produces the small uncertainty
estimate observed above. The major difference compared to full fixed-order
calculations is that the nNLO calculation misses subleading terms in the soft limit 
already at NNLO, so it is interesting to study more conservative ways of 
estimating their size.  
The most relevant of these subleading terms are next-to-leading power logarithms of the form
\begin{align}
\label{eq:sublogs}
\alpha_s^n \ln^m(1-z) \, , \qquad 0 \le m \le 2n -1 \, .
\end{align} 
These logarithms are singular but integrable  in the 
threshold region. In principle an analysis of these next-to-leading-power 
logarithms is possible within SCET \cite{Larkoski:2014bxa}. However, to date, only partial studies of these terms were completed, for the Abelian part of the Drell-Yan process and without employing the SCET framework, see for example \cite{Bonocore:2014wua, Bonocore:2015esa}.

In our case we can easily evaluate the cross section using a factor of  $1/z$ (which was the choice made in \cite{Ahrens:2010zv} and \cite{Broggio:2014yca} for the case of $t\bar{t}$ production) rather than $1/\sqrt{z}$ in the overall prefactor in (\ref{eq:soft-fact}).  When expanded around the limit $z\to1$, each of these 
two choices of prefactor produces next-to-leading-power logarithms of the form (\ref{eq:sublogs}),
but with coefficients which differ by a factor of two.  Therefore, the numerical difference 
between results evaluated with these two choices of prefactor gives additional insight into the generic
size of such subleading terms.
For both of these
choices, one can then evaluate the cross section at $\mu = \mu_0, 
2 \mu_0$, and $\mu_0/2$, as usual.  In this way, one obtains six
different values for the cross section and one can choose the interval
between the smallest and largest value as an estimate of the residual
perturbative uncertainty.

When this procedure is followed in the evaluation of the total cross
section at approximate NLO one obtains the prediction found in the
second column of Table~\ref{tab:LBCS}. The central value of the
approximate NLO cross section is determined by calculating the average
of the maximum and minimum among the six values of the cross section.
For the choice $\mu_0 = 620$~GeV, the central value and the uncertainty interval obtained in this way
are quite close to the complete NLO result shown in the first column
of Table~\ref{tab:LBCS}.  While this can be somewhat accidental, it
shows that, at least for the scale choice $\mu_0 = 620$~GeV, the
terms subleading in the soft limit are numerically of the same size of
the quark-gluon channel contributions, which is neglected in the soft
limit.  The last column in Table~\ref{tab:LBCS} shows the nNLO total
cross section calculated by estimating the residual perturbative
uncertainty as it was done in the third column for the approximate NLO
case. We stress once more that nNLO results are obtained by matching
the NNLO corrections in the soft limit to the complete NLO results. As
such, they include the same quark-gluon channel contribution included
in the NLO result. In this case the nNLO total cross section is larger
than the NLO one by 5 \% and the residual perturbative uncertainty is
roughly half the one found at NLO.  
It is important to keep in mind that we do not want to attribute 
any special value to this way of estimating the effects of the subleading terms. The procedure is simply  motivated by two goals: \emph{i)} to show that scale variation 
alone can lead to an underestimate of the residual perturbative uncertainty affecting approximate formulas, 
\emph{ii)} to take advantage of the fact that this procedure combined 
with the choice 
$\mu_0 = 620$ GeV allows us to obtain approximate NLO uncertainty bands 
which mimic nicely the scale uncertainty bands of the complete NLO 
calculations, as shown in Table~\ref{tab:LBCS}.

  For the smaller choice of the
  reference scale, $\mu_0 = 235$~GeV, the contribution of the
  quark-gluon channel to the scale uncertainty is dominant, and the
  method outlined above does not lead to approximate NLO predictions
  that mimic satisfactorily the complete NLO uncertainty band, as shown in the last row of Table~\ref{tab:LBCS}. However, the last column of Table~\ref{tab:LBCS} shows that the predictions for the nNLO total cross section obtained choosing
   $\mu_0 = 235$~GeV or $\mu_0 = 620$~GeV and by subsequently estimating the residual perturbative uncertainty with the more conservative method described above are  similar. In addition, as it can be seen from the last column of Table~\ref{tab:LBCS}, the nNLO cross section range obtained starting from $\mu_0 = 620$~GeV is larger than the one obtained starting from $\mu_0 = 235$~GeV; we also observe that the former nearly contains the latter.

\begin{figure}[h!]
	\begin{center}
		\begin{tabular}{cc}
			\includegraphics[width=7cm]{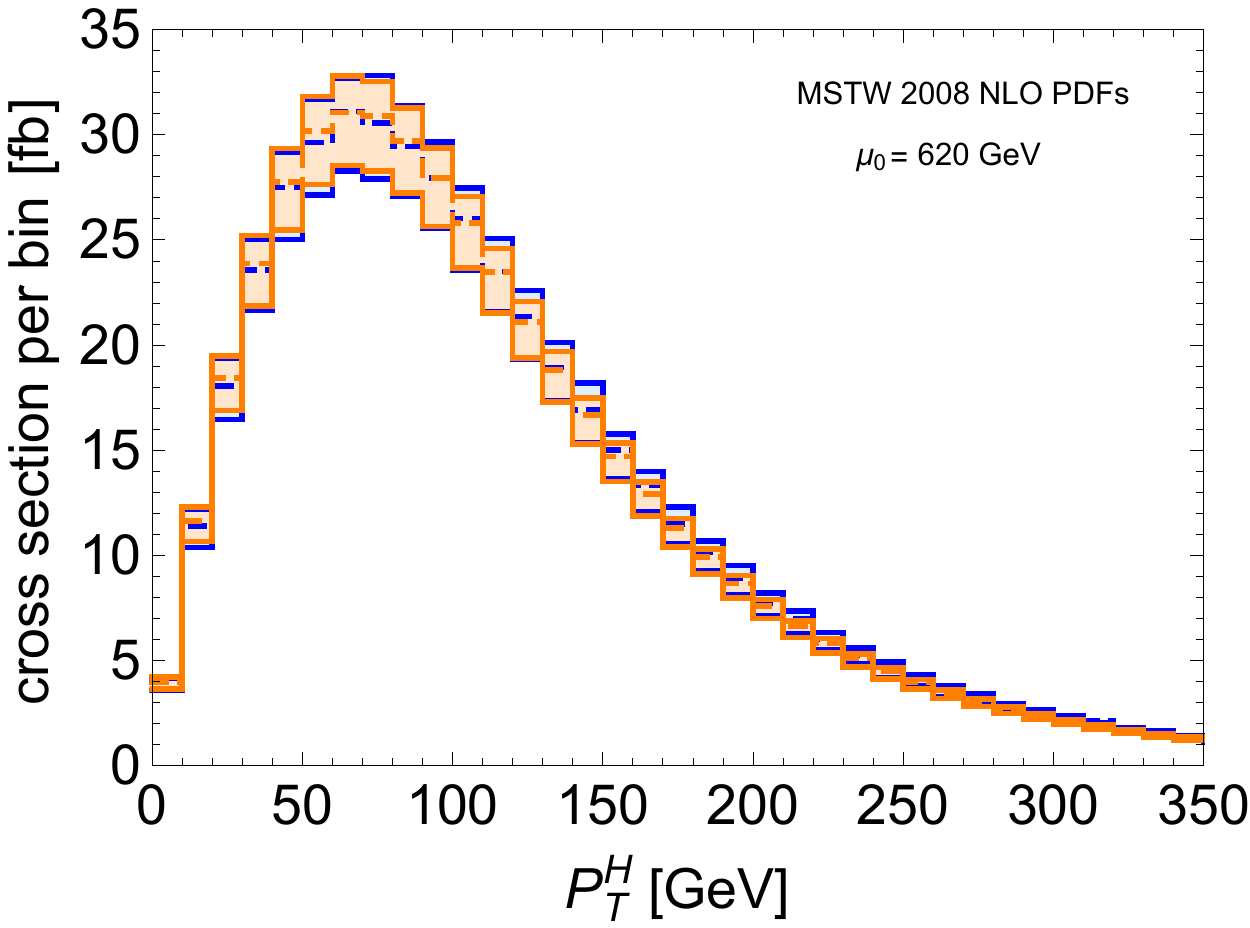} & \includegraphics[width=7cm]{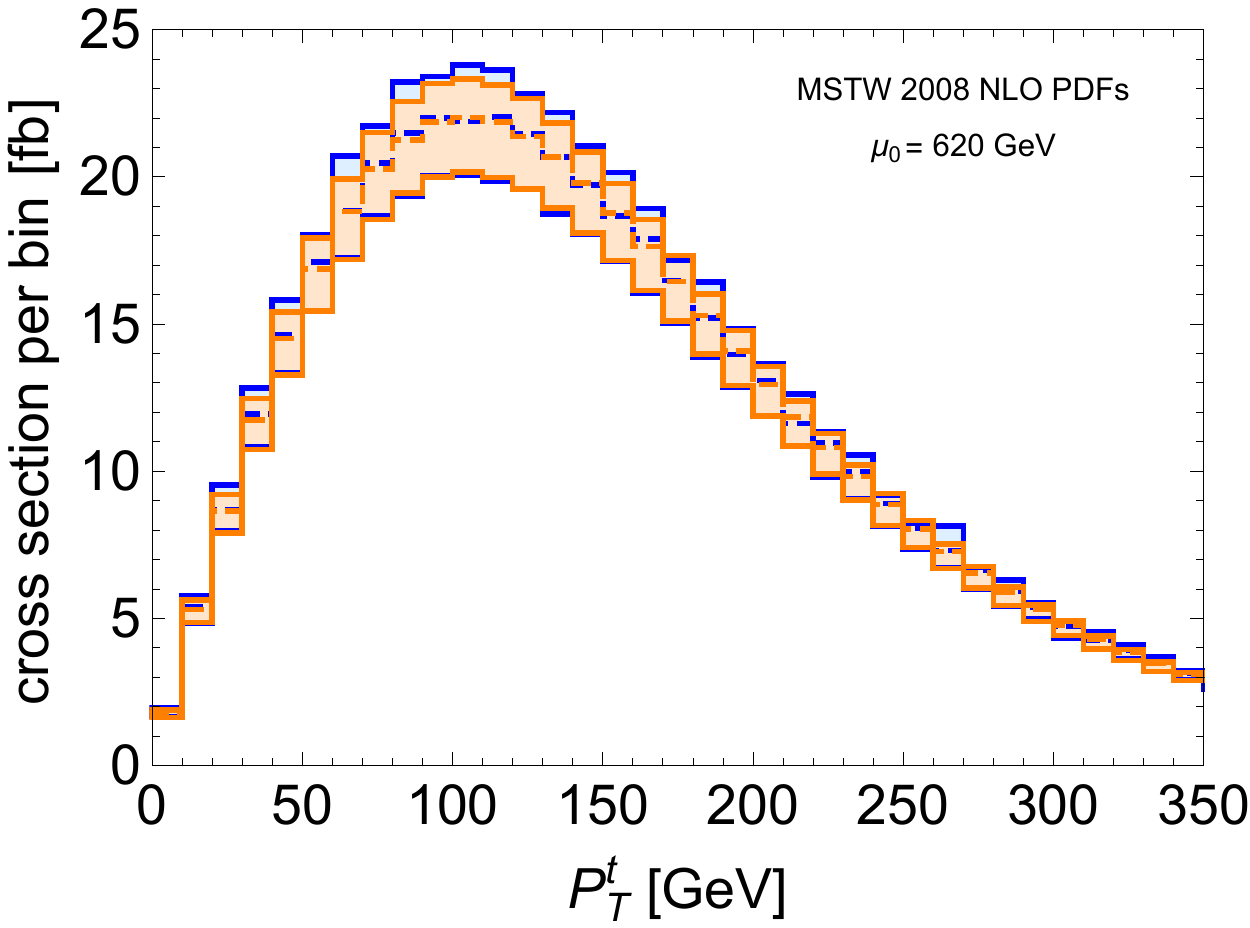} \\
			\includegraphics[width=7cm]{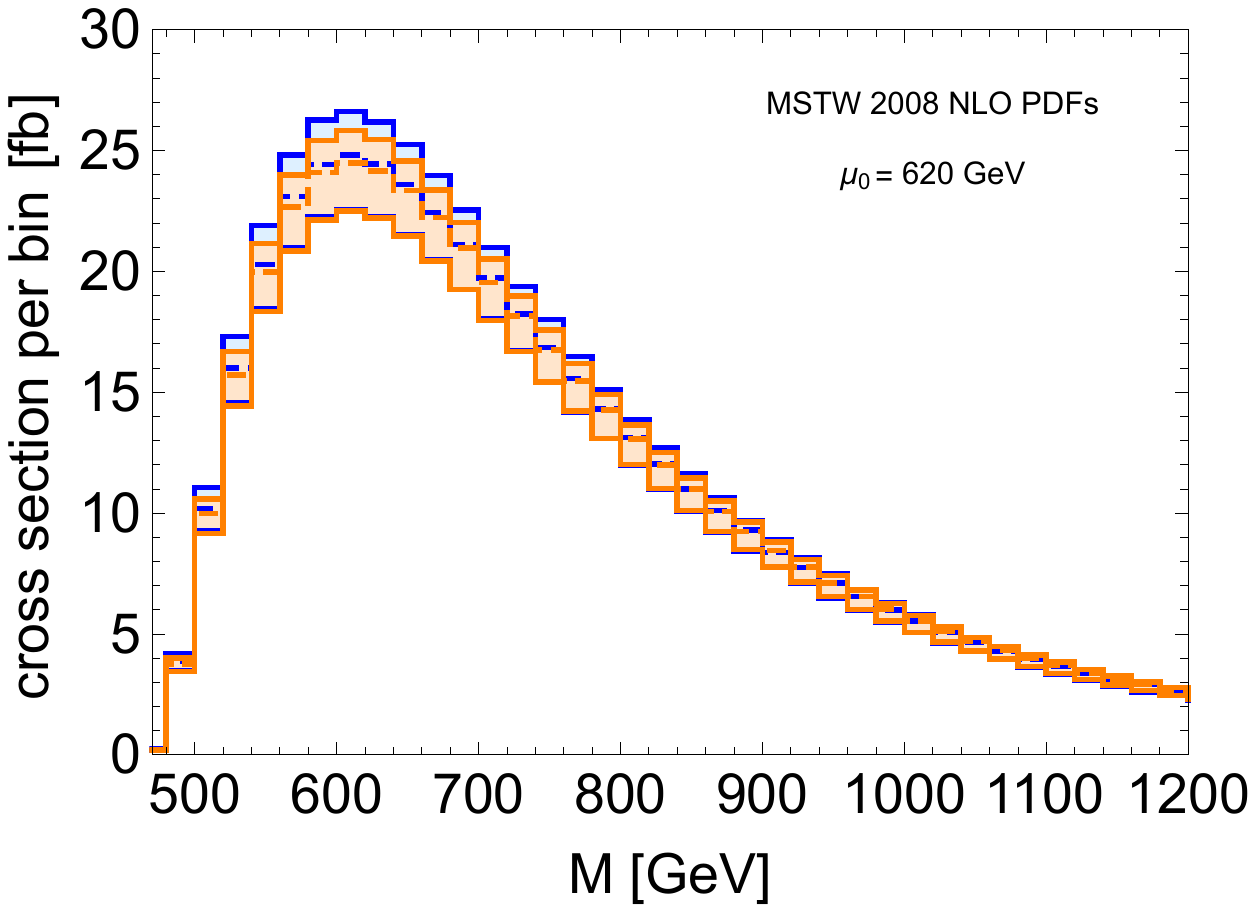} & \includegraphics[width=7cm]{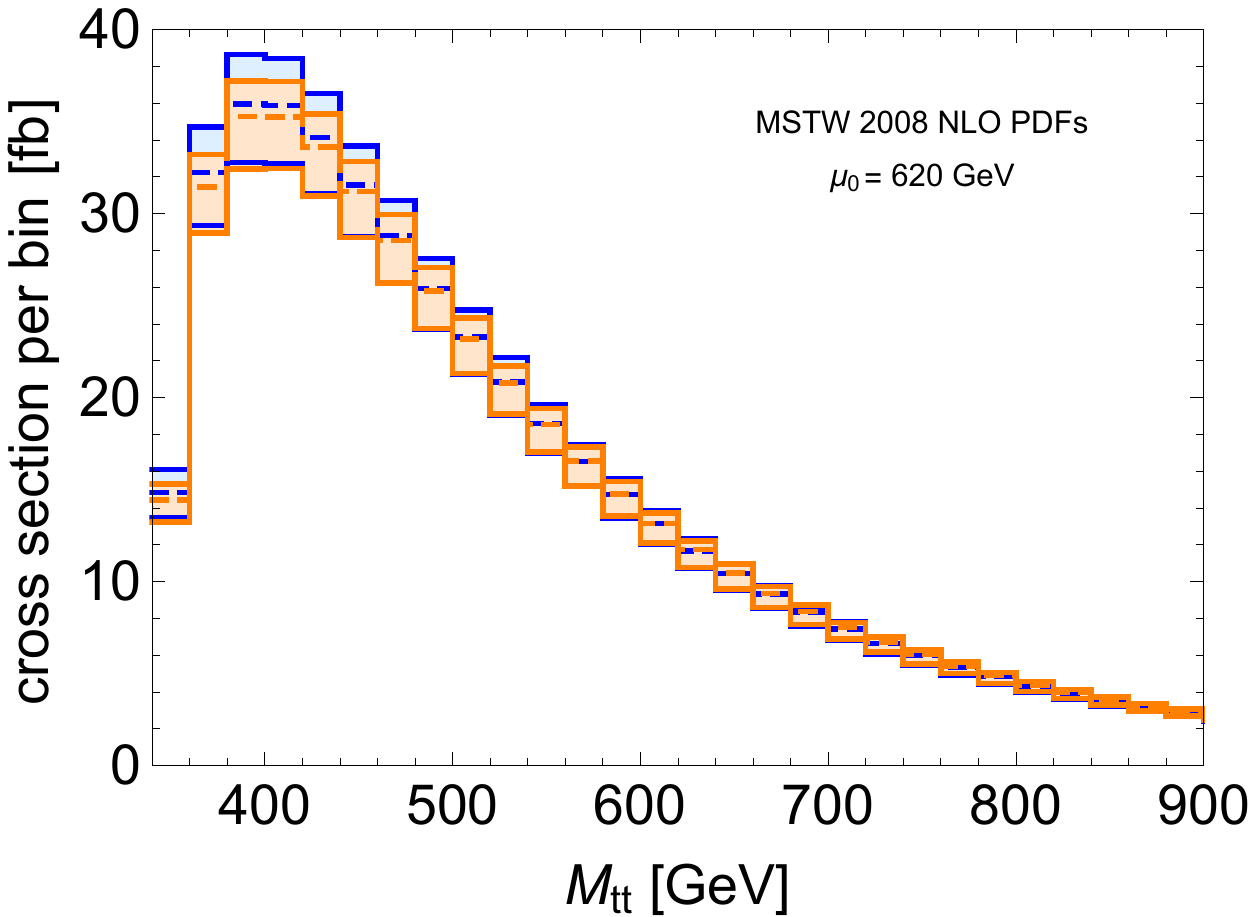} \\
			\includegraphics[width=7cm]{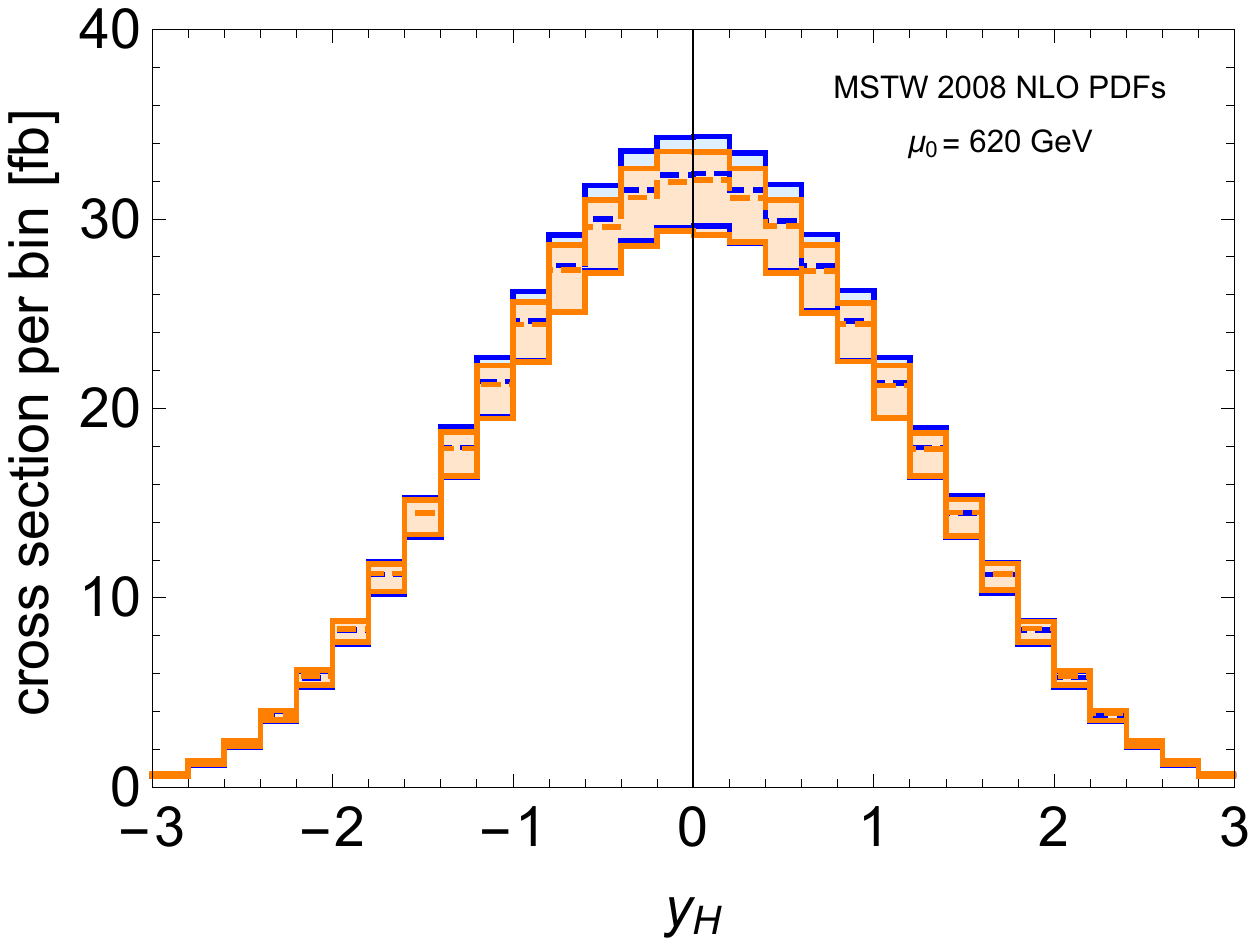} & \includegraphics[width=7cm]{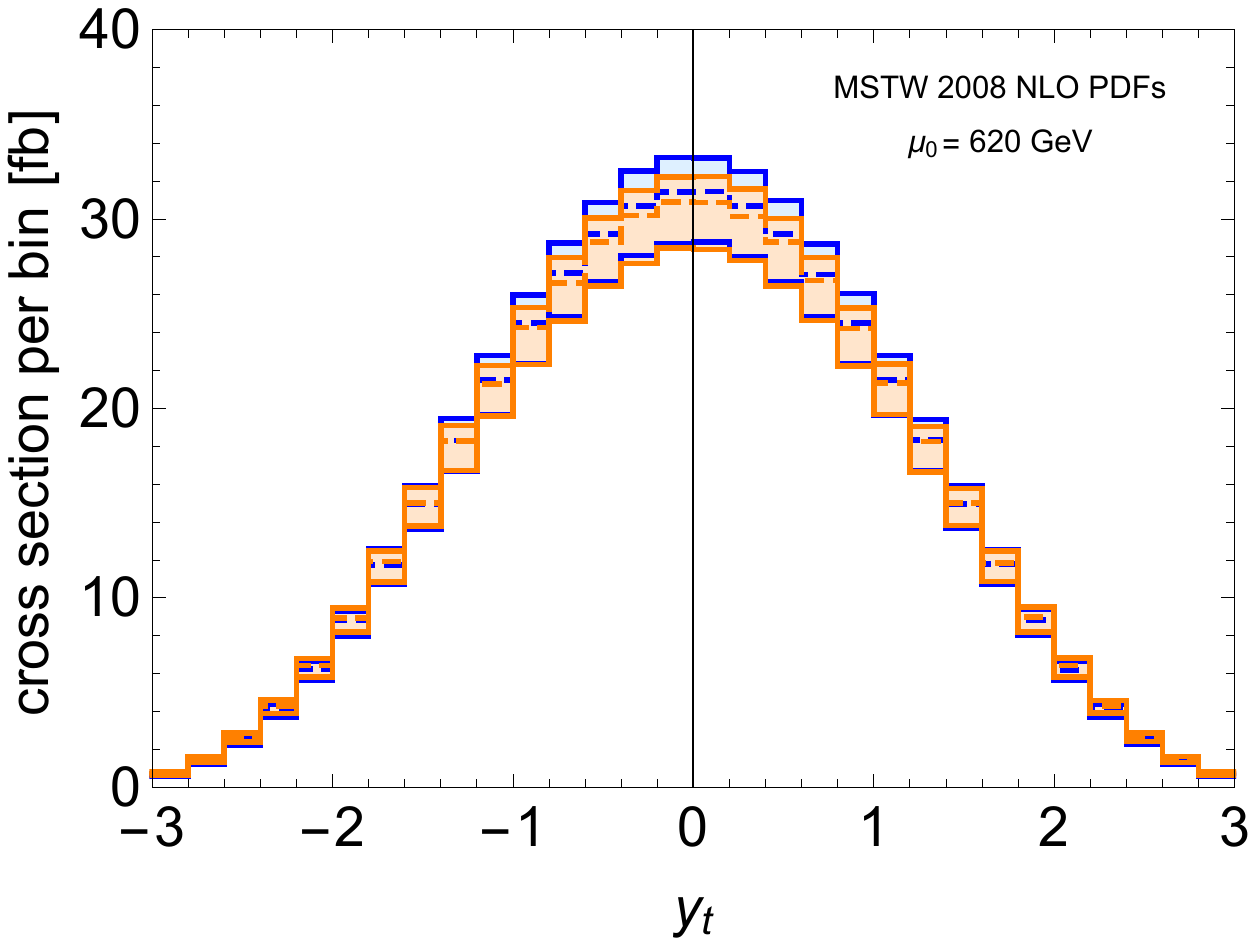} \\
		\end{tabular}
	\end{center}
	\caption{Differential distributions at approximate NLO (orange band) compared to the NLO calculation carried out with \texttt{MG5} by excluding the quark-gluon channel (blue band).
		The central value of the scale is set to $\mu_0 = 620$~GeV and varied in the range $[ \mu_0 /2 ,2 \mu_0 ]$. MSTW 2008 NLO PDFs were used in all cases. The differential distributions shown are, from top left, Higgs transverse momentum $p_T^H$, top-quark transverse momentum $p_T^t$, invariant mass of the top pair and Higgs boson $M$, invariant mass of the top pair $M_{t \bar{t}}$, Higgs rapidity $y_H$, and top-quark rapidity $y_t$. \label{fig:distNLO620}
	}
\end{figure}

\subsection{Differential distributions}
\label{sec:dxs}
An advantage of our approach is that it can be used to calculate any
arbitrary differential cross section.  We do this by employing
standard Monte Carlo methods. In particular, during the evaluation of
the approximate NNLO corrections to the total cross section in
(\ref{eq:soft-fact}), we use the phase-space and four-momenta
parameterizations described in Section~\ref{sec:soft-limit} in order
to create binned distributions.\footnote{We have performed the
  numerical integrations by employing the {\tt Cuba} library
  \cite{Hahn:2004fe}, and are grateful to Thomas Hahn for advice on
  extracting the integration weights needed to obtain correctly
  normalized distributions.}

In order to illustrate this approach we consider six differential
distributions:
\begin{itemize}
        \item distribution differential in the invariant mass $M$ of the 
$t\bar{t}H$ final state;
        \item distribution differential in the invariant mass $M_{t \bar t}$ of 
the $t\bar{t}$ pair;
	\item distribution differential in the transverse momentum of the Higgs boson, $p_T^H$;
	\item distribution differential in the transverse momentum of the top quark, $p_T^t$;
	\item distribution differential in the Higgs boson rapidity, $y_H$;
	\item distribution differential in the top quark rapidity, $y_t$.
\end{itemize} 
All of the distributions are evaluated in the laboratory frame.  We
have validated the results from our Monte Carlo based method by
explicitly changing variables and calculating the first three of the
distributions above by standard numerical integration in bins; the agreement between the two methods gives us confidence of the ability of our Monte Carlo
implementation to calculate arbitrary distributions which are
differential with respect to variables depending on the momenta of the
massive particles in the final state.

\begin{figure}[h]
	\begin{center}
		\begin{tabular}{cc}
			\includegraphics[width=7cm]{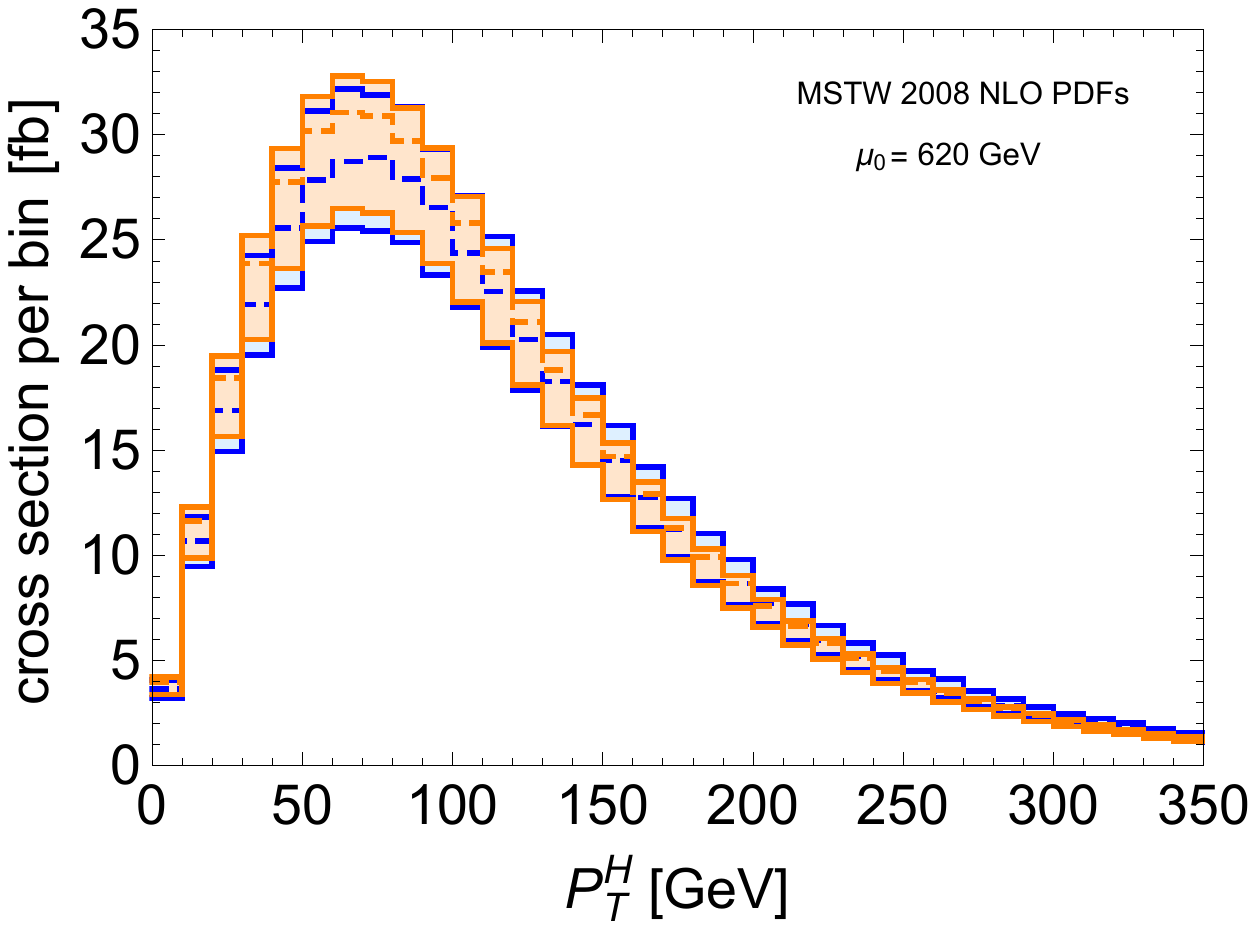} & \includegraphics[width=7cm]{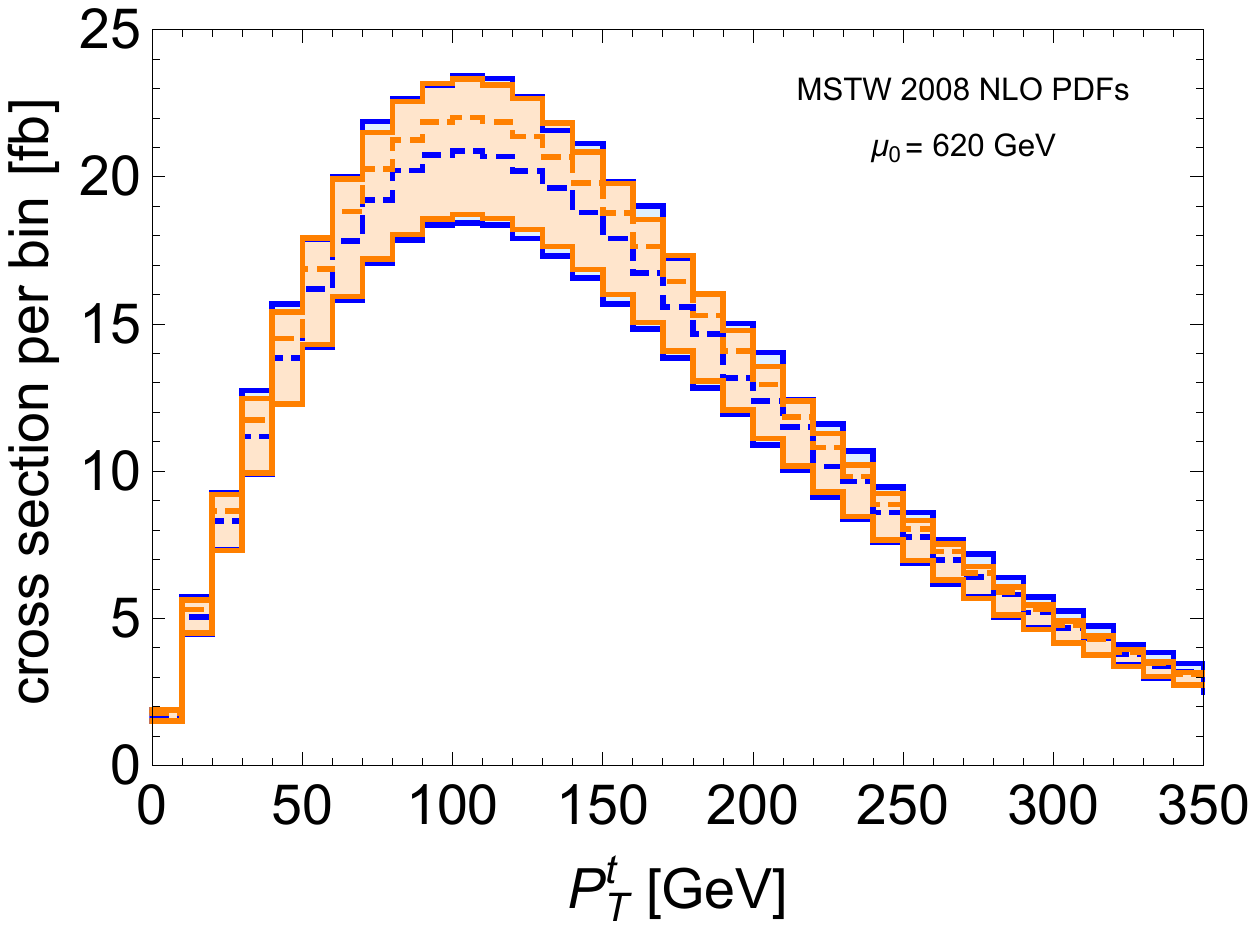} \\
			\includegraphics[width=7cm]{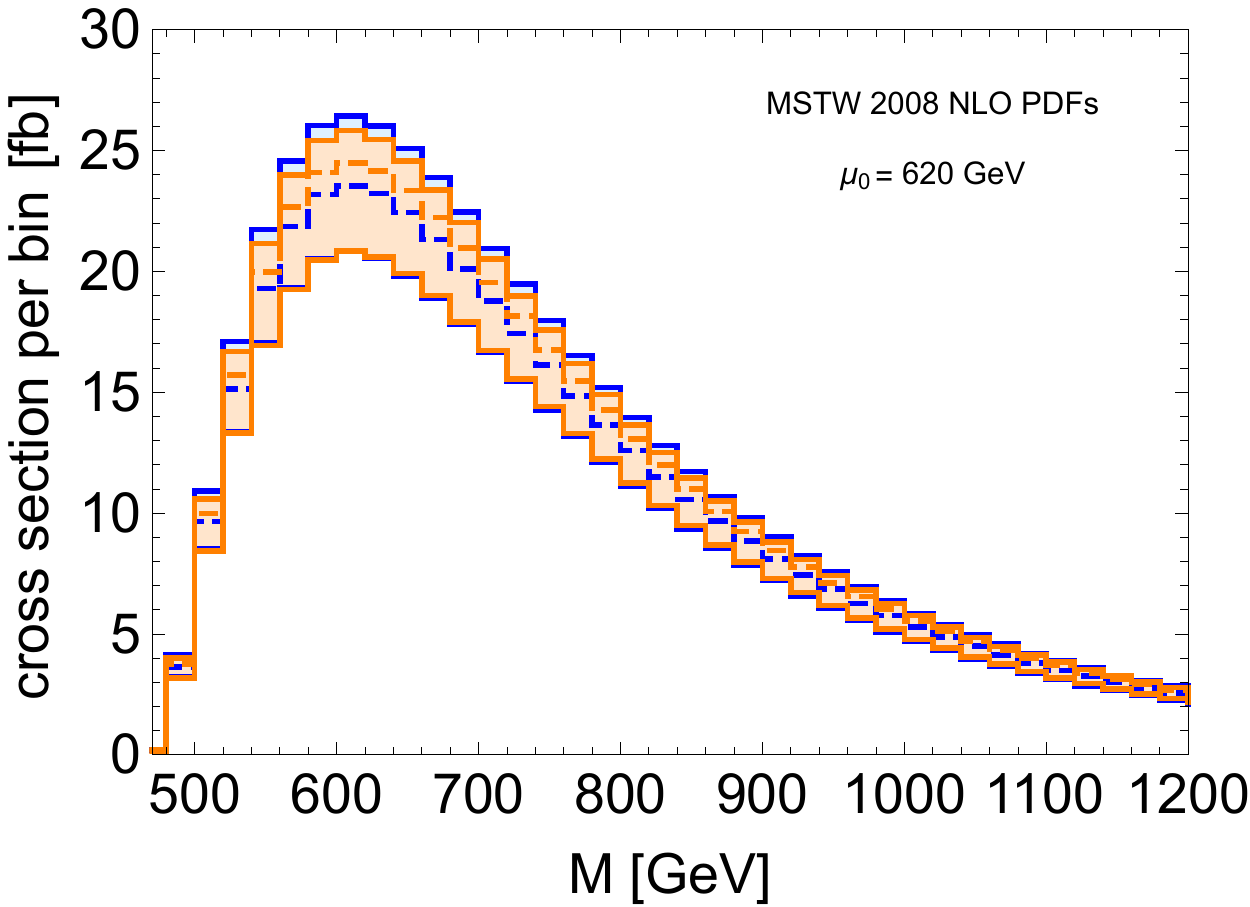} & \includegraphics[width=7cm]{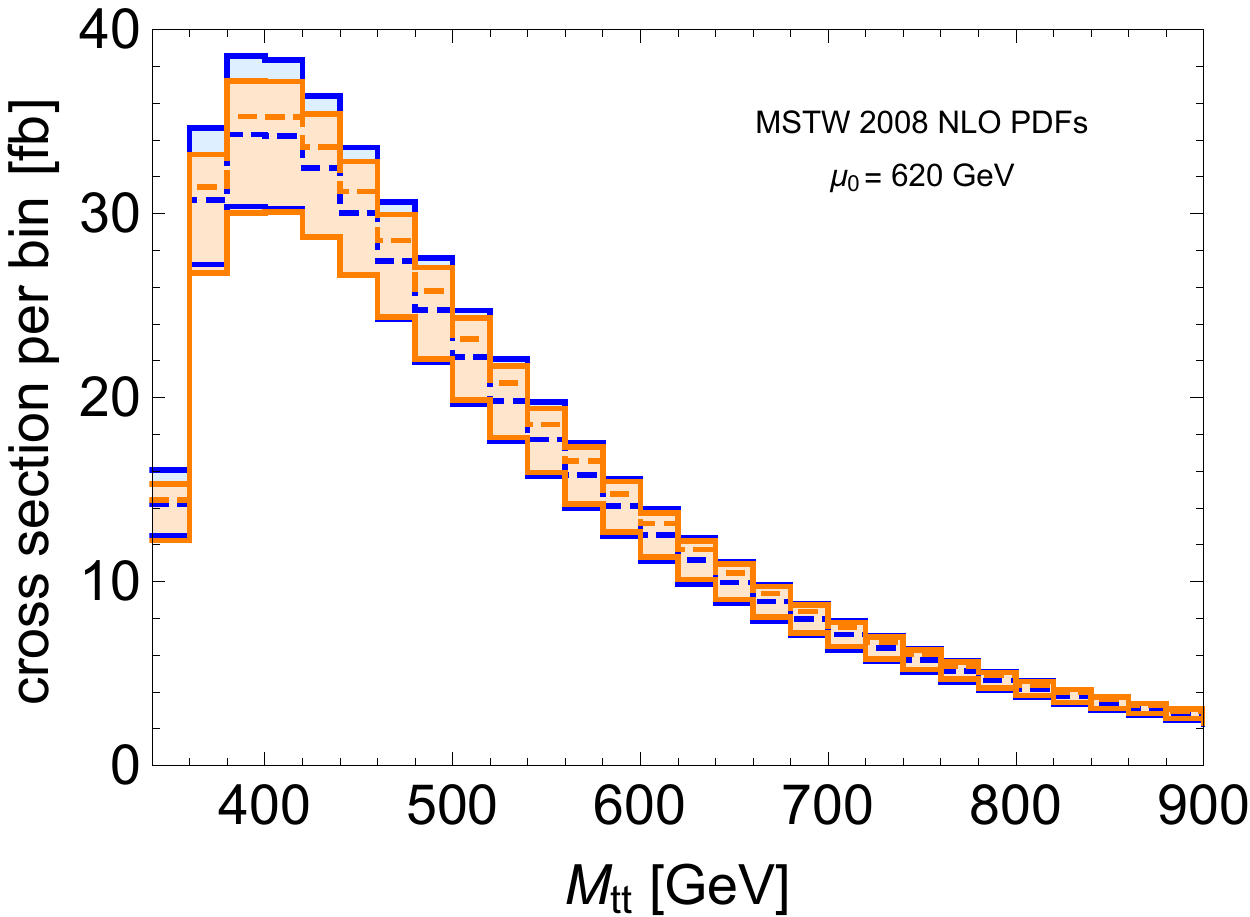} \\
			\includegraphics[width=7cm]{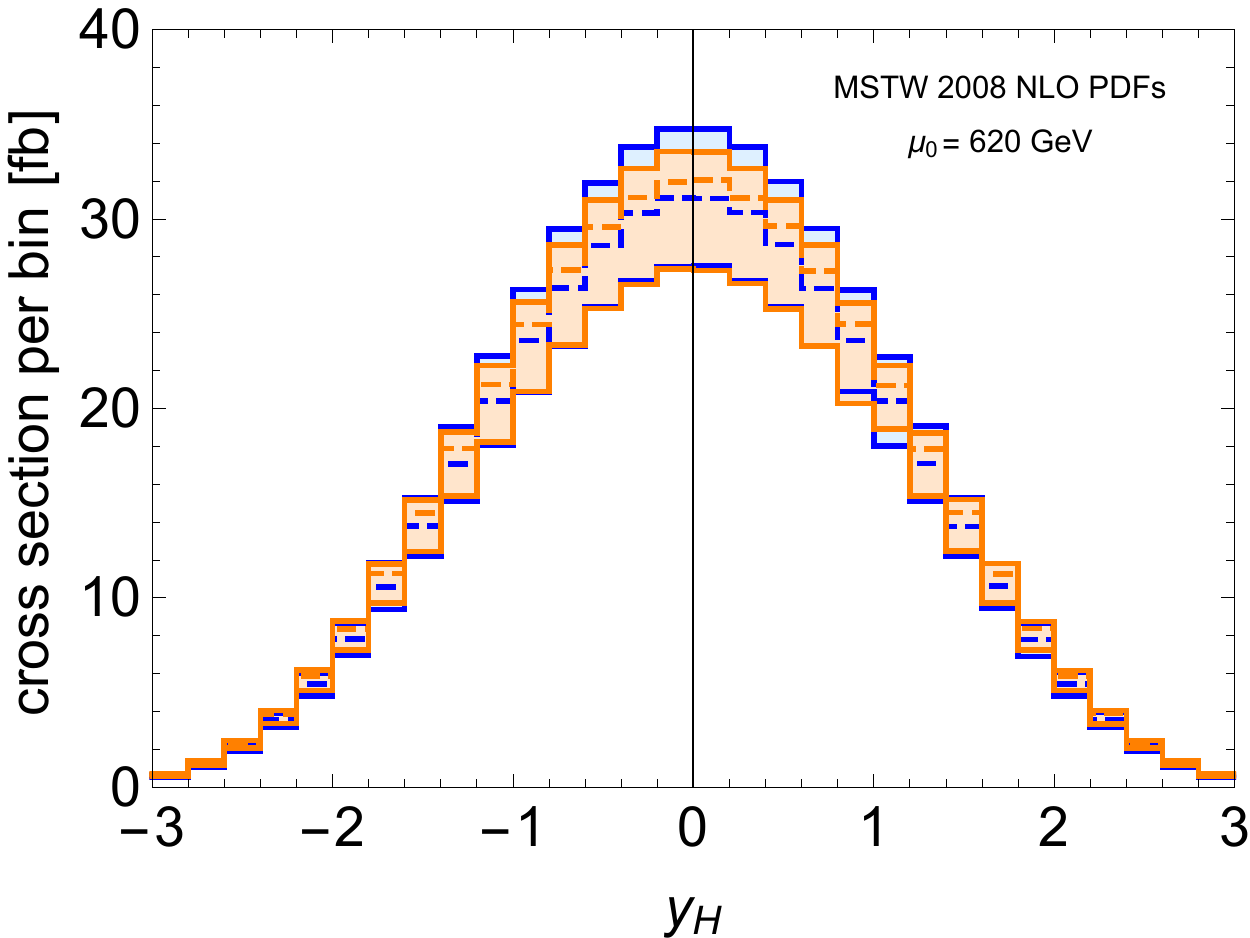} & \includegraphics[width=7cm]{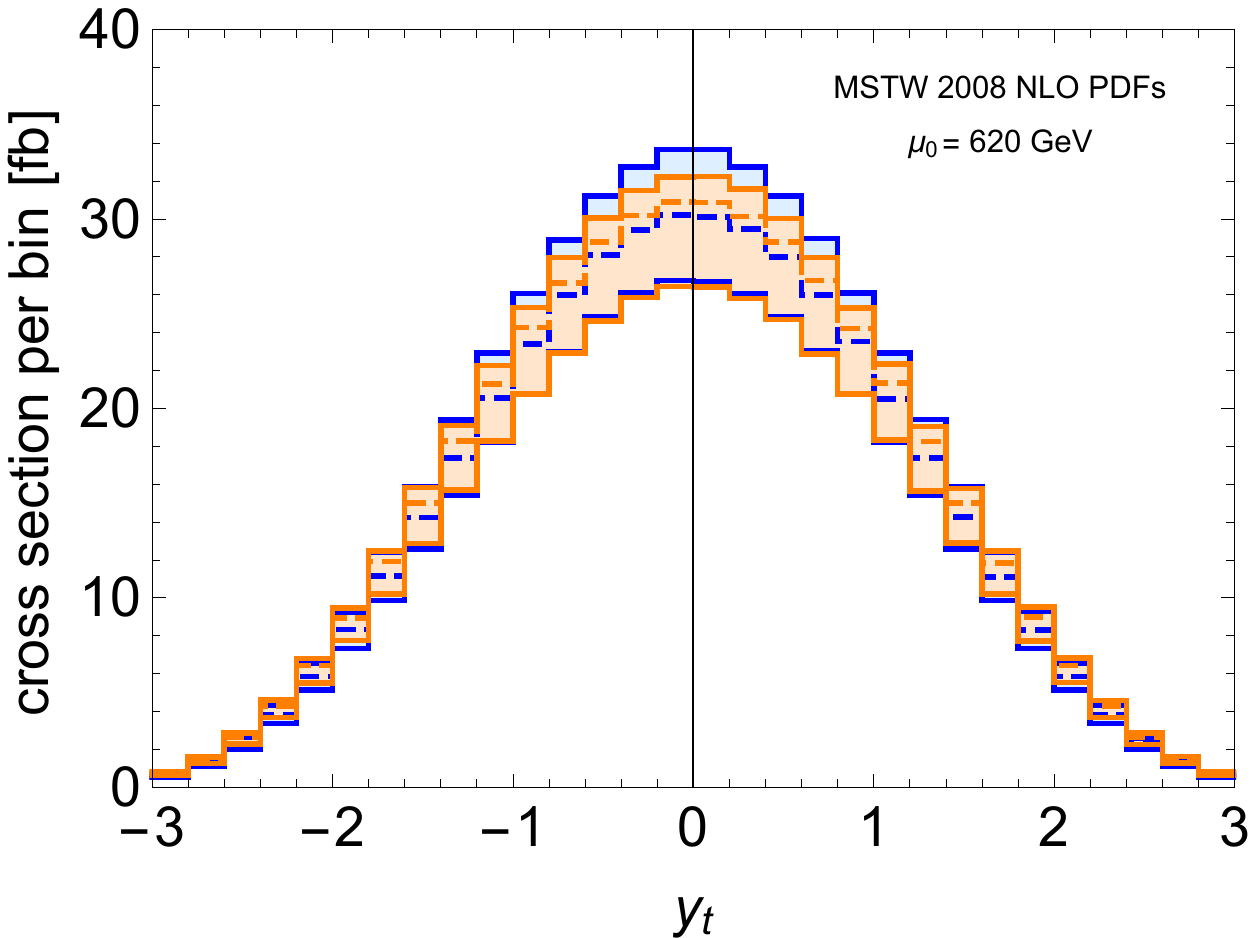} \\
		\end{tabular}
	\end{center}
	\caption{Differential distributions at approximate NLO (orange band) compared to the complete NLO calculation carried out with \texttt{MG5}  (blue band).
	The central value of the scale is set to $\mu_0 = 620$~GeV. The approximate NLO band was obtained  by considering different sets of subleading corrections and by varying the scale in the range $[ \mu_0 /2 ,2 \mu_0 ]$.		
	MSTW 2008 NLO PDFs were used in all cases. 
	\label{fig:distNLO620large}
	}
\end{figure}

As with the total cross section, we begin with a comparison between
the full and approximate NLO results. Figures~\ref{fig:MdistNLO}
and~\ref{fig:pHfig} show this comparison for differential
distributions in $M$ and $p_T^H$ respectively, evaluated at the
default scale choice $\mu_0 = 620$~GeV.  In addition to the full NLO
results, we have also shown NLO results with the quark-gluon channel
omitted.  As seen from the bottom panels of the figures, the
approximate NLO results recover much more than 90\% of the 
NLO result across all bins, if one excludes from the latter the
contributions of the $qg$ channel.  Even if one compares the
approximate NLO distributions with the complete NLO calculation one
finds that the approximate result never differs from the complete one
by more than $10 \%$.  As shown in Figures~\ref{fig:MdistNLO235}
and~\ref{fig:pHfig235}, approximate NLO distributions satisfactorily
reproduce the NLO calculations (without the quark gluon channel) also
in the case in which one employs the traditional scale choice $\mu_0 =
235$~ GeV.  In Figure~\ref{fig:distNLO620} we show a comparison
between the NLO result with the quark-gluon channel excluded and the
approximate NLO results, this time for all six distributions listed
above and including bands from scale variation as described in the
caption.  We see that in all cases the agreement is quite good also at
values of the scale different from our default choice $\mu_0 =
620$~GeV.  
Figure~\ref{fig:distNLO620large} shows the approximate NLO
distributions in the case in which the uncertainty band is estimated
by keeping into account the numerical effect of terms subleading in
the soft limit, according to the method described in
Section~\ref{sec:xs}. The figure also shows the full (i.e. including
the quark-gluon channel contribution) NLO distributions including
their scale variation, evaluated with {\tt MG5}. By looking at
Figure~\ref{fig:distNLO620large} one can see that, as expected, the
approximate NLO bands are larger than the ones in
Figure~\ref{fig:distNLO620}. However, similarly to the case of the
total cross section, one also observes that these larger bands at
approximate NLO reproduce quite well the scale uncertainty of the
complete NLO distributions. This hints to the fact that the
uncertainty bands of the nNLO distributions evaluated in this way
could satisfactorily mimic the scale uncertainty bands of the
(unknown) full NNLO distributions.

\begin{figure}[h]
	\begin{center}
		\begin{tabular}{cc}
			\includegraphics[width=7cm]{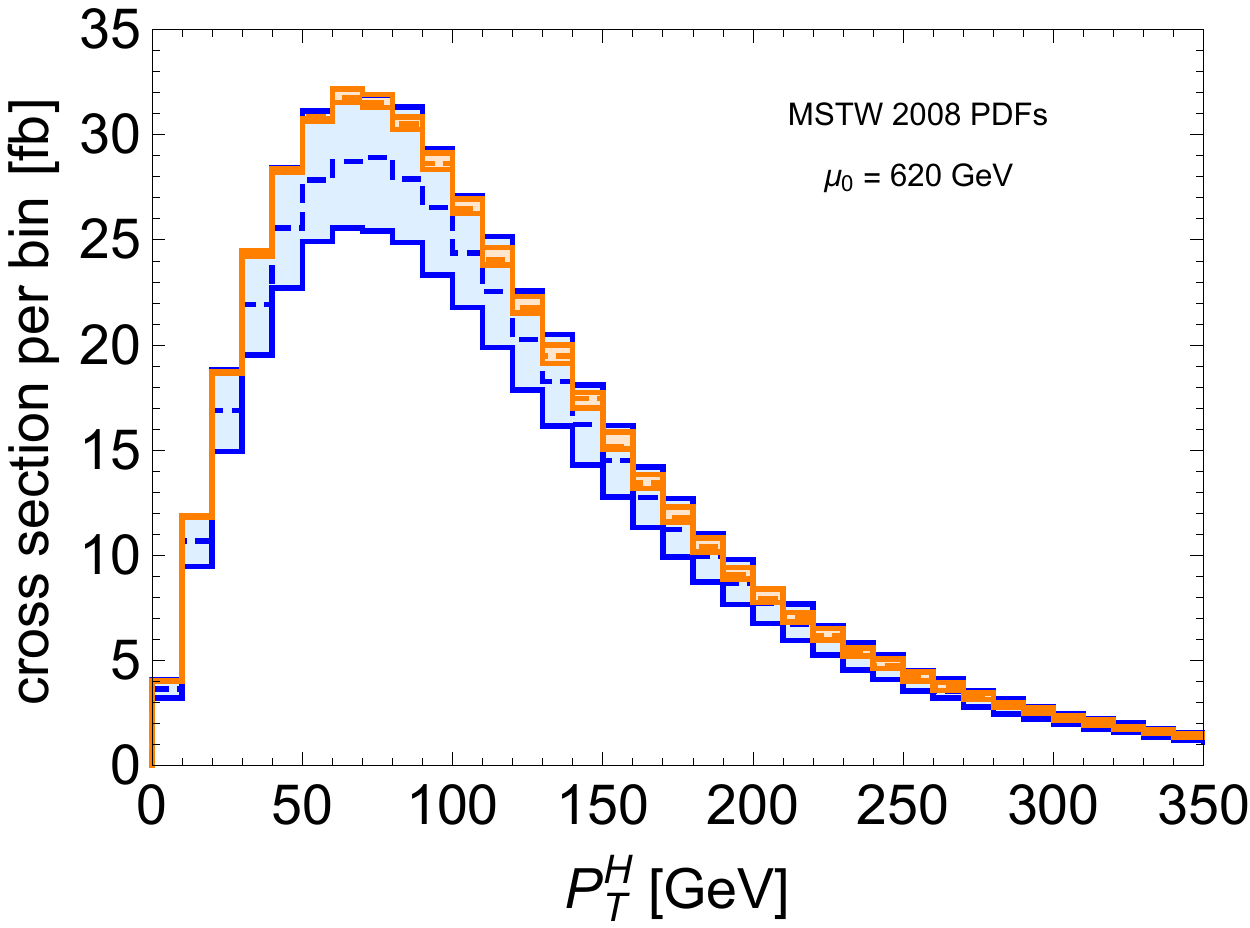} & \includegraphics[width=7cm]{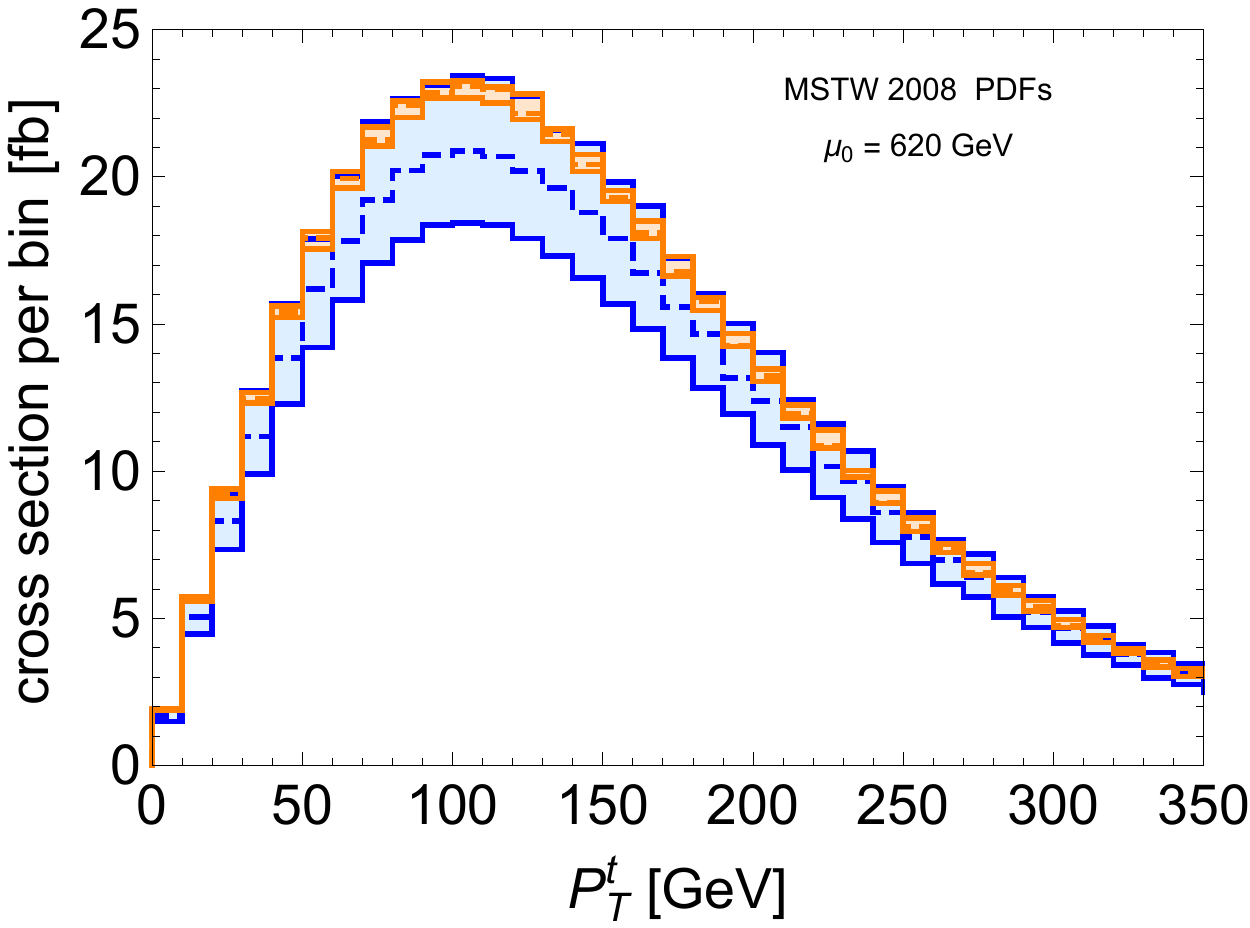} \\
			\includegraphics[width=7cm]{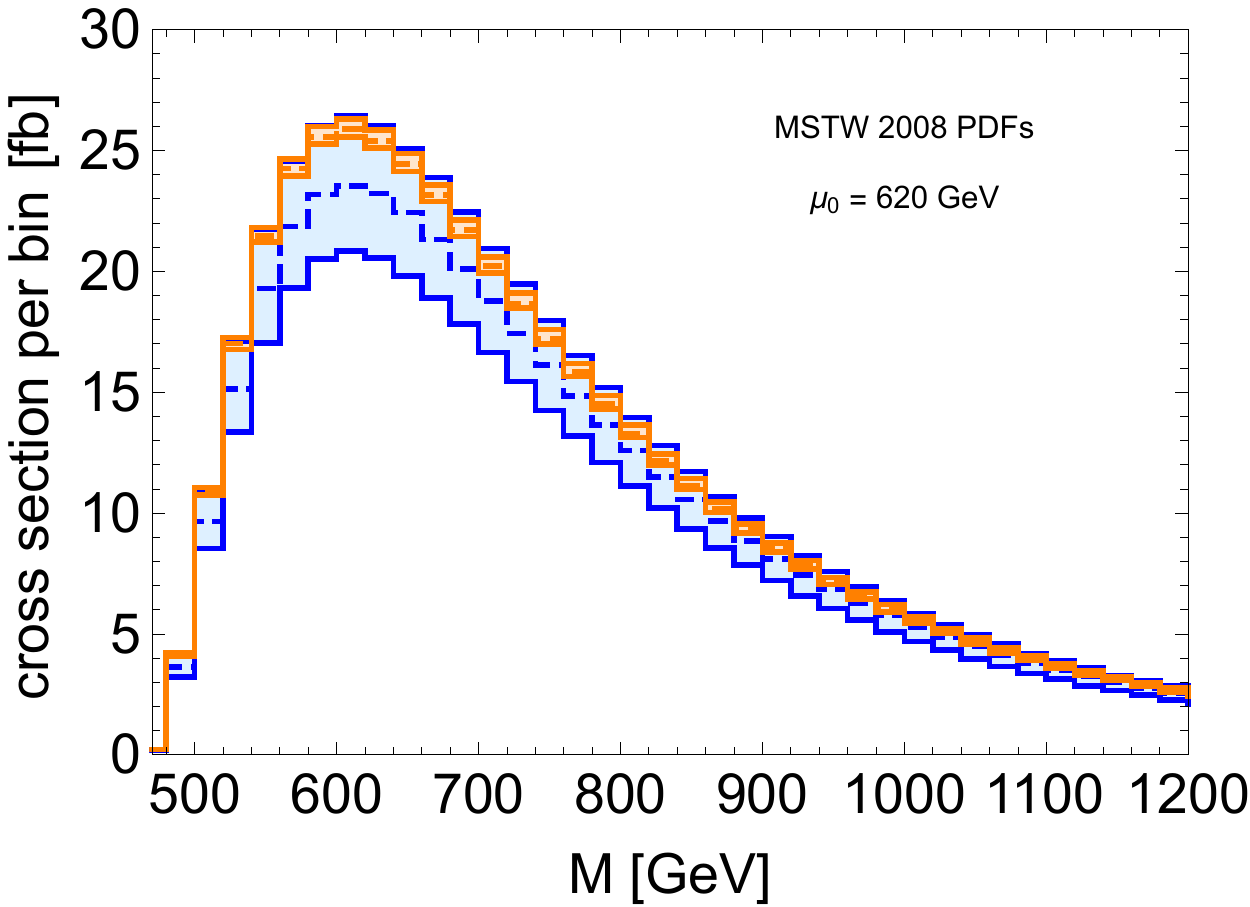} & \includegraphics[width=7cm]{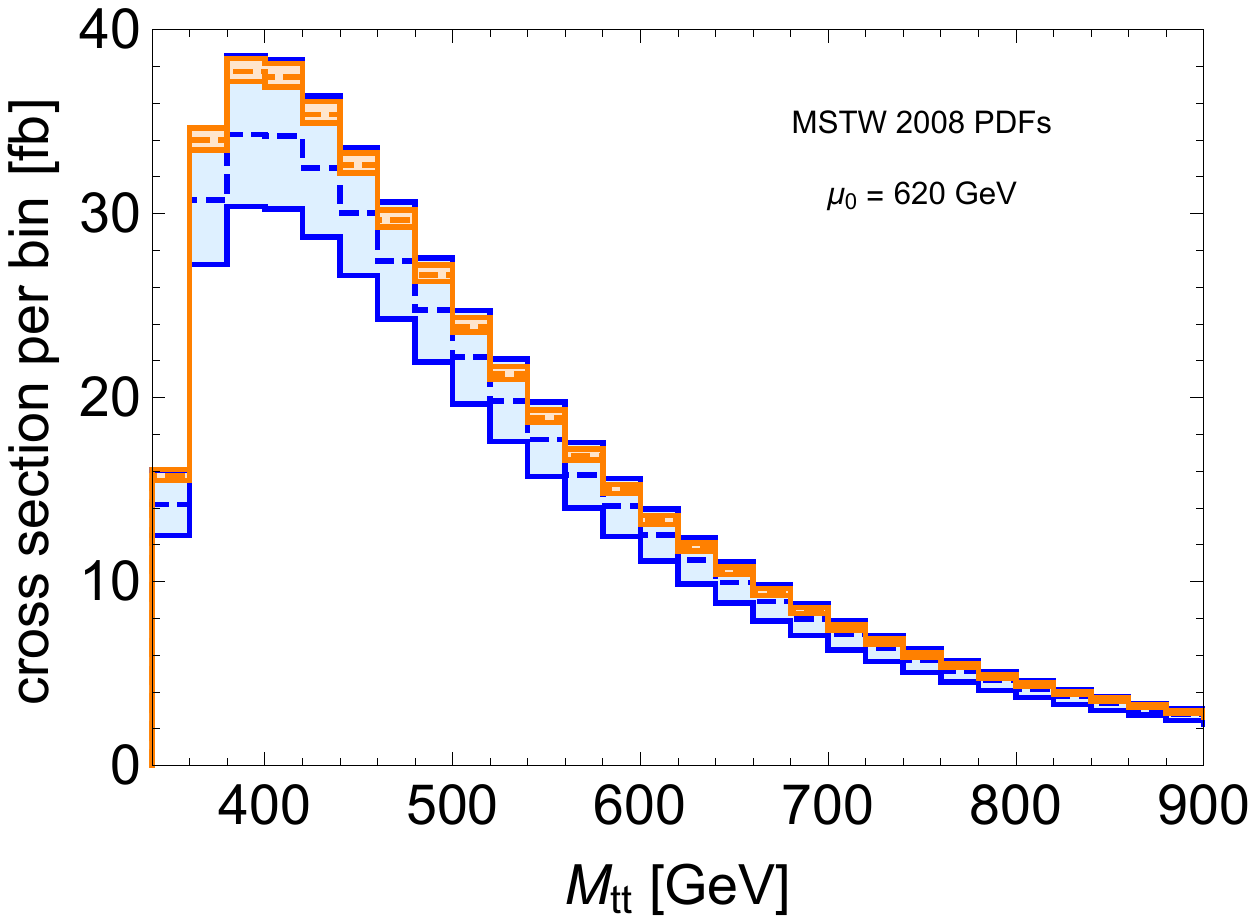} \\
			\includegraphics[width=7cm]{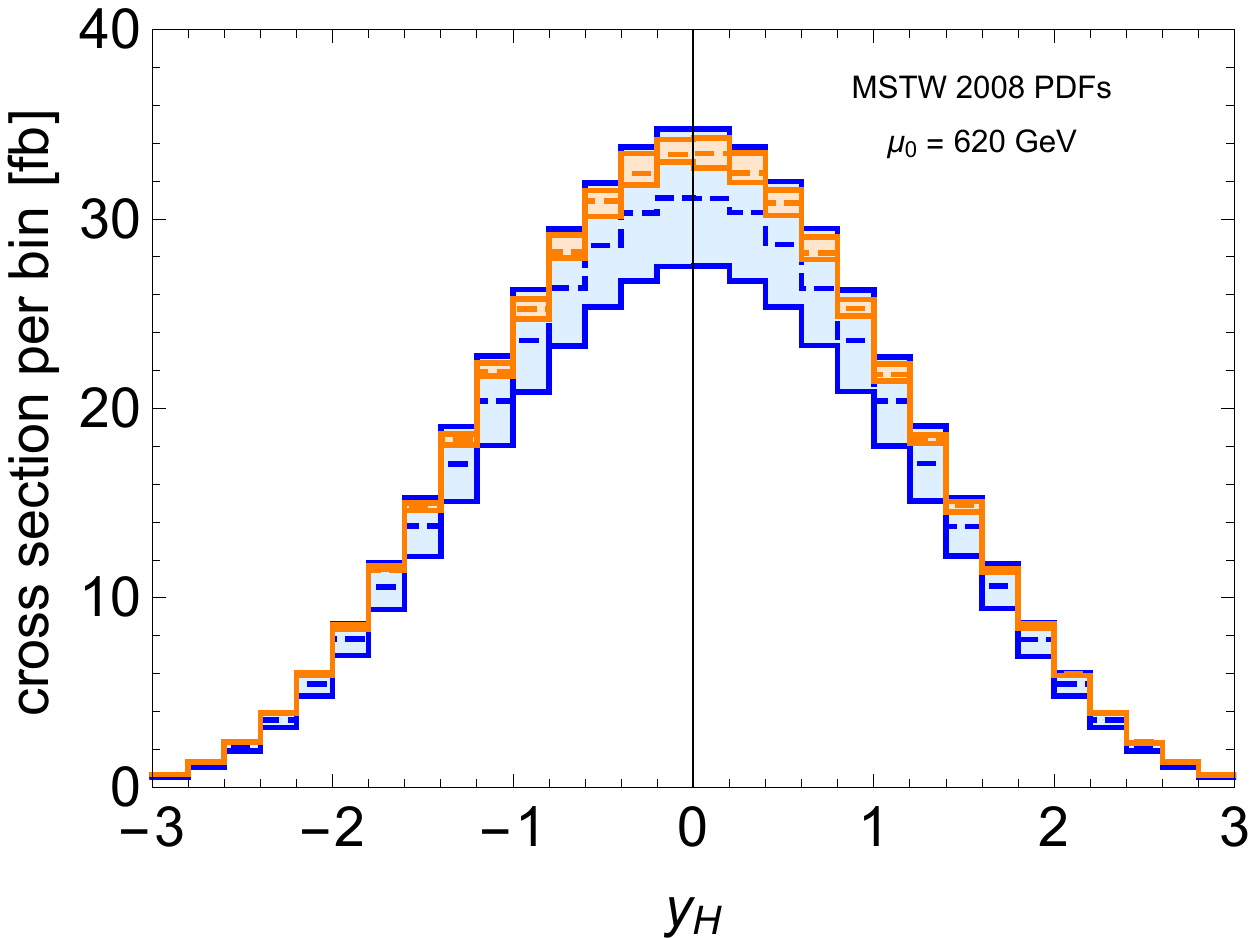} & \includegraphics[width=7cm]{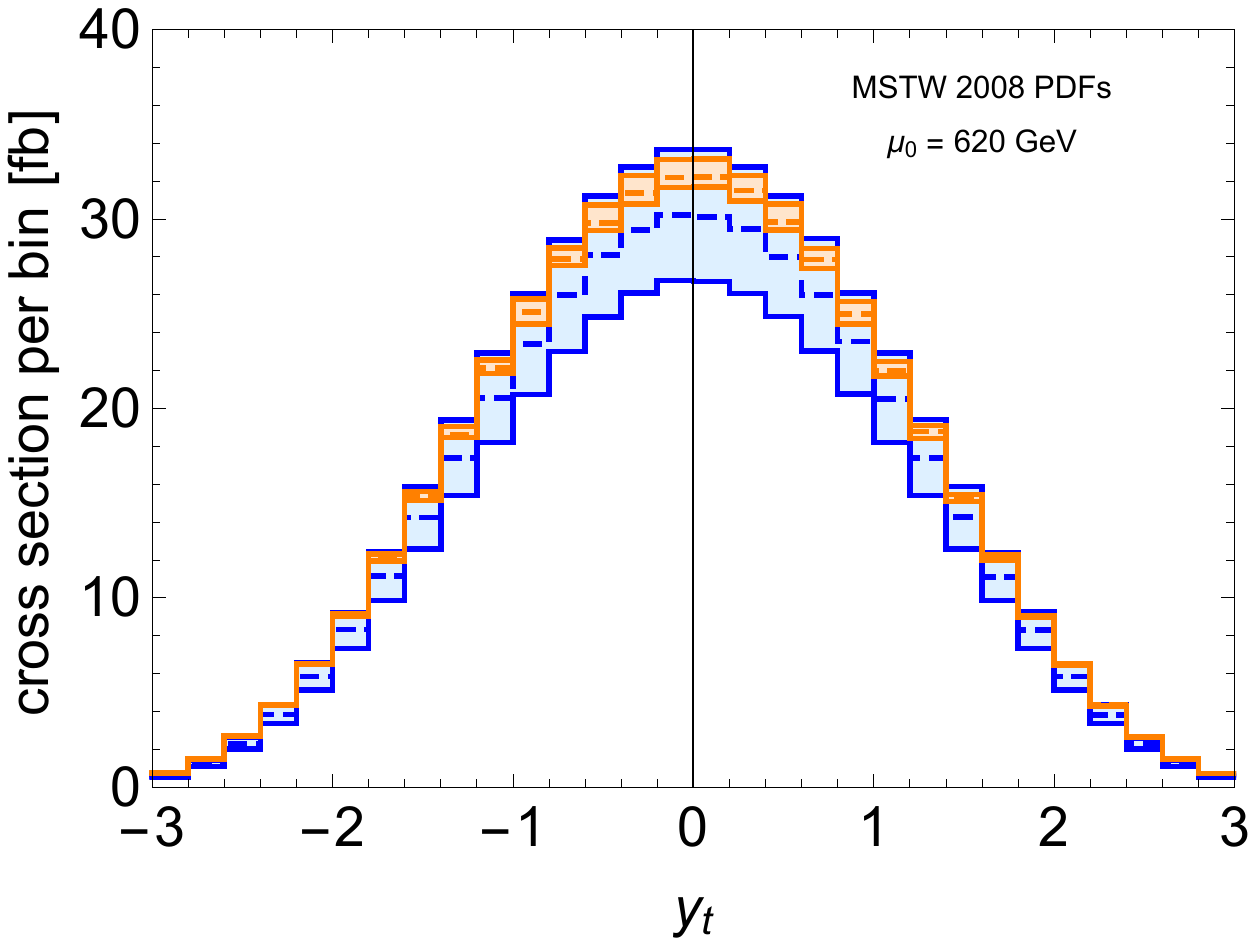} \\
		\end{tabular}
	\end{center}
	\caption{Differential distributions at nNLO (orange band) compared to the NLO calculation carried out with \texttt{MG5}  (blue band). NLO distributions are evaluated with NLO PDFs, nNLO distributions with NNLO PDFs.
		The central value of the scale is set to $\mu_0 = 620$~GeV and varied in the range $[ \mu_0 /2 ,2 \mu_0 ]$.  \label{fig:distnnLOvsNLO}
	}
\end{figure}

\begin{figure}[h]
	\begin{center}
		\begin{tabular}{cc}
			\includegraphics[width=7cm]{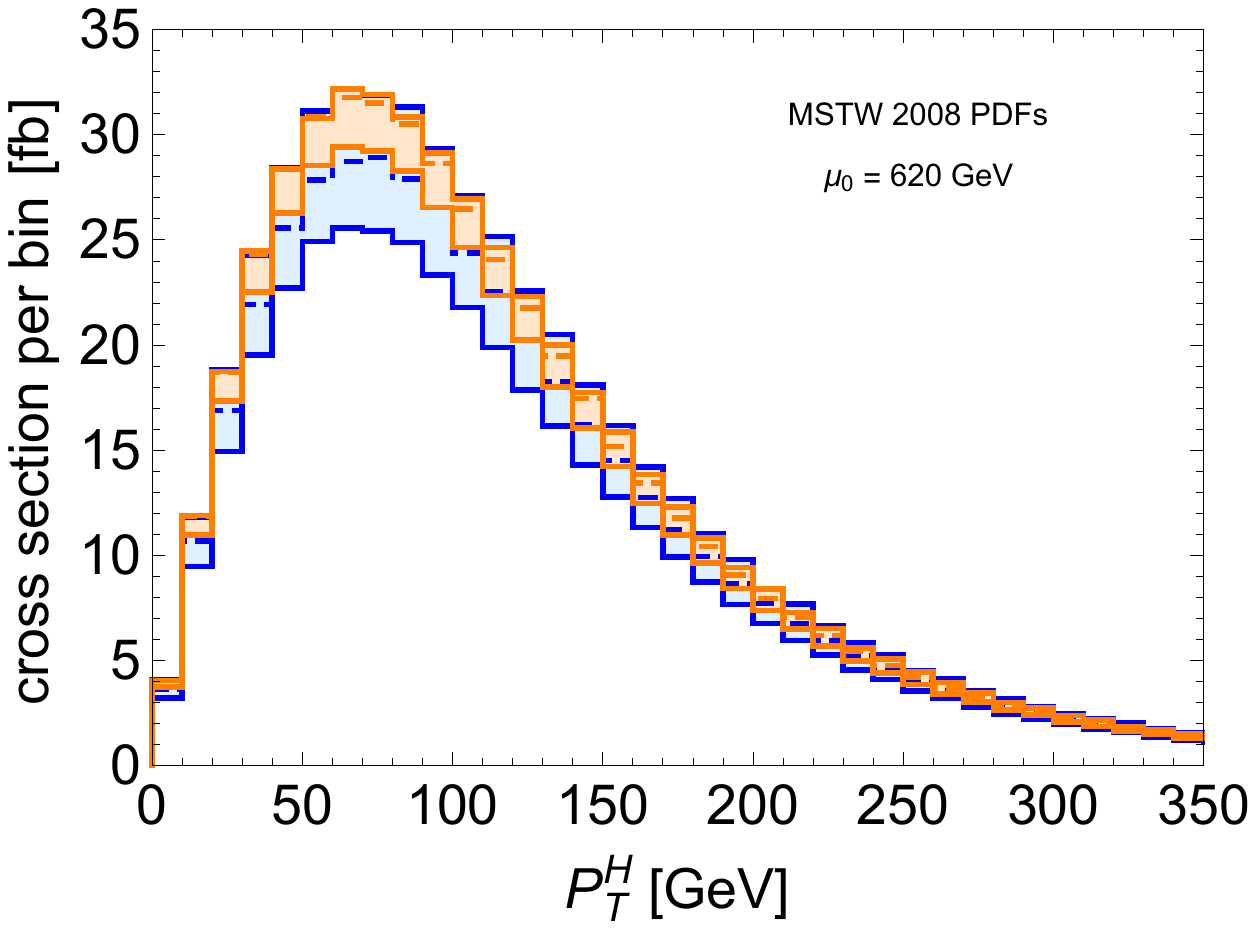} & \includegraphics[width=7cm]{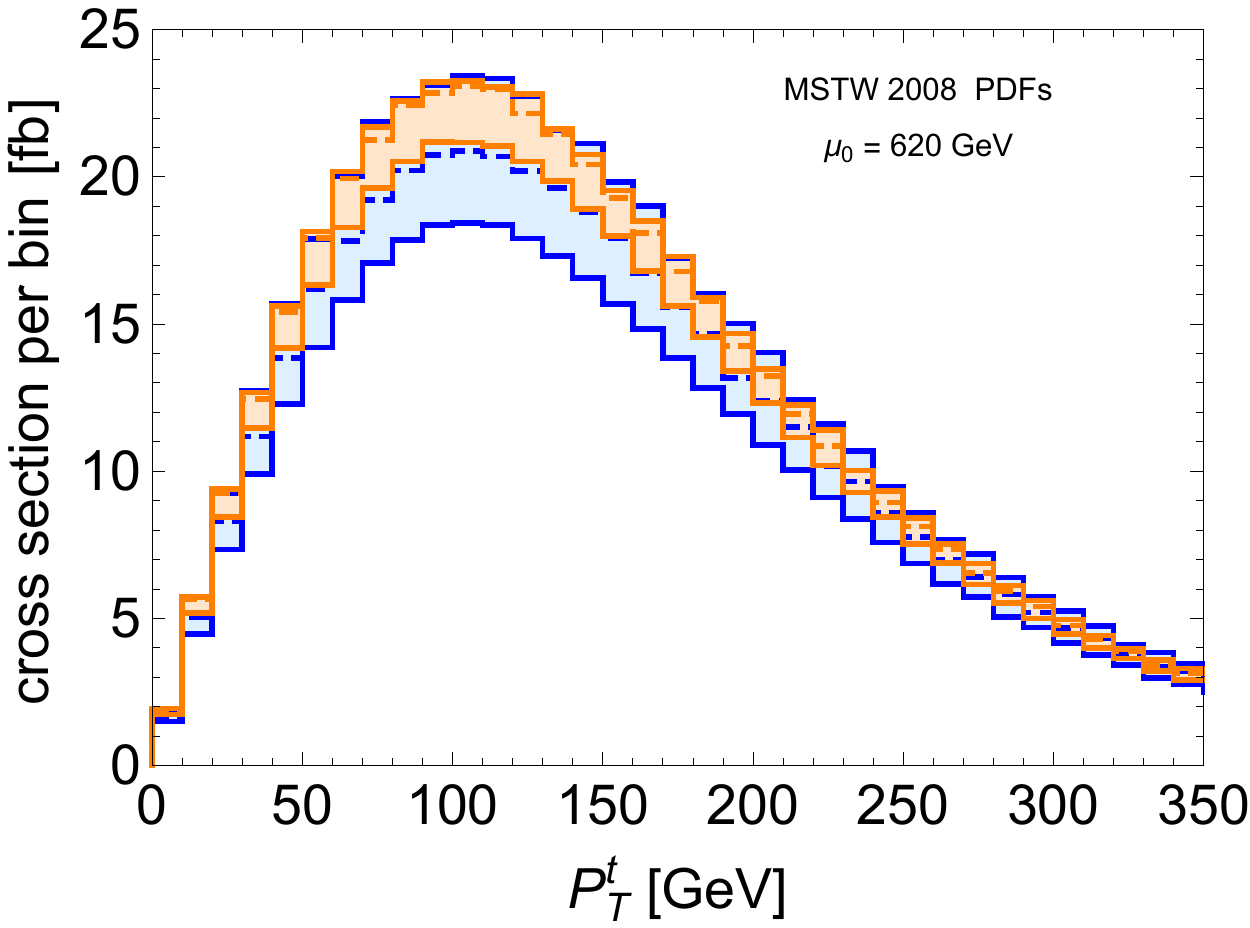} \\
			\includegraphics[width=7cm]{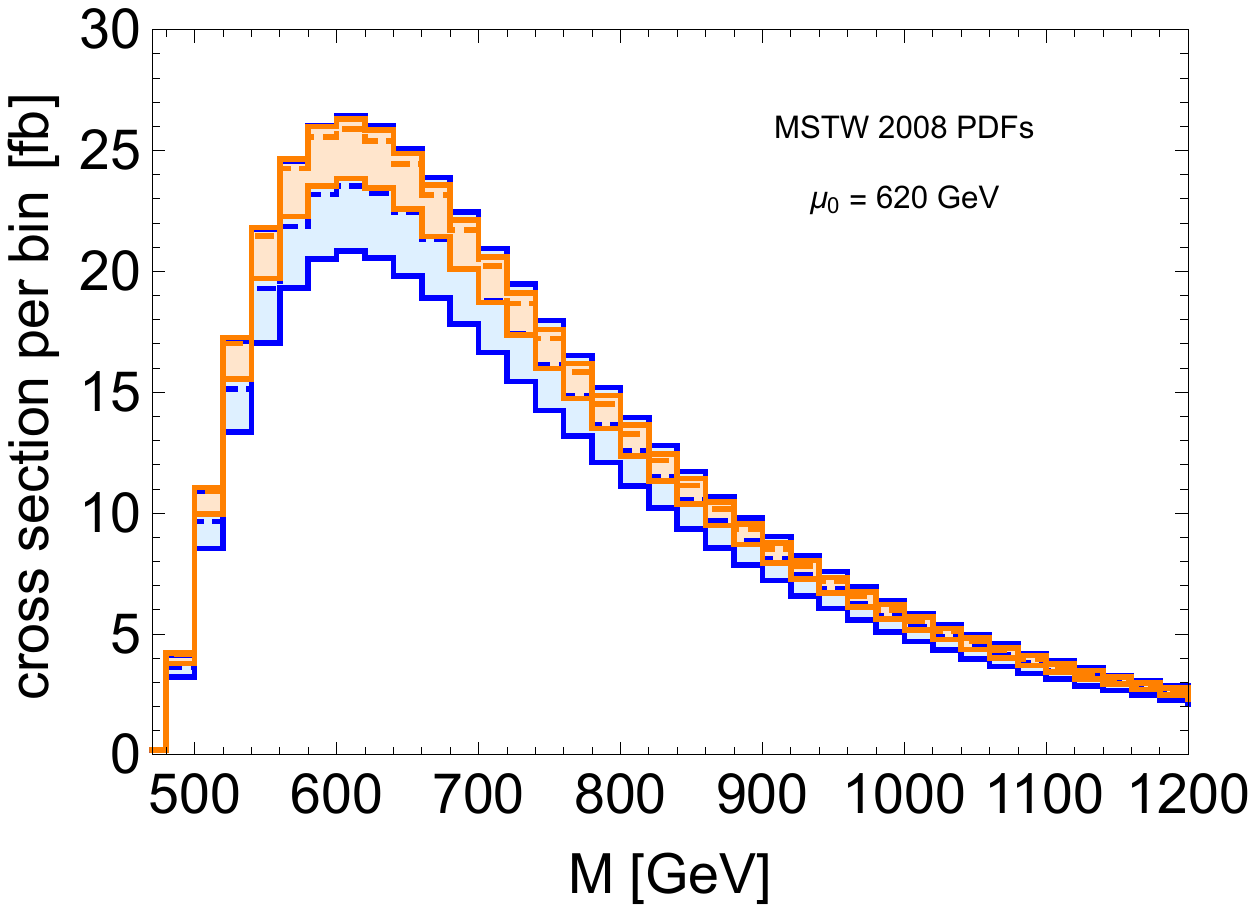} & \includegraphics[width=7cm]{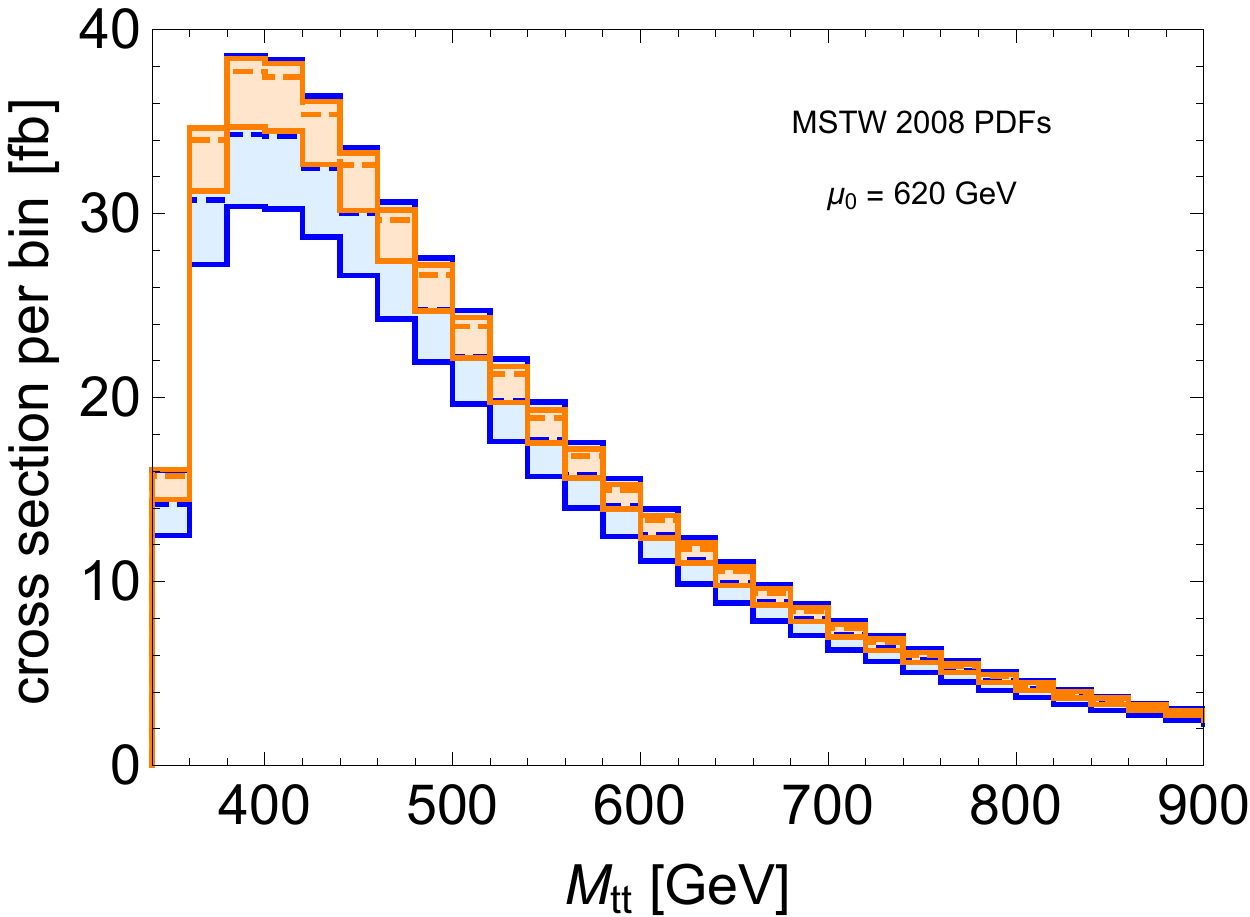} \\
			\includegraphics[width=7cm]{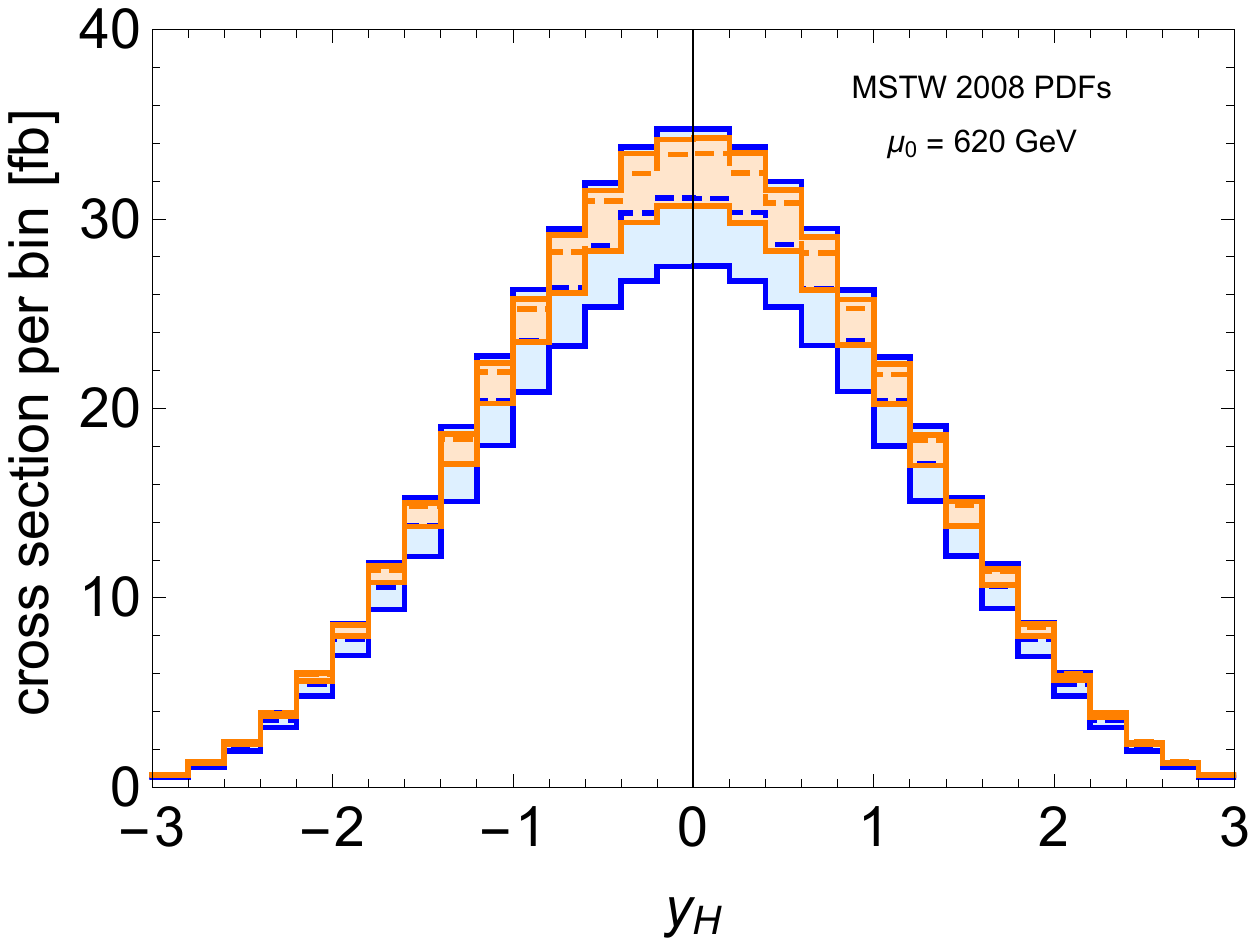} & \includegraphics[width=7cm]{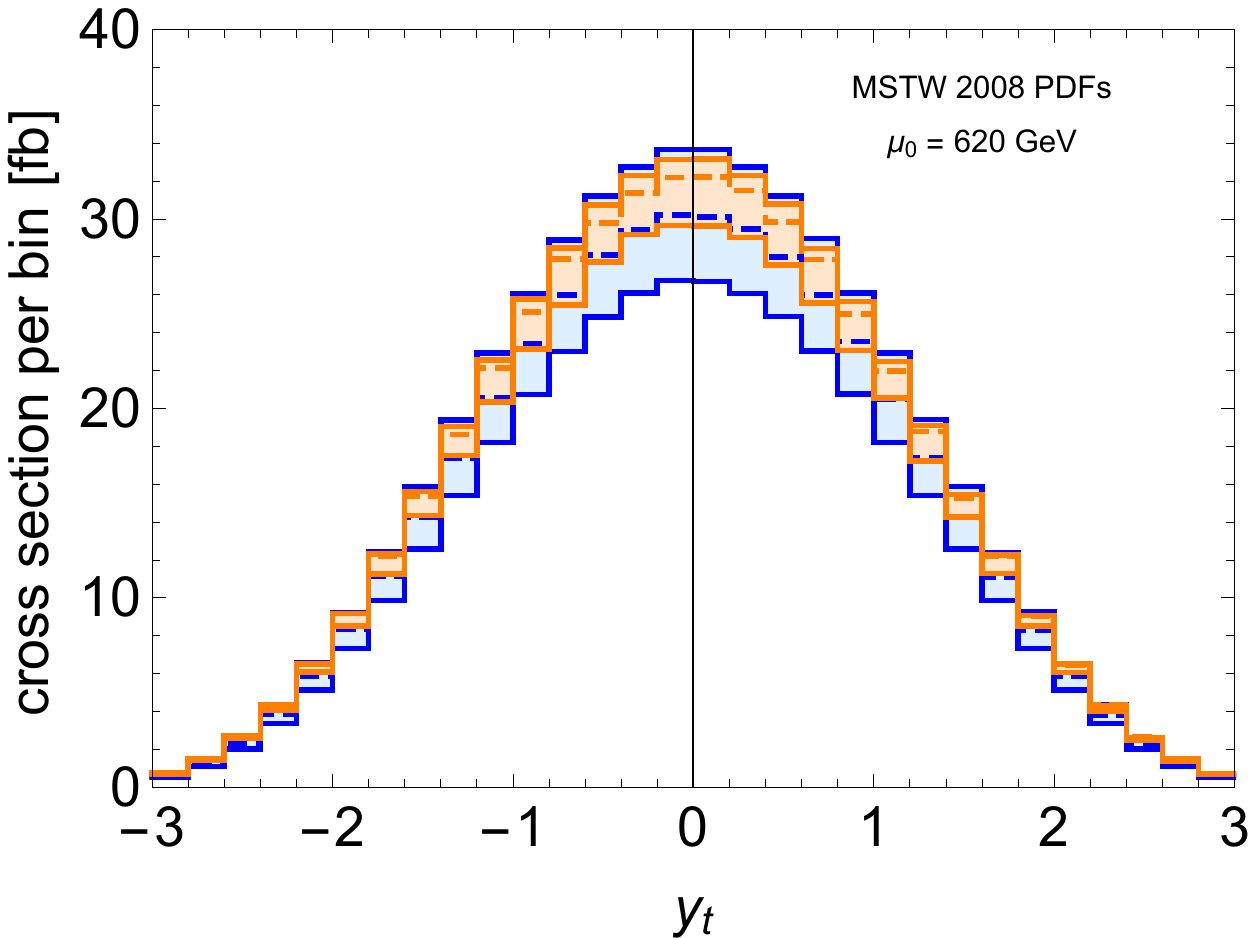} \\
		\end{tabular}
	\end{center}
	\caption{Differential distributions at nNLO (orange band) compared to the NLO calculation carried out with \texttt{MG5}  (blue band). In this case   the uncertainty bands are obtained by considering different sets of subleading corrections and by varying the scale in the range $[ \mu_0 /2 ,2 \mu_0 ]$. NLO distributions are evaluated with NLO PDFs, nNLO distributions with NNLO PDFs.  \label{fig:distnnLOvsNLOLB}
	}
\end{figure}

In Figure~\ref{fig:distnnLOvsNLO} we show nNLO differential
distributions along with the complete NLO results and including
uncertainties from scale variation.  In all cases the effects of the
higher-order corrections contained in the nNLO distributions are quite
similar -- they enhance the results in the individual bins to the
upper portion of the NLO uncertainty band, and greatly reduce the
width of the bands obtained by scale variation.  A more conservative
estimate of the residual perturbative uncertainty affecting our
predictions is shown in Figure~\ref{fig:distnnLOvsNLOLB}. In this case
the uncertainty bands are obtained by following the same procedure
already employed in the calculations in Table~\ref{tab:LBCS} and in
Figure~\ref{fig:distNLO620large}.  The same features observed in
Figure~\ref{fig:distnnLOvsNLO} are found also in
Figure~\ref{fig:distnnLOvsNLOLB}, namely the nNLO band is located in
the upper portion of the NLO uncertainty band in all the distributions
which we considered.  However, in Figure~\ref{fig:distnnLOvsNLOLB} the
nNLO bands are larger than in Figure~\ref{fig:distnnLOvsNLO} and about
half as large as the NLO scale variation bands, which are also shown
in Figure~\ref{fig:distnnLOvsNLOLB}.

\begin{figure}[t]
	\begin{center}
		\begin{tabular}{cc}
			\includegraphics[width=7cm]{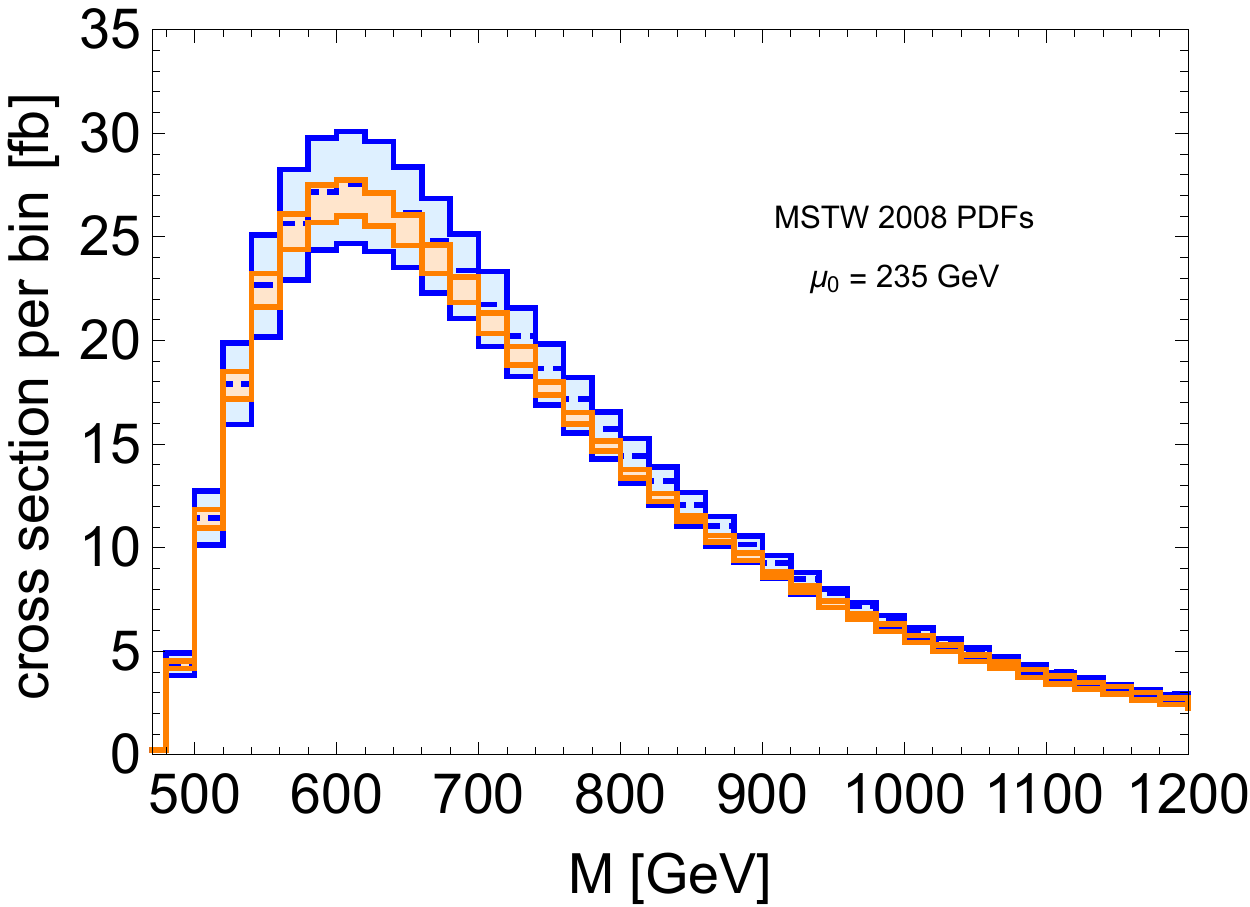} & \includegraphics[width=7cm]{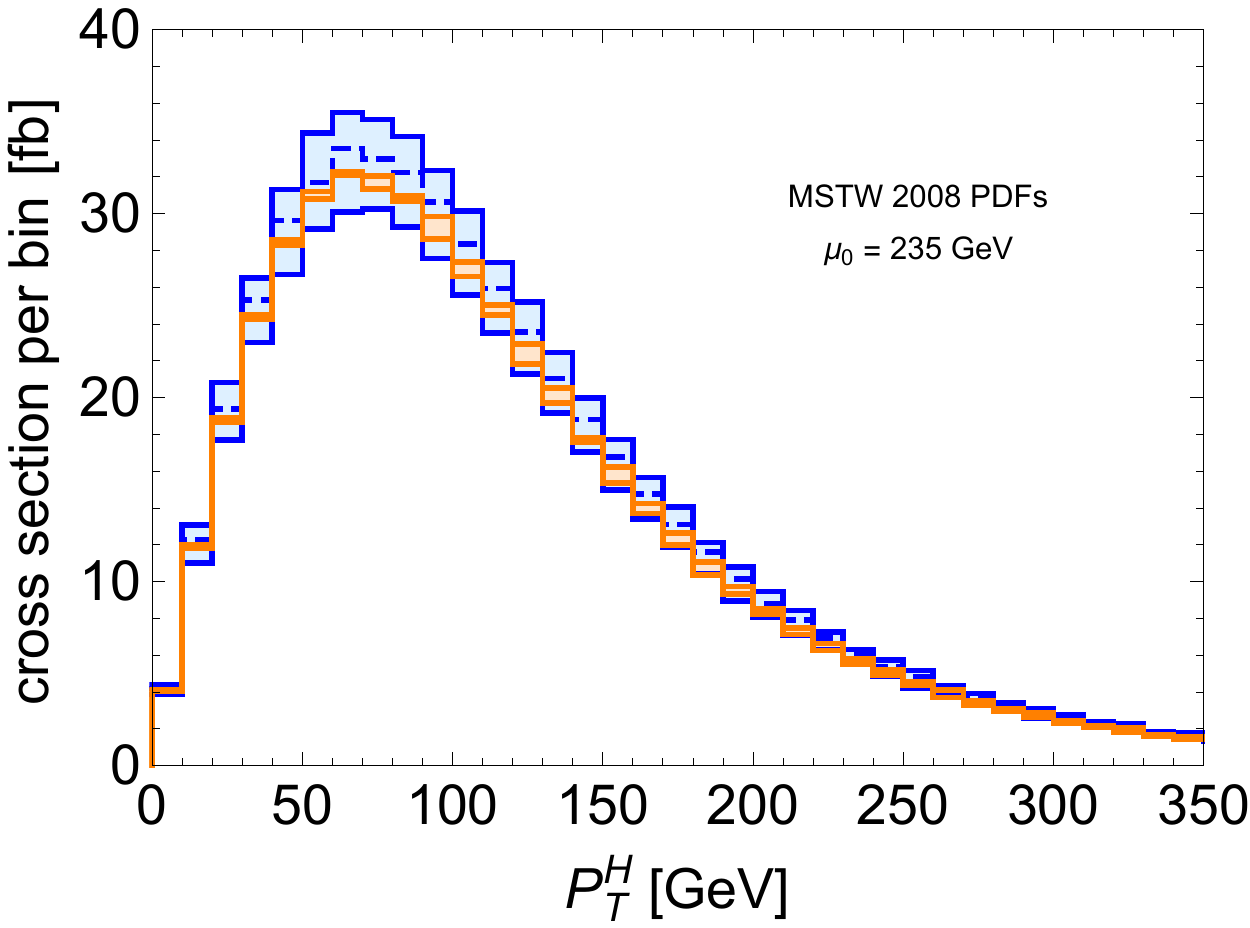} \\
		\end{tabular}
	\end{center}
	\caption{Differential distributions at nNLO (orange band) compared to the complete NLO calculation carried out with \texttt{MG5}  (blue band). The uncertainty bands are obtained  by varying the scale $\mu_0 = 235$~GeV in the range $[ \mu_0 /2 ,2 \mu_0 ]$. NLO distributions are evaluated with NLO PDFs, nNLO distributions with NNLO PDFs.  \label{fig:distnNLOvsNLO235}
	}
\end{figure}

\begin{figure}[t]
	\begin{center}
		\begin{tabular}{cc}
			\includegraphics[width=7cm]{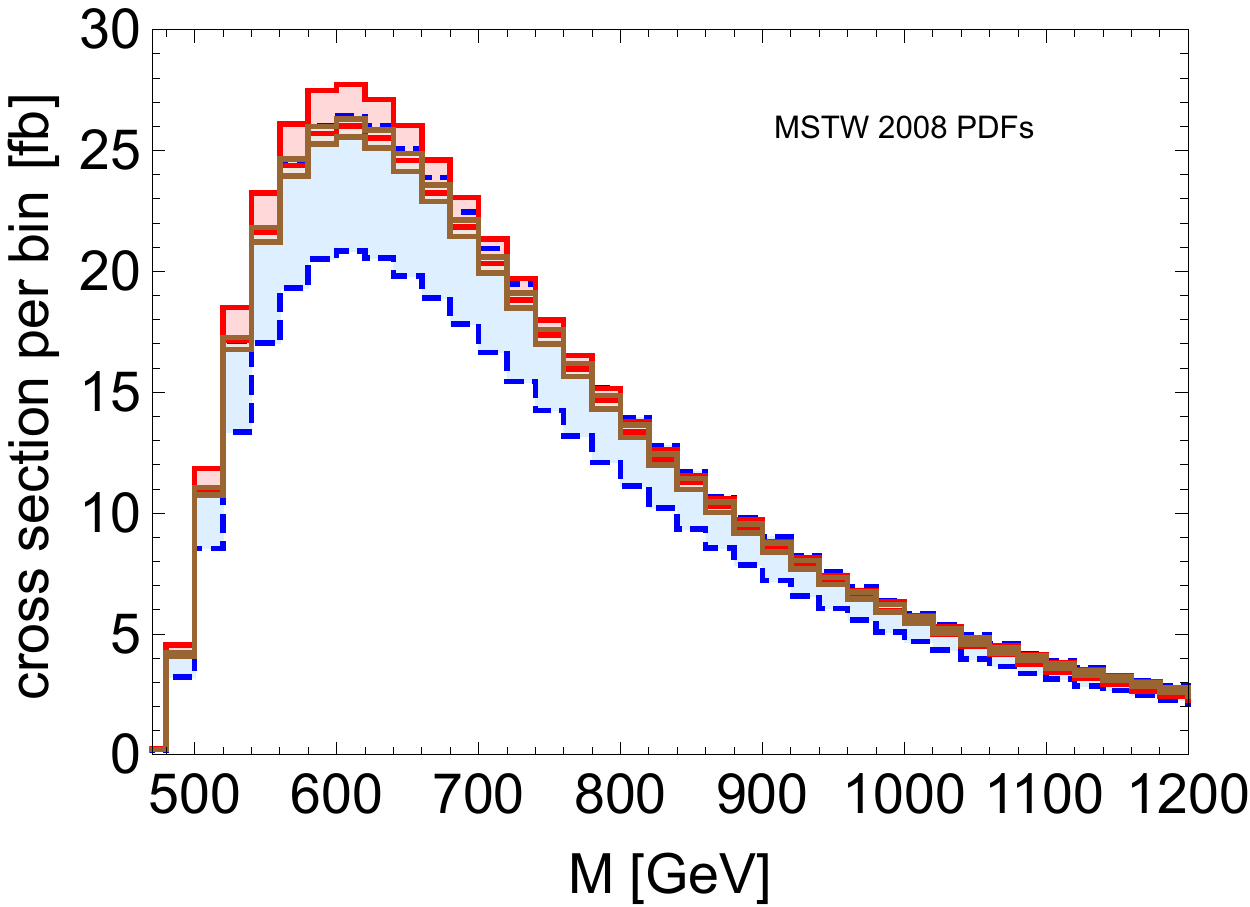} & \includegraphics[width=7cm]{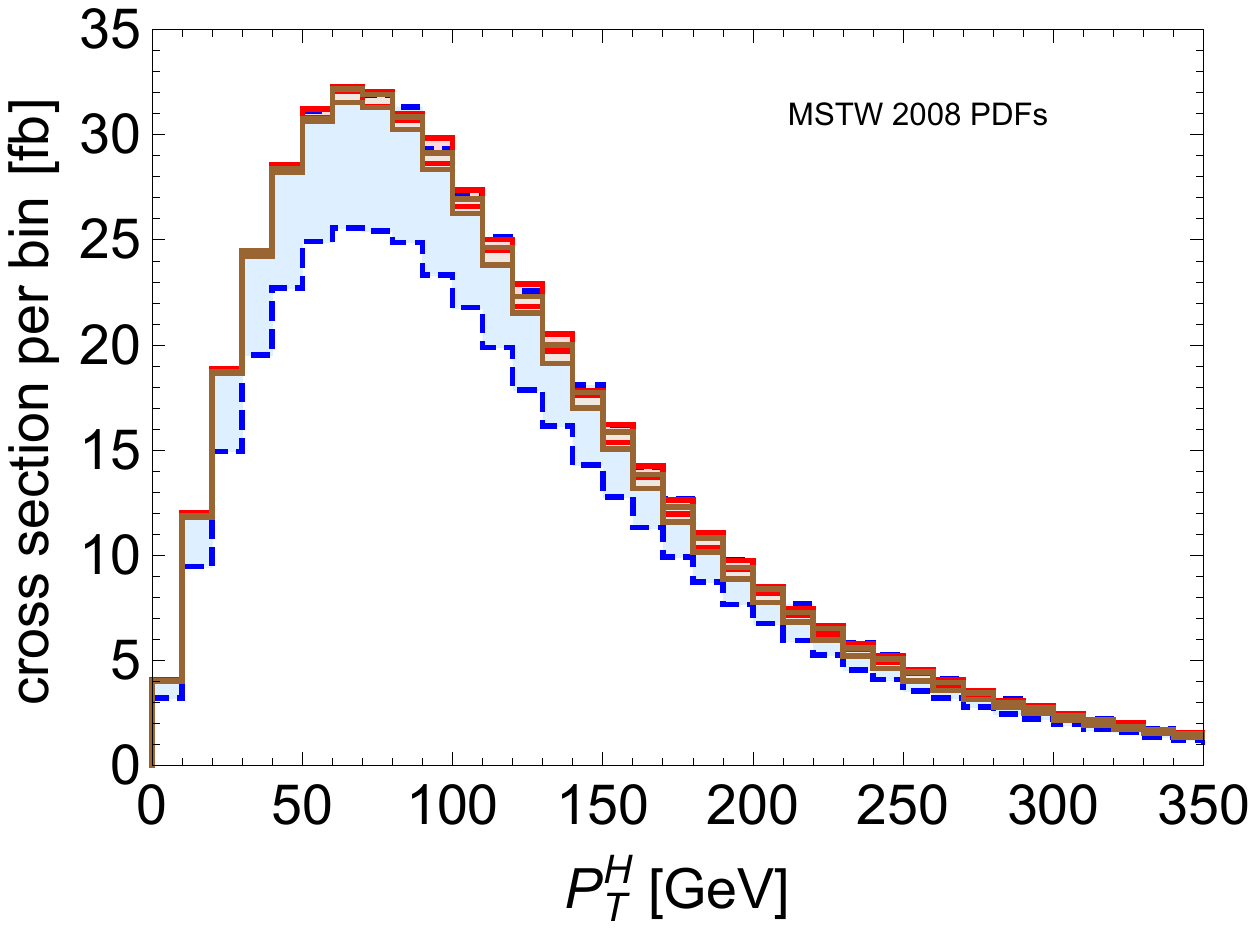} \\
			\includegraphics[width=7cm]{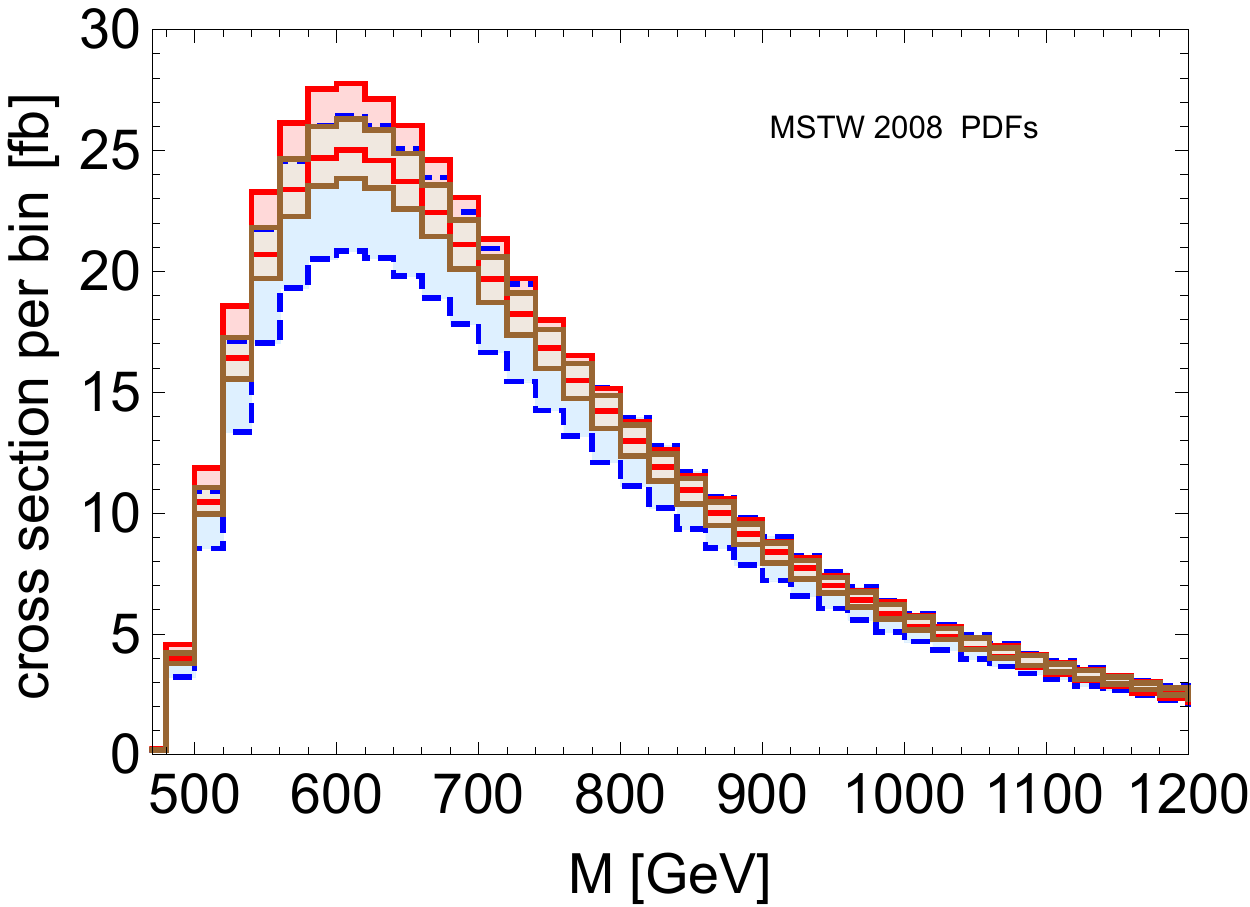} & \includegraphics[width=7cm]{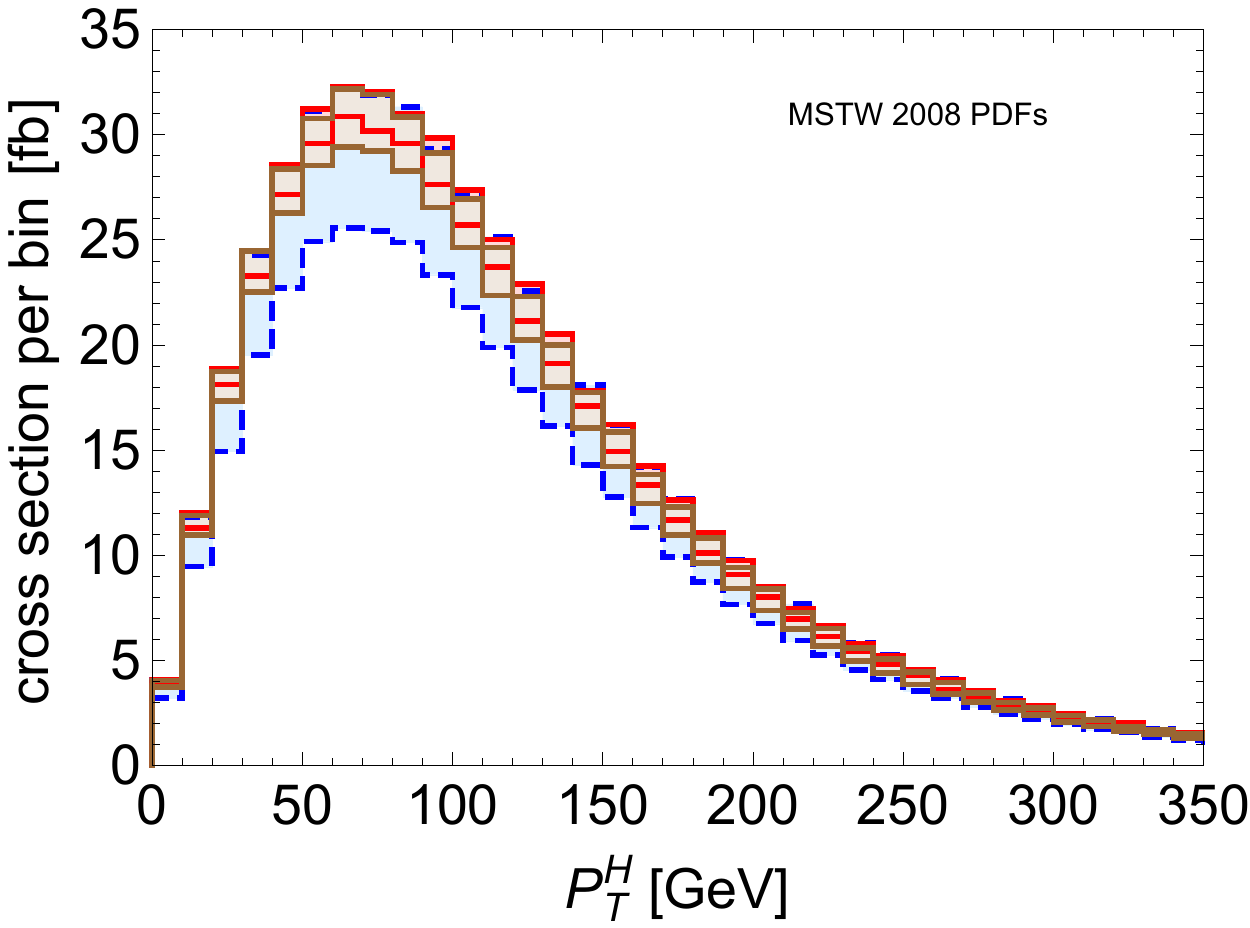} \\
		\end{tabular}
	\end{center}
	\caption{Differential distributions at nNLO calculated by setting $\mu_0 = 235$~GeV (red band)
		and by setting $\mu_0 = 620$~GeV (brown band). The bands in the first row are obtained by scale variation only, while the bands in the second row account for the effects of different sets of subleading terms according to the method explained in the text.
		 The NLO distributions evaluated by setting $\mu_0 = 620$~GeV  (blue band) are shown in the background. NLO distributions are evaluated with NLO PDFs, nNLO distributions with NNLO PDFs. \label{fig:distnNLO620vsnNLO235}
	}
\end{figure}

\begin{figure}[t]
	\begin{center}
		\begin{tabular}{cc}
			\includegraphics[width=7cm]{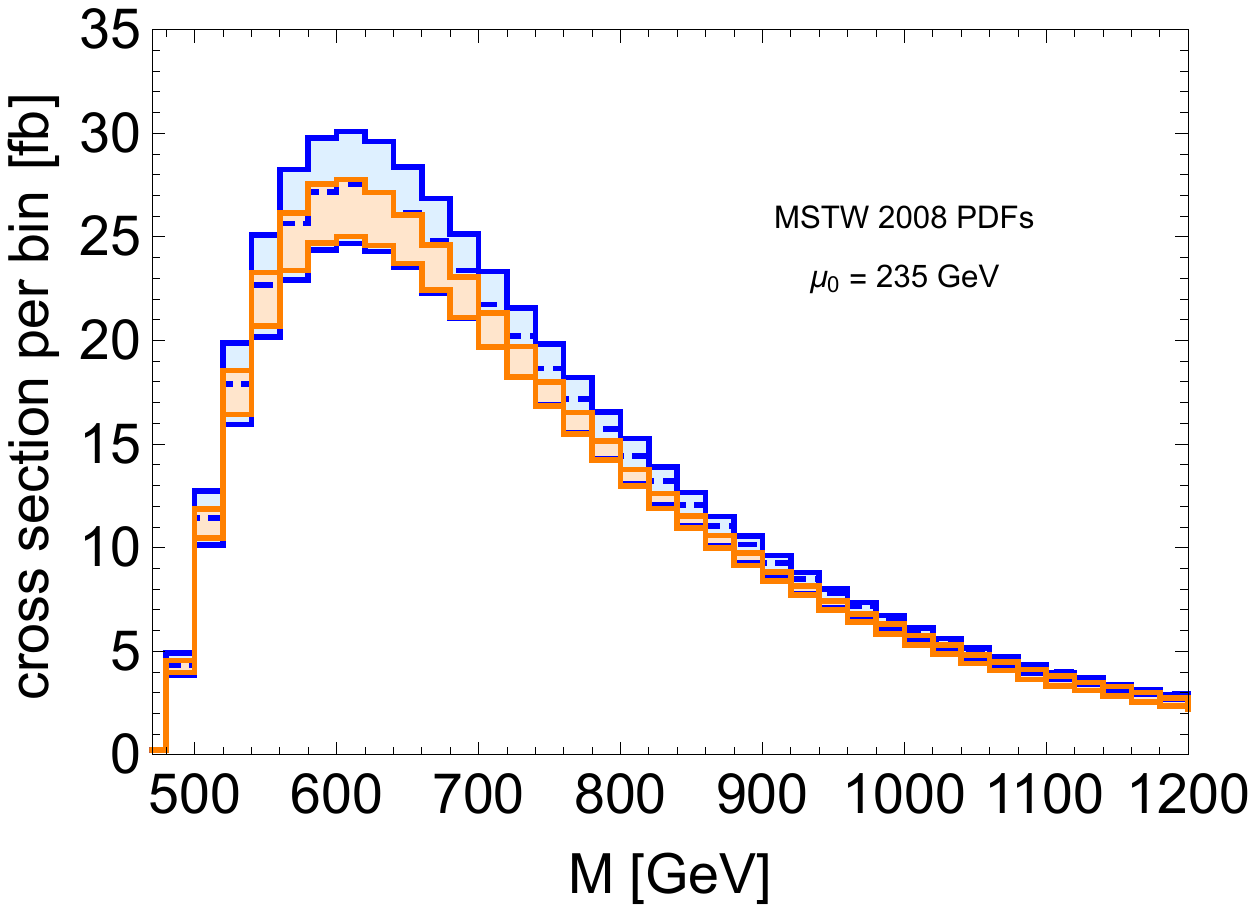} & \includegraphics[width=7cm]{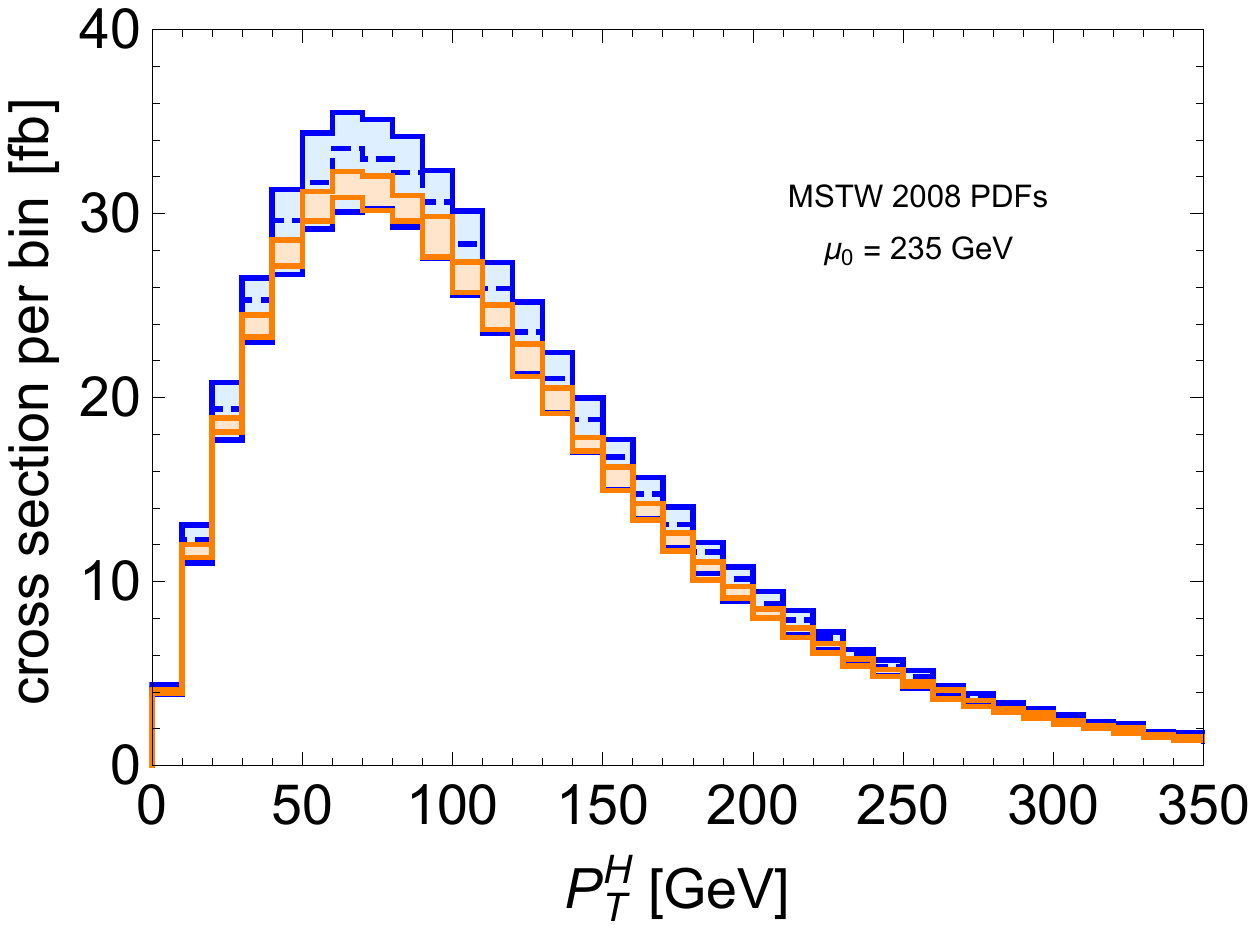} \\
		\end{tabular}
	\end{center}
	\caption{Differential distributions at nNLO (orange band) compared to the NLO calculation carried out with \texttt{MG5}  (blue band) for the choice $\mu_0 = 235$~GeV.   The uncertainty bands are obtained  by varying the scale and by estimating the effect of subleading terms as explained in the text. NLO distributions are evaluated with NLO PDFs, nNLO distributions with NNLO PDFs.  \label{fig:distnNLOvsNLO235LB}
	}
\end{figure}

For completeness, in Figure~\ref{fig:distnNLOvsNLO235} we show the invariant mass and Higgs transverse momentum distributions at nNLO calculated by setting $\mu_0 = 235$~GeV. The uncertainty bands are obtained by varying the scale in the range $[\mu_0/2, 2 \mu_0]$. While this particular choice of the default scale is not ideal because of the issues mentioned above, we see that the nNLO bands fall in the middle of the NLO uncertainty band obtained by setting $\mu_0 = 235$~GeV. It is however interesting to compare the nNLO scale uncertainty bands obtained by choosing $\mu_0 = 235$~GeV with the ones obtained by choosing $\mu_0 = 620$~GeV. This is done in the upper panels of Figure~\ref{fig:distnNLO620vsnNLO235} for the invariant mass and the Higgs transverse momentum distribution. One can see that the thin nNLO scale variation bands obtained by choosing $\mu_0 = 235$~GeV and $\mu_0 = 620$~ GeV have a large overlap, which means that the predictions for the nNLO differential distributions show little sensitivity to the choice of $\mu_0$. The NLO scale variation band for $\mu_0=620$~GeV is shown in the background for reference. However, we stress that, as discussed above, the nNLO bands obtained by scale variation alone are likely to underestimate the residual perturbative uncertainty affecting our result. Consequently, we regard the results at $\mu_0 = 620$~GeV with the conservative estimate of the residual perturbative uncertainty shown in 
Figure~\ref{fig:distnnLOvsNLOLB}  as our best estimates for the nNLO differential distributions. In the lower panels of Figure~\ref{fig:distnNLO620vsnNLO235} we repeat the analysis shown in the upper panels, but we show the larger uncertainty bands obtained by considering the effects of two different sets of subleading corrections, as explained above. Also in this case we find that the band corresponding to the choice $\mu_0 = 620$~GeV has a significant overlap with the band corresponding to the choice $\mu_0 = 235$~GeV. This fact again indicates that the nNLO predictions have little sensitivity to the choice of $\mu_0$. Finally, for reference, we compare the 
nNLO differential distributions with larger bands and $\mu_0 = 235$~GeV to the corresponding complete NLO differential distributions in Figure~\ref{fig:distnNLOvsNLO235LB}.

We would like to conclude the
discussion of the results presented here by emphasizing that the power
of our approach is that it can be used to calculate arbitrary
differential distributions at nNLO accuracy, a fact that was
demonstrated in this section. Furthermore, cuts on the momenta of the
final-state particles can easily be applied, allowing for a more
direct comparison with experimental results.

\section{Conclusions}
\label{sec:conclusions}

We have studied soft-gluon corrections to the total and differential
cross sections for $t\bar{t}H$ hadroproduction. The starting point was
a factorization formula for the differential cross section in the PIM
threshold limit $z\to 1$, which we derived by adapting results 
from $t\bar{t}$ production.  We then collected the perturbative
ingredients needed to use this formula to perform soft-gluon resummation
to NNLL order: namely the hard functions, soft functions, and soft
anomalous dimensions in the quark annihilation and gluon fusion
channels to NLO.  While the soft functions and soft anomalous
dimensions could be obtained quite easily from results in the
literature, our calculation of the hard function to NLO is new. We
performed and cross-checked this calculation by customizing the
automated one-loop amplitude providers {\tt GoSam}, {\tt MadLoop}, and
{\tt Openloops} to extract the color-decomposed amplitudes which
define the hard function. The customized programs could easily be
applied to calculate the NLO hard functions for any other $2\to 3$
process involving four colored partons, and thus provide an essential 
building block for NNLL resummations for such processes.  

As a first application of our formalism to phenomenology, we studied
the soft-gluon corrections in the form of approximate NNLO formulas,
which are a fixed-order truncation of the resummed results. In
particular, we implemented the NNLO corrections obtained from the
soft-gluon resummation formalism into a bespoke parton-level Monte
Carlo program, which can be used to calculate the total cross section
along with arbitrary differential distributions depending on the
momenta of the massive final-state particles.  We illustrated the
functionality of this tool in Section~\ref{sec:pheno} by studying
numerically the soft emission NNLO corrections to six different
differential distributions at the LHC with collider energy 13~TeV, and
matched these corrections with the exact NLO distributions from the
event generator \verb|MadGraph5_aMC@NLO| in order to obtain the
above-mentioned approximate NNLO results, which we labeled nNLO.  For
the choice of the factorization scale to $\mu=\mu_0=620$~GeV, which is
roughly at the peak of the distribution in the invariant mass of the
$t\bar{t}H$ final state, we observed that approximate NNLO corrections
enhance the (differential) cross sections compared to NLO. The same is
true for results evaluated with $\mu= \mu_0/2$ and $\mu=2\mu_0$,
although the nNLO results are much less sensitive to the choice of
factorization scale than the NLO ones.  This is best seen through an
examination of the total cross sections at different perturbative
orders listed in Table~\ref{tab:CSat620}, and through the plots of
binned distributions in Figure~\ref{fig:distnnLOvsNLO}. Although
kinematic cuts were not applied in our calculations, they can be
implemented in the Monte-Carlo code in a straightforward way.

The current paper is a significant step forward in the study of
higher-order QCD corrections in $t\bar{t}H$ production, but it is
important to emphasize that there are still open issues. The
overarching question is to what extent the NNLO corrections generated
from the NNLL soft-gluon resummation formula approximate the true, as
yet unknown NNLO corrections.  We showed in Section~\ref{sec:pheno}
that the NLO corrections in the soft limit approximate quite well the
full NLO results, and while this speaks in favor of estimating NNLO
corrections using the soft limit there is no guarantee that the level
of agreement seen at NLO persists at higher orders. For this reason, a
conservative use of the nNLO results calculated here would require
uncertainty estimates beyond the very small dependence on the
factorization scale which we observed.

We addressed this is issue in Section~\ref{sec:pheno}. We obtained a
more conservative estimate of the residual perturbative uncertainty
affecting the nNLO predictions by evaluating the total and
differential cross sections by including two different sets of terms
which are formally subleading in the soft limit, which correspond to
choose two different forms for the $z$ dependence of the integrand in
(\ref{eq:soft-fact}). For each of these two choices we evaluated the
scale variation in the usual way. This provided us with six
evaluations for each physical quantity. For the total cross section
and for each bin in each distribution, the uncertainty was then
determined by looking at the interval between the largest and smallest
value obtained in the six calculations. This procedure, when applied
to approximate NLO calculations, leads to predictions which are very
close to the complete NLO predictions. When applied to nNLO this
procedure leads to predictions which are slightly larger than the NLO
ones both in the case of the total cross section and in the case of
the six differential distributions that we considered in this
work. The nNLO uncertainty interval for the total cross section and
the bands for the differential distribution obtained with this method
are roughly half as large as the ones obtained by evaluating these
quantities at NLO.

Especially for the differential distributions, it
would  be very useful to gain further insight into the
soft-gluon corrections by implementing the true NNLL resummation
instead of its NNLO approximation.  We plan to return to this
computationally expensive task in future work.
Finally, our code has the potential to be upgraded to include  the decays of the top quarks and Higgs boson, so that it could be used to evaluate observables with cuts on momenta of the detected particles.

\section*{Acknowledgments}
The in-house Monte Carlo code which we developed and employed to
evaluate the (differential) cross sections presented in this paper was
run on the computer cluster of the Center for Theoretical Physics at
the Physics Department of New York City College of Technology.

We thank N. Greiner, T. Hahn, V. Hirschi, P. Maierhofer, G. Ossola and
the {\tt Collier} authors for useful discussions and support in the
use of their codes.  We are grateful to A. Papanastasiou for his help
in excluding the $qg$-channel from the calculations carried out with
{\tt MG5}, and to M. Spira for providing us with values of the NLO
total cross section which we used to debug and test our calculations
at the beginning of this project.

The research activity of
A.F. is supported in part by the National Science Foundation Grant 
No. PHY-1417354 and by the PSC-CUNY Research Award 67616-00 45.
A.B. and A.F. would like to thank M. Passera and INFN Padova 
for their kind hospitality during their visit at Universit\`a di Padova in July 2015, where part of the work presented here was carried out.


\begin{thebibliography}{99}
\bibliographystyle{JHEP}
\bibitem{Beenakker:2001rj} 
  W.~Beenakker, S.~Dittmaier, M.~Kramer, B.~Plumper, M.~Spira and P.~M.~Zerwas,
  Phys.\ Rev.\ Lett.\  {\bf 87}, 201805 (2001)
  [hep-ph/0107081].



\bibitem{Beenakker:2002nc} 
  W.~Beenakker, S.~Dittmaier, M.~Kramer, B.~Plumper, M.~Spira and P.~M.~Zerwas,
  Nucl.\ Phys.\ B {\bf 653}, 151 (2003)
  [hep-ph/0211352].



\bibitem{Reina:2001sf} 
  L.~Reina and S.~Dawson,
  Phys.\ Rev.\ Lett.\  {\bf 87}, 201804 (2001)
  [hep-ph/0107101].



\bibitem{Reina:2001bc} 
  L.~Reina, S.~Dawson and D.~Wackeroth,
  Phys.\ Rev.\ D {\bf 65}, 053017 (2002)
  [hep-ph/0109066].



\bibitem{Dawson:2002tg} 
  S.~Dawson, L.~H.~Orr, L.~Reina and D.~Wackeroth,
  Phys.\ Rev.\ D {\bf 67}, 071503 (2003)
  [hep-ph/0211438].



\bibitem{Dawson:2003zu} 
  S.~Dawson, C.~Jackson, L.~H.~Orr, L.~Reina and D.~Wackeroth,
  Phys.\ Rev.\ D {\bf 68}, 034022 (2003)
  [hep-ph/0305087].



\bibitem{Frederix:2011zi} 
  R.~Frederix, S.~Frixione, V.~Hirschi, F.~Maltoni, R.~Pittau and P.~Torrielli,
  Phys.\ Lett.\ B {\bf 701}, 427 (2011)
  [arXiv:1104.5613 [hep-ph]].



\bibitem{Garzelli:2011vp} 
  M.~V.~Garzelli, A.~Kardos, C.~G.~Papadopoulos and Z.~Trocsanyi,
  Europhys.\ Lett.\  {\bf 96}, 11001 (2011)
  [arXiv:1108.0387 [hep-ph]].



\bibitem{Yu:2014cka} 
  Y.~Zhang, W.~G.~Ma, R.~Y.~Zhang, C.~Chen and L.~Guo,
  Phys.\ Lett.\ B {\bf 738}, 1 (2014)
  [arXiv:1407.1110 [hep-ph]].



\bibitem{Frixione:2014qaa} 
  S.~Frixione, V.~Hirschi, D.~Pagani, H.~S.~Shao and M.~Zaro,
  JHEP {\bf 1409}, 065 (2014)
  [arXiv:1407.0823 [hep-ph]].



\bibitem{Frixione:2015zaa} 
  S.~Frixione, V.~Hirschi, D.~Pagani, H.-S.~Shao and M.~Zaro,
  JHEP {\bf 1506}, 184 (2015)
  [arXiv:1504.03446 [hep-ph]].



\bibitem{Hartanto:2015uka} 
  H.~B.~Hartanto, B.~Jager, L.~Reina and D.~Wackeroth,
  Phys.\ Rev.\ D {\bf 91}, no. 9, 094003 (2015)
  [arXiv:1501.04498 [hep-ph]].

\bibitem{Denner:2015yca}
  A.~Denner and R.~Feger,
  arXiv:1506.07448 [hep-ph].


\bibitem{Kulesza:2015vda} 
  A.~Kulesza, L.~Motyka, T.~Stebel and V.~Theeuwes,
  arXiv:1509.02780 [hep-ph].



\bibitem{Ahrens:2010zv} 
  V.~Ahrens, A.~Ferroglia, M.~Neubert, B.~D.~Pecjak and L.~L.~Yang,
  JHEP {\bf 1009}, 097 (2010)
  [arXiv:1003.5827 [hep-ph]].



\bibitem{Becher:2014oda} 
  T.~Becher, A.~Broggio and A.~Ferroglia,
  arXiv:1410.1892 [hep-ph].



\bibitem{Ahrens:2011mw} 
  V.~Ahrens, A.~Ferroglia, M.~Neubert, B.~D.~Pecjak and L.~L.~Yang,
  JHEP {\bf 1109}, 070 (2011)
  [arXiv:1103.0550 [hep-ph]].



\bibitem{Ahrens:2011uf} 
  V.~Ahrens, A.~Ferroglia, M.~Neubert, B.~D.~Pecjak and L.~L.~Yang,
  Phys.\ Rev.\ D {\bf 84}, 074004 (2011)
  [arXiv:1106.6051 [hep-ph]].



\bibitem{Beneke:2010da} 
  M.~Beneke, P.~Falgari and C.~Schwinn,
  Nucl.\ Phys.\ B {\bf 842}, 414 (2011)
  [arXiv:1007.5414 [hep-ph]].



\bibitem{Beneke:2011mq} 
  M.~Beneke, P.~Falgari, S.~Klein and C.~Schwinn,
  Nucl.\ Phys.\ B {\bf 855}, 695 (2012)
  [arXiv:1109.1536 [hep-ph]].



\bibitem{Beneke:2012wb} 
  M.~Beneke, P.~Falgari, S.~Klein, J.~Piclum, C.~Schwinn, M.~Ubiali and F.~Yan,
  JHEP {\bf 1207}, 194 (2012)
  [arXiv:1206.2454 [hep-ph]].



\bibitem{Ahrens:2011px} 
  V.~Ahrens, A.~Ferroglia, M.~Neubert, B.~D.~Pecjak and L.~L.~Yang,
  Phys.\ Lett.\ B {\bf 703}, 135 (2011)
  [arXiv:1105.5824 [hep-ph]].



\bibitem{Li:2014ula} 
  H.~T.~Li, C.~S.~Li and S.~A.~Li,
  Phys.\ Rev.\ D {\bf 90}, no. 9, 094009 (2014)
  [arXiv:1409.1460 [hep-ph]].



\bibitem{Ferroglia:2009ep} 
  A.~Ferroglia, M.~Neubert, B.~D.~Pecjak and L.~L.~Yang,
  Phys.\ Rev.\ Lett.\  {\bf 103}, 201601 (2009)
  [arXiv:0907.4791 [hep-ph]].



\bibitem{Ferroglia:2009ii} 
  A.~Ferroglia, M.~Neubert, B.~D.~Pecjak and L.~L.~Yang,
  JHEP {\bf 0911}, 062 (2009)
  [arXiv:0908.3676 [hep-ph]].



\bibitem{Cullen:2011ac} 
  G.~Cullen, N.~Greiner, G.~Heinrich, G.~Luisoni, P.~Mastrolia, G.~Ossola, T.~Reiter and F.~Tramontano,
  Eur.\ Phys.\ J.\ C {\bf 72}, 1889 (2012)
  [arXiv:1111.2034 [hep-ph]].



\bibitem{Cullen:2014yla} 
  G.~Cullen {\it et al.},
  Eur.\ Phys.\ J.\ C {\bf 74}, no. 8, 3001 (2014)
  [arXiv:1404.7096 [hep-ph]].



\bibitem{Cascioli:2011va} 
  F.~Cascioli, P.~Maierhofer and S.~Pozzorini,
  Phys.\ Rev.\ Lett.\  {\bf 108}, 111601 (2012)
  [arXiv:1111.5206 [hep-ph]].



\bibitem{Hirschi:2011pa} 
  V.~Hirschi, R.~Frederix, S.~Frixione, M.~V.~Garzelli, F.~Maltoni and R.~Pittau,
  JHEP {\bf 1105}, 044 (2011)
  [arXiv:1103.0621 [hep-ph]].



\bibitem{Farhi:2015jca} 
  D.~Farhi, I.~Feige, M.~Freytsis and M.~D.~Schwartz,
  arXiv:1507.06315 [hep-ph].



\bibitem{Becher:2007ty} 
  T.~Becher, M.~Neubert and G.~Xu,
  JHEP {\bf 0807}, 030 (2008)
  [arXiv:0710.0680 [hep-ph]].


\bibitem{Alwall:2014hca} 
  J.~Alwall {\it et al.},
  JHEP {\bf 1407}, 079 (2014)
  [arXiv:1405.0301 [hep-ph]].



\bibitem{Broggio:2014yca} 
  A.~Broggio, A.~S.~Papanastasiou and A.~Signer,
  JHEP {\bf 1410}, 98 (2014)
  [arXiv:1407.2532 [hep-ph]].



\bibitem{Ahrens:2009uz} 
  V.~Ahrens, A.~Ferroglia, M.~Neubert, B.~D.~Pecjak and L.~L.~Yang,
  Phys.\ Lett.\ B {\bf 687}, 331 (2010)
  [arXiv:0912.3375 [hep-ph]].



\bibitem{Ossola:2006us} 
  G.~Ossola, C.~G.~Papadopoulos and R.~Pittau,
  Nucl.\ Phys.\ B {\bf 763}, 147 (2007)
  [hep-ph/0609007].



\bibitem{Ossola:2007ax} 
  G.~Ossola, C.~G.~Papadopoulos and R.~Pittau,
  JHEP {\bf 0803}, 042 (2008)
  [arXiv:0711.3596 [hep-ph]].



\bibitem{vanDeurzen:2013saa} 
  H.~van Deurzen, G.~Luisoni, P.~Mastrolia, E.~Mirabella, G.~Ossola and T.~Peraro,
  JHEP {\bf 1403}, 115 (2014)
  [arXiv:1312.6678 [hep-ph]].



\bibitem{Peraro:2014cba} 
  T.~Peraro,
  Comput.\ Phys.\ Commun.\  {\bf 185}, 2771 (2014)
  [arXiv:1403.1229 [hep-ph]].



\bibitem{Binoth:2008uq} 
  T.~Binoth, J.-P.~Guillet, G.~Heinrich, E.~Pilon and T.~Reiter,
  Comput.\ Phys.\ Commun.\  {\bf 180}, 2317 (2009)
  [arXiv:0810.0992 [hep-ph]].



\bibitem{Mastrolia:2010nb} 
  P.~Mastrolia, G.~Ossola, T.~Reiter and F.~Tramontano,
  JHEP {\bf 1008}, 080 (2010)
  [arXiv:1006.0710 [hep-ph]].


\bibitem{Denner:2002ii} 
  A.~Denner and S.~Dittmaier,
  Nucl.\ Phys.\ B {\bf 658}, 175 (2003)
  [hep-ph/0212259].


\bibitem{Denner:2005nn} 
  A.~Denner and S.~Dittmaier,
  Nucl.\ Phys.\ B {\bf 734}, 62 (2006)
  [hep-ph/0509141].
  
  \bibitem{Denner:2010tr} 
    A.~Denner and S.~Dittmaier,
    Nucl.\ Phys.\ B {\bf 844}, 199 (2011)
    [arXiv:1005.2076 [hep-ph]].


\bibitem{Denner:2014gla} 
  A.~Denner, S.~Dittmaier and L.~Hofer,
  PoS LL {\bf 2014}, 071 (2014)
  [arXiv:1407.0087 [hep-ph]].



\bibitem{Broggio:2013uba} 
  A.~Broggio, A.~Ferroglia, M.~Neubert, L.~Vernazza and L.~L.~Yang,
  JHEP {\bf 1307}, 042 (2013)
  [arXiv:1304.2411 [hep-ph]].
  

\bibitem{Martin:2009iq} 
A.~D.~Martin, W.~J.~Stirling, R.~S.~Thorne and G.~Watt,
Eur.\ Phys.\ J.\ C {\bf 63}, 189 (2009)
[arXiv:0901.0002 [hep-ph]].

  
\bibitem{Larkoski:2014bxa} 
A.~J.~Larkoski, D.~Neill and I.~W.~Stewart,
JHEP {\bf 1506}, 077 (2015)
doi:10.1007/JHEP06(2015)077
[arXiv:1412.3108 [hep-th]].
  
\bibitem{Bonocore:2014wua} 
D.~Bonocore, E.~Laenen, L.~Magnea, L.~Vernazza and C.~D.~White,
Phys.\ Lett.\ B {\bf 742}, 375 (2015)
doi:10.1016/j.physletb.2015.02.008
[arXiv:1410.6406 [hep-ph]].


\bibitem{Bonocore:2015esa} 
D.~Bonocore, E.~Laenen, L.~Magnea, S.~Melville, L.~Vernazza and C.~D.~White,
JHEP {\bf 1506}, 008 (2015)
doi:10.1007/JHEP06(2015)008
[arXiv:1503.05156 [hep-ph]].


\bibitem{Hahn:2004fe} 
  T.~Hahn,
  Comput.\ Phys.\ Commun.\  {\bf 168}, 78 (2005)
  [hep-ph/0404043].


\end{thebibliography}
\end{document}